\documentclass[11pt,a4paper]{amsart}

\usepackage{fullpage}
\usepackage{mathptmx}
\usepackage{amssymb}
\allowdisplaybreaks
\usepackage{mathrsfs}
\usepackage{booktabs}
\usepackage{longtable}
\usepackage{array}
\usepackage[hidelinks]{hyperref}
\usepackage{bookmark}
\usepackage{aliascnt}
\usepackage[nameinlink,capitalize]{cleveref}

\numberwithin{equation}{section}

\theoremstyle{plain}
\newtheorem{theorem}{Theorem}[section]
\newcommand{\sharedthm}[2]{\newaliascnt{#1}{theorem}\newtheorem{#1}[#1]{#2}\aliascntresetthe{#1}}
\sharedthm{lemma}{Lemma}
\sharedthm{corollary}{Corollary}
\sharedthm{proposition}{Proposition}
\sharedthm{fact}{Fact}

\theoremstyle{definition}
\sharedthm{definition}{Definition}
\sharedthm{problem}{Problem}
\newtheorem{openproblem}{Open Problem}
\sharedthm{example}{Example}

\theoremstyle{remark}
\sharedthm{remark}{Remark}

\crefname{fact}{Fact}{Facts}
\crefname{problem}{Problem}{Problems}
\crefname{openproblem}{Open Problem}{Open Problems}
\crefname{remark}{Remark}{Remarks}
\crefname{equation}{Eq.}{Eqs.}
\crefname{table}{Table}{Tables}

\newcommand{\PSPACE}{\mathsf{PSPACE}}
\newcommand{\NPSPACE}{\mathsf{NPSPACE}}
\newcommand{\NP}{\mathsf{NP}}
\newcommand{\Ptime}{\mathsf{P}}

\newcommand{\zo}{\{0,1\}}
\newcommand{\eps}{\varepsilon}
\newcommand{\enc}[1]{\langle #1 \rangle}
\newcommand{\wt}{\mathrm{wt}}
\newcommand{\blk}{\mathrm{blk}}
\newcommand{\parity}{\mathrm{par}}
\newcommand{\sd}{\mathrm{sd}}
\newcommand{\bin}{\mathrm{bin}}

\newcommand{\K}{\mathsf{K}}
\newcommand{\Kt}{\mathsf{K}^{t}}

\newcommand{\Land}{\mathscr{L}}
\newcommand{\Gland}{\mathscr{G}}
\newcommand{\Eland}{\mathscr{E}}

\title{\textbf{Reconfiguration of Time-Bounded Descriptions}}
\author[Dimitrios Myrisiotis]{Dimitrios Myrisiotis\\
Great Bay University\\
China}
\hypersetup{pdfauthor={Dimitrios Myrisiotis},pdftitle={Reconfiguration of Time-Bounded Descriptions}}

\begin{document}

\begin{abstract}
Time-bounded Kolmogorov complexity measures a string by the length of its shortest program that prints it within a deadline.
An adjunction-universal machine (Bauwens, Kozachinskiy, and Shen, arXiv, 2025) is a Turing machine that simulates every Turing machine $M$ on every input $z$ with polynomial overhead when given a code of $M$, computable from $M$ in polynomial time, followed by $z$ unchanged.
Any two adjunction-universal machines give the same measure up to an additive constant and a polynomial change of the deadline.
However, this measure records only the shortest program for a string.
We study all programs of one length that print a string within one deadline, join two of them when they differ in one bit, and ask when this graph is connected.
Fix any adjunction-universal machine.
For almost all $m$-bit strings and every deadline beyond some polynomial, the graph is disconnected at program length $m+O(1)$ and has at least $2^{m-O(1)}$ components at length $2m+3$.
In contrast, we build an adjunction-universal machine in which every such graph is connected or empty, at every deadline, once the program length reaches a computable threshold that is at most exponential in the length of the string.
Exponential thresholds are unavoidable: for every adjunction-universal machine with such thresholds and every large $m$, more than three quarters of the $m$-bit strings need a threshold exponential in $m$.
For our machine, above the threshold, every program for the string is within three bit flips of a connected set of fastest programs for the string, along a path inside the graph, at every deadline by which that program halts.
For this machine, deciding whether two given programs are connected in this graph is $\mathsf{PSPACE}$-complete when the input is the string, the deadline in unary and the two programs; for each fixed string it takes polynomial time.
\end{abstract}

\maketitle


\begingroup
\makeatletter
\def\@pnumwidth{2.5em}
\def\@tocrmarg{3.5em}
\def\@dotsep{4.5}
\def\l@section#1#2{%
\@dottedtocline{1}{0pt}{2.8em}{\bfseries\numberline{}#1}{#2}}
\def\l@subsection#1#2{%
\@dottedtocline{2}{2.8em}{3.3em}{\numberline{}#1}{#2}}
\renewcommand{\tocsection}[3]{%
\@ifnotempty{#2}{\llap{\hb@xt@\@tempdima{#2.\hfil}}}#3}
\let\tocsubsection\tocsection
\makeatother
\tableofcontents
\endgroup


\section{Introduction}\label{sec:1}

Time-bounded Kolmogorov complexity measures a string by the length of its shortest program on which a fixed universal machine outputs the string within a given number of steps.
Suppose that each of two efficient onto machines can run any machine $M$ by reading a code of $M$ followed by an unchanged program $z$ of $M$, with polynomial overhead; we call this simulation by adjunction.
Then their measures agree up to an additive constant in length and a polynomial change of the deadline (\Cref{fact:2.15}).
That comparison concerns only the shortest description.
It says nothing about how the other descriptions of the same string are arranged, or whether one of them can be turned into another by small changes.

This paper studies that arrangement.
Fix a machine, a target string, a program length and a deadline.
The \emph{fiber} of the target is the set of all programs of that length that halt with the target as output within the deadline, where two programs are joined when they differ in one bit.
We ask whether the fiber is connected, how many components it has, and how hard it is to decide whether two given programs lie in the same component.
In this decision problem the input is the target, the deadline in unary and two programs, and a pair that does not lie in the fiber is a no-instance.

Two features of the question shape the answers.
First, the fiber contains every program of the given length, not only the programs that begin with one particular code.
A path may pass through programs that begin differently, so an argument about the programs beginning with one code, which we call its \emph{cylinder}, does not decide connectivity of the fiber.
Second, connectivity of the fiber is not monotone in the deadline.
A larger deadline keeps every old program and every old path, but newly admitted programs can form new components, so a connected fiber can become disconnected at a larger deadline (for the first machine of \Cref{thm:H}, by \Cref{thm:uniform-reference,thm:eventual-fragmentation}).
A larger program length gives a different cube altogether, rather than a larger subgraph of the old one, so the fragmentation results choose program lengths as functions of the target length.

Graphs of equal-output genotypes under local mutation are studied in the theory of neutral networks.
Reidys, Stadler and Schuster study equal-image classes in genotype--phenotype maps \cite{RSS97}, and Hu, Banzhaf and Ochoa study finite systems whose genotypes are programs \cite{HBO25}.
The questions here differ in their setting: they concern adjunction-universal machines, deadlines counted in the machine's own transitions, and every description of the given length.
Dependence on the choice of machine is also known for other structural properties of descriptions: for the number of descriptions of a string \cite{Tad10}, for reductions to the set of random strings \cite{ABK06a}, and in the distinction between optimality and universality by adjunction \cite{BKS25}.

The results concern three conditions on an efficient machine $V$, stated precisely in \Cref{sec:2.2}.
Besides its program, $V$ may read an auxiliary input; the fibers above use the empty auxiliary input.
The machine is \emph{onto}: every string is the output of some program on the empty auxiliary input.

Next, the machine has the \emph{simulation property}, the precise form of simulation by adjunction: every machine $M$ has a \emph{code} $\enc{M}_V$ for $V$, computable from $M$ in polynomial time.
For every string $z$, the program $\enc{M}_Vz$ outputs $M(z)$, and it diverges when $M$ diverges on $z$.
The simulation takes time polynomial in the running time of $M$ and the lengths of $z$ and $\enc{M}_V$, with one polynomial for all $M$.
Finally, the machine has the \emph{interpretation property}: given as auxiliary input a second code $\enc{M}'_V$ of $M$, also computable from $M$ in polynomial time, it runs $M$ on the program with the same kind of overhead.

We call a machine with the first two properties an \emph{adjunction-universal machine}, and the three properties together the \emph{reference conditions}.
The name follows the term universal by adjunction, which Bauwens, Kozachinskiy and Shen record, quoting Downey and Hirschfeldt, for machines universal in their sense \cite[Sections 2 and 4]{BKS25}.
In their definition the string placed before the unchanged programs of a machine need only exist, and no time bound is imposed; here the code is computable from $M$ in polynomial time and the simulation has polynomial overhead.
The rigid reference $U$ of \Cref{def:2.10}, built on a parity-protected encoding of machines, satisfies all three.
The programs $\enc{M}_Vz$ that an adjunction-universal machine must run are its \emph{mandatory descriptions}.
Deadlines always count transitions of $V$ itself, which we call native transitions, and a subscript $V$, as in $O_V(1)$, marks a constant that may depend on $V$.

In short, adjunction forces the fibers of most targets to fragment permanently at suitable lengths.
Yet it allows a machine whose fibers are all empty or connected, at every deadline, once the program length reaches a threshold depending on the target, and for most targets the length needed is necessarily exponential in the target length.
The same machine shows that deciding connectivity can be hard when the target is part of the input and easy for every fixed target.

\subsection{Our Results}\label{sec:1.1}

\Cref{tab:s3-results} summarizes the results, the program lengths at which they hold, and the distinction between the shape of a fiber and the complexity of querying it.
In the table, ``one polynomial deadline'' means a deadline polynomial in the target length that depends only on the machine, and the exceptional targets are the same at every later deadline.
The informal statements after the table present the main results in words, each with a pointer to its precise counterpart in the body.

\begin{longtable}{@{}>{\raggedright\arraybackslash}p{.20\linewidth}>{\raggedright\arraybackslash}p{.52\linewidth}>{\raggedright\arraybackslash}p{.20\linewidth}@{}}
\caption{Selected guarantees for full program fibers.
Constants in statements about every adjunction-universal machine may depend on the machine.}\label{tab:s3-results}\\
\toprule
Setting & Guarantee and regime & Precise statements\\
\midrule
\endfirsthead
\toprule
Setting & Guarantee and regime & Precise statements\\
\midrule
\endhead
Every adjunction-universal machine & At $N=m+O_V(1)$, a fraction $1-O_V(1/m)$ of $m$-bit targets has nonempty disconnected fibers at every deadline after one polynomial deadline. & \Cref{thm:linear-overhead-fragmentation}\\
Every adjunction-universal machine & At $N=2m+3$, a fraction $1-O_V(\log(m+1)/m)$ of $m$-bit targets has at least $2^{m-O_V(1)}$ components, for every sufficiently large $m$ and every deadline beyond one polynomial deadline. & \Cref{thm:eventual-fragmentation}\\
Every adjunction-universal machine & For $m=2^k$ with $k\ge2$, some length $N<2m+O_V(1)$ has at least $2^{m/2-1}$ components for all but an $O_V(m^{-1/2})$ fraction of its targets, at every deadline beyond one polynomial deadline. & \Cref{thm:component-multiplicity,cor:exponential-components}\\
Two constructed machines satisfying the reference conditions & At one common early linear deadline and length $2m+3$, every sufficiently long target has a connected exponential-size fiber for one machine and an edgeless one for the other. & \Cref{thm:uniform-reference}\\
One constructed machine satisfying the reference conditions & Every fixed target has a computable threshold $N_x\le2^{|x|+2}+1$; all fibers at or above it are empty or connected at every deadline, with diameter at most $2N+4$. & \Cref{thm:eventual-connected-fibers}\\
Any adjunction-universal machine with such thresholds & For every sufficiently large $m$, more than three quarters of the $m$-bit targets require $N_x\ge2^{c_Vm-2}$. & \Cref{thm:exponential-fragmentation-length,cor:exponential-eventual-threshold}\\
Uniform versus fixed-target complexity & The eventual-connectivity machine has a $\PSPACE$-complete full connectivity language, with every fixed-target restriction in $\Ptime$.
It is fixed-parameter tractable in target length; the rigid reference $U$ has a $\PSPACE$-complete fixed-target connectivity problem. & \Cref{thm:uniform-hard-fixed-easy-reference,cor:target-parameter-reference}\\
Globally connected targets & Every finite set is exactly the set of globally connected targets of some machine satisfying the reference conditions.
For every adjunction-universal machine, that set is co-c.e., has computably bounded c.e.\ subsets, and has intrinsic density zero. & \Cref{thm:exact-finite-connected-targets,cor:effective-connected-target-bound,thm:connected-target-list-density,cor:connected-target-intrinsic-density,prop:connected-target-coce}\\
\bottomrule
\end{longtable}

\paragraph{\emph{Mandatory descriptions and persistent fragmentation.}}

An adjunction-universal machine must run every program of every source machine, so a suitable source forces descriptions that the machine cannot avoid.
We choose sources that give each target several mandatory descriptions far apart in Hamming distance, and we count how many other programs, anywhere in the cube, would be needed to connect them.
Here and below, ``almost all'' refers to the fraction of the $m$-bit targets as $m$ tends to infinity.

\begin{theorem}[Informal; \Cref{thm:linear-overhead-fragmentation,thm:eventual-fragmentation,cor:exponential-components}]
\label{thm:preview-fragmentation}
Let $V$ be an adjunction-universal machine.
Each of the following holds at every deadline beyond a deadline polynomial in $m$, where the polynomial depends only on $V$, with the same exceptional targets at all those deadlines.
\begin{enumerate}
\item At program length $m+O_V(1)$, every $m$-bit target has a nonempty fiber, since it has mandatory descriptions there, and all but an $O_V(1/m)$ fraction of these fibers are disconnected.
\item At program length exactly $2m+3$, for every sufficiently large $m$, all but an $O_V(\log(m+1)/m)$ fraction of the $m$-bit targets have fibers with at least $2^{m-O_V(1)}$ components.
\item When $m\ge4$ is a power of two, there is a program length below $2m+O_V(1)$, depending only on $V$ and $m$, that gives at least $2^{m/2-1}$ components for all but an $O_V(m^{-1/2})$ fraction of the $m$-bit targets.
\end{enumerate}
\end{theorem}

The constants hidden in $O_V(1)$ come from the length of the code of a fixed source machine, chosen separately for each part.
The bounds therefore hold for every adjunction-universal machine, whatever its other programs do.

These bounds start only once the mandatory descriptions have halted.
At an earlier deadline a fiber can look entirely different, and there the answer depends on the machine.

\begin{theorem}[Informal; \Cref{thm:uniform-reference,thm:eventual-fragmentation}]
\label{thm:H}
There are two machines satisfying the three reference conditions that disagree about connectivity at program length $2m+3$ and one common deadline linear in $m$.
For every sufficiently large $m$, every $m$-bit target has a connected fiber with $2^{m+1}$ vertices for the first machine and an edgeless fiber with $2^m$ vertices for the second.
By contrast, for every adjunction-universal machine, almost all $m$-bit targets have disconnected fibers at length $2m+3$ at every deadline beyond one polynomial deadline.
\end{theorem}

The two statements are compatible.
A path present at the early deadline is still present later, but programs admitted afterwards form further components.
The later statement concerns connectivity of the whole fiber, not the loss of a path that existed earlier.

\paragraph{\emph{Eventual connectivity.}}

The fragmentation bounds hold at lengths chosen from the target length.
The fibers of a target may be disconnected at those lengths and still be empty or connected at every length beyond a threshold that depends on the target.
Our main construction provides this for all targets and all deadlines at once, and it remains hard to query.

\begin{theorem}[Informal; \Cref{thm:eventual-connected-fibers,thm:uniform-hard-fixed-easy-reference,cor:reference-fixed-target-easy,cor:exponential-eventual-threshold}]
\label{thm:preview-eventual}
There is a machine satisfying the three reference conditions with the following properties.
\begin{enumerate}
\item For every target $x$ one can compute a threshold $N_x\le2^{|x|+2}+1$ such that, at every program length $N\ge N_x$ and every deadline, the fiber of $x$ is empty or connected, with diameter at most $2N+4$.
\item Deciding whether two programs lie in the same fiber component is $\PSPACE$-complete, yet for each fixed target it is decidable in polynomial time.
\end{enumerate}
Conversely, for every adjunction-universal machine with such thresholds and every sufficiently large $m$, more than three quarters of the $m$-bit targets $x$ have $N_x\ge2^{c_Vm-2}$, for a constant $c_V>0$.
\end{theorem}

The lower bound concerns program length, not decision time: it says how long programs must be before every fiber of the target is guaranteed to be empty or connected.
It comes from the fragmentation argument run at an exponentially large program length, where a disconnected fiber persists at every later deadline.

\paragraph{\emph{Fixed and varying targets.}}

The previous theorem separates the complexity of the whole problem from that of each fixed target.
Measuring how the running time may depend on the target length gives a finer comparison, even between machines whose connectivity problems are complete for the same class.

\begin{theorem}[Informal; \Cref{cor:target-parameter-reference,cor:no-single-exponential-target,cor:7.18}]
\label{thm:preview-parameter}
There are two machines satisfying the three reference conditions whose connectivity problems are both $\PSPACE$-complete.
For the first, the machine of \Cref{thm:preview-eventual}, one algorithm runs in time $2^{2^{O(k)}}\ell^{O(1)}$ on queries of length $\ell$ with targets of length $k$.
If $\Ptime$ differs from $\PSPACE$, no algorithm for this machine runs in time $2^{O(k)}\ell^{O(1)}$.
For the second, the problem is already $\PSPACE$-complete at the fixed target $00$, so if $\Ptime$ differs from $\PSPACE$ it is not fixed-parameter tractable in the target length.
\end{theorem}

The single-exponential exclusion holds because the hardness reduction uses targets of logarithmic length (\Cref{cor:no-single-exponential-target}).
No matching double-exponential lower bound on the running time is claimed.

\paragraph{\emph{Targets connected at every length and deadline.}}

Write $G(V)$ for the set of \emph{globally connected} targets, those whose fibers are empty or connected at every program length and every deadline.
This is much stronger than eventual connectivity: one disconnected fiber, at any length and any deadline, excludes a target from $G(V)$.

\begin{theorem}[Informal; \Cref{thm:ce-connected-target-packing,cor:effective-connected-target-bound,thm:connected-target-list-density,cor:connected-target-intrinsic-density,prop:connected-target-coce,thm:exact-finite-connected-targets,thm:finite-target-reference}]
\label{thm:preview-connected-targets}
For every adjunction-universal machine $V$, the set $G(V)$ is co-c.e., its c.e.\ subsets obey a computable bound on their size, and hence it has no infinite c.e.\ subset, and it has intrinsic density zero.
Conversely, for every finite set $S$ of strings there is a machine $V$ satisfying the three reference conditions with $G(V)=S$.
For every such $S$ there is also a machine satisfying the three reference conditions for which the fixed-target connectivity problem is polynomial-time decidable for every target in $S$ and $\PSPACE$-complete for every target outside $S$.
\end{theorem}

If $\Ptime$ differs from $\PSPACE$, the last machine has polynomial-time fixed-target problems exactly for the targets in $S$.
Intrinsic density zero is the notion characterized by Astor through computable injective sampling \cite{Ast15}; the contribution here is that $G(V)$ has it for every adjunction-universal machine $V$, with an explicit rate (\Cref{thm:connected-target-list-density,cor:connected-target-intrinsic-density}).
Fix such a $V$ and a total computable injective list of strings, and let $k$ be the length of one code, which depends on the list.
Then for every $q\ge2$, at most a fraction $O(2^k/\log q)$ of the first $q$ entries lies in $G(V)$.
For an adjunction-universal machine $V$ with $G(V)$ infinite, the computable bound makes $G(V)$ effectively immune, meaning that a computable function of the index of a c.e.\ set bounds the size of every c.e.\ subset of $G(V)$ (\Cref{prop:connected-target-coce}).
Whether $G(V)$ can be infinite for some adjunction-universal machine $V$ is open (\Cref{op:infinite-connected-targets}).

\subsection{Our Techniques}\label{sec:1.2}

The negative and positive results use the mandatory descriptions in opposite ways.
The negative arguments count what every adjunction-universal machine must provide.
The positive construction keeps those programs but controls when the other programs become available and how they attach.

\paragraph{\emph{Counting paths in the full cube.}}

An adjunction-universal machine gives no control over programs outside the cylinder of a code, and any of them might help connect two descriptions.
The negative arguments therefore count those programs instead of excluding them.
A fixed linear-time source maps $0x$ and $1\overline x$ to $x$, and its code has some length $d$.
At program length $m+d+1$, inside the cylinder of that code, its two mandatory descriptions of an $m$-bit target $x$ are the only descriptions of $x$.
A simple path joining them therefore leaves the cylinder and has at least $m+2$ vertices outside it.

Paths for different targets share no vertex, because one deterministic program has one output.
There are only $(2^d-1)2^{m+1}$ programs outside the cylinder of a code of length $d$, so only a fraction $O(2^d/m)$ of the targets can have connected endpoints.
The exceptional targets are defined through the untimed fiber, the graph with no deadline, so one exceptional set serves every later deadline.

To force many components rather than one disconnection, each target needs many mandatory descriptions that are pairwise far apart.
A balanced linear map with well-separated preimage sets, built from finite-field parity checks adapted from the codes of Bose and Ray-Chaudhuri and of Hocquenghem \cite{BRC60,Hoc59}, supplies them.
A traversal of a spanning tree shows that a connected subgraph of the cube containing $k\ge2$ terminals at pairwise distance at least $D$ has at least $kD/2$ vertices.
Applied to each component in turn, this bound shows that most targets must have exponentially many components at length $2m+3$.

To reach exactly length $2m+3$, the source must know the length of its own code, which is chosen only after the source is fixed.
The construction therefore supplies that length as a unary prefix of the raw input, so no source ever needs to compute its own code.
Reed--Muller evaluation codes \cite[Section 2.3]{ASW15} give a second family with an explicit tradeoff among program length, component count and the exceptional fraction.

The same balanced construction, tuned to a program length exponential in the target length, gives the threshold lower bound.
The code of the source is still fixed across all those lengths, so most targets have a disconnected fiber at that length at every later deadline.
Any threshold beyond which every fiber is empty or connected must therefore exceed that length.

\paragraph{\emph{Releasing a connected fiber at every deadline.}}

The construction must make the fiber empty or connected at every deadline, not only when time is unlimited.
A program needed on a connecting path might halt later than the descriptions it is meant to join, and at the intermediate deadlines the fiber could then be disconnected.
The construction therefore admits a connected core first and lets simulated descriptions in only after the core is present.

The core rests on an explicit coloring of the cube of program bodies in which every color class is connected and every vertex has a neighbor of every color.
It combines two syndrome registers with a selector bit; within each color class, paths to a fixed representative have linear length, and the coloring is computed on a schedule depending only on the body length.
The syndrome part is the established coupon-coloring construction \cite{CKTV15}; the connectedness of the classes and the fixed schedule are proved here.
Each color is printed as a short string, and the strings printable at a given body length form its \emph{palette}.

Programs have two mode bits.
Core programs print the color of their body.
In the two helper modes, the parity of the body decides whether a program prints its color, as a router, or runs the simulation, as a source helper.
Every core and router program of one body length halts at one common time.
Base programs and their matching source helpers run the same native simulation of a fixed base machine after a common delay, so they halt together, after the core and the routers are present.
Every admitted program whose output lies in the current palette of colors therefore attaches to the core in at most three steps, which gives empty or connected fibers at every deadline and the diameter bound.
The palette grows with the body length, and every fixed target eventually enters it, which gives a computable threshold exponential in the target length.

Uniform hardness uses targets outside the palette.
No core or router program prints such a target, so its fiber contains only base programs and their source helpers, and each helper is a leaf attached to its base.
A rigid code of a formula verifier in the parity-protected encoding then makes the base components with that code, after deleting their helper leaves, correspond exactly to the components of the solution graph of a formula.
The reduction starts from the $\PSPACE$-complete problem of connecting two satisfying assignments by single-variable changes \cite{GKMP09}.

The reduction picks the target before building the code, although the code length depends on the target.
For a formula instance of encoding length $L$, a target of length a fixed multiple of $\lceil\log_2(L+2)\rceil$ is long enough to lie outside the palette at the resulting body length, so only one verifier and one code are built.
Membership in $\PSPACE$ uses Savitch's recursive reachability method \cite{Sav70}, with the native graph tests proved explicitly here.

\paragraph{\emph{Effective restrictions and finite hubs.}}

To restrict $G(V)$, the path count must be applied to many targets at once, although for general $V$ the paper gives no membership test for $G(V)$, so the argument cannot rely on one.
For a c.e.\ family of targets, one source machine waits for the indexed distinct elements of its enumeration and assigns each an antipodal pair of raw inputs.
At each suffix length, the finitely many successful endpoints halt by a common deadline, whether or not that deadline can be computed.
If all these targets lie in $G(V)$, the outside-path count applies to all of them together.
Choosing the suffix length from the code turns this into a computable bound on the size of c.e.\ subsets, and running the same source on a total injective list gives the density rate.

A fixed finite set $S$ reverses the order of quantifiers, since the machine may now depend on $S$.
Each target of $S$ gets a control header with a single one bit, which prints that target whatever suffix follows.
The fiber of such a target therefore contains a whole cube of these hub programs before any base program halts, and every base program of the target differs from a hub program in one bit.

For targets outside $S$, the fiber is empty at short lengths and early deadlines, and otherwise deleting the common base prefix is an exact isomorphism of fibers.
The fiber of the base machine at deadline $t$ corresponds to the new fiber at deadline $B+Lt$, for constants $B$ and $L$, so the fiber structure and the hardness of the base machine carry over.
A sentinel keeps the shifted program's left boundary during simulation, and all constants are computable from the base machine and $S$.
Interpretation is added last, by a new first transition that dispatches on the auxiliary input; on the empty auxiliary input the machine performs the old computation step for step, so no fiber changes (\Cref{lem:interpretation-completion}).

\subsection{Other Related Work}\label{sec:1.3}

The relevant precedents concern local-move output fibers, structural properties of decompressors, and the graph constructions used in the proofs.
This subsection says what each establishes and how its setting differs from the one here.

\paragraph{\emph{Neutrality and description sets.}}

Reidys, Stadler and Schuster develop configuration graphs and equal-image neutral sets \cite[Sections 2 and 4 of the author preprint]{RSS97}.
Their RNA compatibility graph also permits changes in two paired positions, so it does not coincide with the one-bit program graph in every case.
Hu, Banzhaf and Ochoa use literal programs in a finite register-based model and study neutrality, search and phenotype transitions \cite{HBO25}.
Schaper, Johnston and Louis study disconnected neutral components in RNA sequence maps under single point mutations \cite[Section~2]{SJL12}.
Equal-output mutation graphs of programs are therefore not new objects.
The results here instead quantify over all adjunction-universal machines, or construct specific ones, with guarantees for the full set of descriptions at stated native deadlines, most of them holding at every later deadline.

Tadaki proves that the entire prefix-complexity function can stay unchanged while output-description cardinalities change \cite[Theorems 4 and 9 of the final manuscript]{Tad10}.
That is a close structural precedent, in a prefix-free optimal-machine model.
Allender, Buhrman and Kouck\'y construct machines retaining a universal machine by adjunction behind a fixed prefix while varying efficient reducibility properties of random-string sets \cite[Theorem 14 of the author draft]{ABK06a}, so dependence on the machine under adjunction is also a precedent.

Bauwens, Kozachinskiy and Shen distinguish optimality from universality by adjunction and preserve plain complexity while installing the latter \cite[Section 4, Proposition 1]{BKS25}; their questions concern plain complexity, which has no deadline.
In that universality the string placed before the programs of a machine need only exist, with no time bound, whereas for an adjunction-universal machine the code is computable in polynomial time and the simulation has polynomial overhead.
Milovanov shows that, for a suitable optimal plain decompressor, an oracle for description-length thresholds does not give a polynomial-time decision of the halting problem \cite[Theorem~1.1]{Mil26}.
These works concern minima, cardinalities or reductions to complexity predicates, whereas the results here concern connectivity of full fibers.

\paragraph{\emph{Coding, domination, and reachability.}}

The finite-field checks adapt those of Bose and Ray-Chaudhuri and of Hocquenghem \cite{BRC60,Hoc59}.
The specific rank, balance and deterministic-construction properties needed in the packing proof are proved here.
Reed--Muller evaluation codes provide the separated terminal family \cite[Section 2.3]{ASW15}.

Coupon coloring asks that every open neighborhood contain every color; Chen, Kim, Tait and Verstraete give finite-field constructions for cube colorings \cite{CKTV15}, and Goddard and Henning study the corresponding total-domatic and fractional formulations \cite{GH18}.
Connected domatic partitions are also established \cite{Zel86}, and Shaw constructs linearly many completely independent spanning trees of the cube with linear diameter \cite[Theorem~1.2]{Sha24}.
These general partition notions are known; the selector construction here is used for its connected color classes, its representative paths and its fixed native schedule.

The hardness reductions begin with the satisfying-assignment st-connectivity theorem of Gopalan, Kolaitis, Maneva and Papadimitriou \cite[Theorem 2.9 and Section 3.3 of the full version]{GKMP09}.
That theorem concerns two supplied satisfying assignments, which is distinct from asking whether the whole solution graph is connected.
General reconfiguration and its implicit-graph perspective are discussed by Ito, Demaine, Harvey, Papadimitriou, Sideri, Uehara and Uno \cite{IDH+11} and surveyed by Nishimura \cite{Nis18}.

\paragraph{\emph{Computability and parameterization.}}

Astor characterizes intrinsic density through computable injective sampling \cite[Corollary 4.3 of the accepted preprint]{Ast15}.
Case discusses effective-immunity bounds and minimal-program sets in acceptable systems \cite{Case25}.
The statements about globally connected targets apply these established notions to a different set of outputs, with an explicit packing bound from one code.
That $G(V)$ is co-c.e.\ follows directly by enumerating finite witnesses of disconnected fibers.

Fixed-parameter tractability is standard, and so is the observation that a $\PSPACE$-complete fixed-target connectivity problem at a constant parameter value rules it out unless $\Ptime$ equals $\PSPACE$.
The content of the comparison here is the pair of machines satisfying the same reference conditions: one has a uniform target-parameter algorithm despite full $\PSPACE$-completeness, while the rigid reference $U$ has a $\PSPACE$-complete fixed-target connectivity problem.

\subsection{Discussion}\label{sec:1.4}

The numerical invariance theorem and fiber connectivity concern different data.
A code controls what each program beginning with it computes and bounds its running time.
It says nothing about adjacency between programs beginning with different codes, and it does not guarantee any path between the programs for one output.

Persistent fragmentation and eventual connectivity concern different program lengths.
The fragmentation bounds hold at lengths chosen from the target length, while eventual connectivity holds from a threshold that depends on the target.
For a machine with such thresholds, every length at which a fiber of a target is disconnected lies below the threshold of that target.
The exponential lower bound on that threshold follows from a single disconnected fiber at an exponentially large length, which persists at every later deadline.

The shape of a fiber and the complexity of querying it are also separate.
A disconnected fiber can still have a simple component test: for the machine of \Cref{thm:preview-eventual}, most targets have disconnected fibers at some lengths (\Cref{thm:exponential-fragmentation-length}), yet every fixed target has a polynomial-time connectivity test (\Cref{cor:reference-fixed-target-easy}).
Moreover, unless $\Ptime$ equals $\PSPACE$, polynomial-time connectivity for every fixed target need not give a uniform polynomial-time algorithm when the target varies.
The construction of \Cref{thm:preview-eventual} exhibits both under adjunction and interpretation.
It does not give a machine satisfying the reference conditions whose full connectivity language fails to be $\PSPACE$-hard, and it does not classify such machines.

For globally connected targets, every finite set can be prescribed, with construction constants depending on that set.
The packing obstruction instead uses one fixed code across unbounded suffix lengths.
These quantifiers explain why finite prescriptions do not combine into an infinite computable family of globally connected targets for one adjunction-universal machine.
The necessary conditions, co-c.e., computably bounded c.e.\ subsets and intrinsic density zero, are not known to suffice for an infinite realization.

\subsection{Open Problems}\label{sec:1.5}

We single out three questions.
The first two stay within this native-fiber model and concern the full connectivity language under the reference conditions and the size of the set of globally connected targets.
The third lets a path change the program length.

\paragraph{\emph{Complexity under the reference conditions.}}

The fixed-target and parameterized separations do not decide whether the complexity of the full connectivity language is the same for every machine satisfying the reference conditions.
The rigid reference and the machines of \Cref{thm:uniform-hard-fixed-easy-reference,thm:finite-target-reference} have $\PSPACE$-complete full connectivity languages.

\begin{openproblem}\label{op:reference-full-hardness}
Is $\mathrm{Conn}(V)$ $\PSPACE$-complete for every machine $V$ satisfying all three reference conditions?
\end{openproblem}

Here $\mathrm{Conn}(V)$ is the language of \Cref{prob:7.1}, which always lies in $\PSPACE$ (\Cref{prop:7.9}).
A negative answer therefore requires a machine meeting all three conditions whose full connectivity language is not $\PSPACE$-hard.

\paragraph{\emph{Infinite globally connected-target sets.}}

For every adjunction-universal machine $V$, the packing bounds exclude every infinite c.e.\ subset of $G(V)$ and force intrinsic density zero.
Finite-hub completion realizes each finite set exactly, while later bridges cannot repair a disconnected fiber at an earlier deadline.

\begin{openproblem}\label{op:infinite-connected-targets}
Does an adjunction-universal machine $V$ exist for which $G(V)$ is infinite?
\end{openproblem}

By \Cref{prop:connected-target-coce}, such a set would be co-c.e.\ and effectively immune.

\paragraph{\emph{Programs of all lengths.}}

Every fiber studied in this paper has the programs of one length as vertices and one-bit flips as edges.
\Cref{def:3.2} also defines the edit graph, whose vertices are the programs of length at most a budget $\ell$ that print $x$ within the deadline, and whose edges are flips, insertions and deletions of a single bit.
Apart from its being an implicit graph (\Cref{lem:3.3}), so that its $st$-connectivity problem lies in $\PSPACE$, no result of this paper concerns it.
Every fiber at a length at most $\ell$ is an induced subgraph of that edit graph, but a path in the edit graph may change length, so neither fragmentation nor connectivity of the fibers carries over by itself.

The packing argument of \Cref{thm:linear-overhead-fragmentation} bounds the length of a path below by the Hamming distance of its endpoints, and in the edit graph only the edit distance bounds it.
The two mandatory descriptions $h0x$ and $h1\overline x$ of that proof are at Hamming distance $|x|+1$, yet they can be at edit distance two as strings.
For example, when $x=(10)^k$ with $k\ge1$, deleting the bit after $h$ and then appending a $1$ carries $h0x$ to $h1\overline x$.
This defeats only the lower bound used in that proof; it gives no path in the edit graph, since the intermediate string need not print $x$.

\begin{openproblem}\label{op:edit-graph}
Do an adjunction-universal machine $V$ and a computable function $b$ exist such that, for every target $x$, every budget $\ell\ge b(x)$ and every deadline $t$, the edit graph of the programs of length at most $\ell$ that print $x$ within $t$ native steps of $V$ is empty or connected?
\end{openproblem}

A positive answer needs paths between programs of different lengths, and the proof of \Cref{thm:eventual-connected-fibers} does not address them, since it connects the programs of each length separately.

\phantomsection\addcontentsline{toc}{subsection}{AI Disclosure}
\begingroup
\makeatletter
\let\addcontentsline\@gobblethree
\makeatother
\subsection*{AI Disclosure}

This work was developed with substantial assistance from Codex and Claude Code.
These tools were used to discover almost all of the proofs in this work, find relevant literature, and write parts of the prose; the author wrote the rest.
The author posed the problem, guided the development, and is responsible for any errors.

\endgroup

\subsection{Paper Organization}\label{sec:1.6}

\Cref{sec:2} gives the preliminaries: strings, Hamming distance and the Boolean cube, native machines, adjunction, interpretation and full fibers.
\Cref{sec:3} supplies the explicit encoding and computational framework, including the numerical comparison of description lengths, defines the full connectivity language, and proves rigid-reference hardness and pointwise interpretation completion.
\Cref{sec:fragmentation} proves the first fragmentation bound for every adjunction-universal machine, the early separation between two machines satisfying the reference conditions, and sharper persistent fragmentation and component-multiplicity bounds.
\Cref{sec:eventual-target-connectivity} constructs eventual connectivity, proves uniform hardness, establishes the exponential threshold obstruction, and derives the target-parameter consequences.
\Cref{sec:connected-target-sets} prescribes finite sets of easy targets and of globally connected targets, and proves effective restrictions on $G(V)$ for every adjunction-universal machine.

\section{Preliminaries}\label{sec:2}

This section fixes the notation for strings, distances, graphs and the Boolean cube, the native machine model with adjunction and interpretation, and the full native fiber studied throughout.

\subsection{Strings, Distances, and Cubes}\label{sec:2.1}

Natural numbers include zero.
Write $\zo=\{0,1\}$, $\eps$ for the empty string, $|z|$ for a string's length, and $uv$ for concatenation.
The sets $\zo^n$ and $\zo^{\le n}$ contain strings of exactly and at most $n$ bits, and $\zo^*$ contains all finite binary strings.
We also write $\zo^{=n}=\zo^n$.
For a bit $b$, the string $b^n$ consists of $n$ copies of $b$, with $b^0=\eps$.
Complementing every bit of $z$ gives $\overline z$.
The binary value of $z=z_1\cdots z_n$ is $\operatorname{val}(z)=\sum_{i=1}^n z_i2^{n-i}$, with value zero for the empty string.
The canonical binary representation $\bin(j)$ has no leading zeros, except $\bin(0)=0$.
Exclusive-or of bits is addition modulo two, with empty sum zero.

\begin{definition}[Hamming weight, Hamming distance, and codes]
\label{def:2.4}
Let $n \ge 0$ and let $a, b \in \zo^{n}$, the case $n = 0$ giving $\wt(\eps) = \Delta(\eps,\eps) = 0$.
The \emph{Hamming weight} of $a$ is the number of its coordinates equal to one,
\[
\wt(a) = \sum_{i=1}^{n} a_{i},
\]
and the \emph{Hamming distance} between $a$ and $b$ is $\Delta(a,b) = |\{ i \mid a_{i} \neq b_{i} \}|$, the number of coordinates on which they differ, so that $\wt(a) = \Delta(a, 0^{n})$.

The function $\Delta$ is a metric on $\zo^{n}$: it is symmetric, it vanishes exactly on equal pairs, and it satisfies the triangle inequality $\Delta(a,c) \le \Delta(a,b) + \Delta(b,c)$.
For the last, any coordinate on which $a$ and $c$ differ is one on which $a$ and $b$ differ, or one on which $b$ and $c$ differ.

For $\delta_{0} \in \mathbb{N}$, a set $C \subseteq \zo^{n}$ is a \emph{code of minimum distance} $\delta_{0}$ if $\Delta(a,b) \ge \delta_{0}$ for all distinct $a, b \in C$.
The lowercase $\delta_{0}$ is used for the parameter so that it is not confused with the distance $\Delta$ itself.

When $C$ has at least two elements the admissible values of $\delta_{0}$ are exactly $0, 1, \dots, \min_{a \neq b} \Delta(a,b)$, so \emph{the} minimum distance of $C$, meaning the largest admissible $\delta_{0}$, exists and equals $\min_{a \neq b} \Delta(a,b)$.
A set with at most one element is a code of every minimum distance, since the condition $\Delta(a,b) \ge \delta_{0}$ is quantified over pairs of distinct elements and there are none.
The word \emph{code} also names the ambient encoding $\enc{M}$ of a machine (\Cref{def:2.2}) and the code $\enc{M}_{V}$ of a machine for a machine $V$ (\Cref{def:2.7}), which for the reference machine is $\enc{M}$ itself; each use below says which meaning is intended.
\end{definition}

A local transformation is a walk through feasible objects.
We use simple undirected graphs and include the empty graph in our connectivity convention.

\begin{definition}[Graphs, walks, components, and the Boolean cube]
\label{def:2.16}
A \emph{graph} $G=(V(G),E(G))$ consists of a set $V(G)$ of vertices and a set $E(G)$ of unordered pairs of distinct vertices, called \emph{edges}.
The letter $V$ also names machines; as a vertex-set operator it is always applied to a graph, as in $V(G)$, so the two uses cannot be confused.

Two vertices $u\neq v$ are \emph{adjacent}, written $u\sim_G v$, when $\{u,v\}\in E(G)$.
We set $u\sim_G u$ false for every $u$, and $u\sim_G v$ false whenever $u$ or $v$ is not a vertex, so that $\sim_G$ is defined on all of $\zo^*\times\zo^*$.

A \emph{walk} of length $k\geq 0$ from $u$ to $v$ is a sequence $u=z_0,z_1,\ldots,z_k=v$ of vertices with $z_{j-1}\sim_G z_j$ for every $1\leq j\leq k$, and a \emph{path} is a walk whose vertices are pairwise distinct.

Two vertices are \emph{connected} when some walk joins them.
This is an equivalence relation on $V(G)$, and its classes are the \emph{connected components} of $G$.
The graph is \emph{connected} when it has at most one component; a vertex $v$ is \emph{isolated} when its component is $\{v\}$; and $G$ is \emph{edgeless} when $E(G)=\emptyset$, that is, when every vertex is isolated.

A walk of least length joining two connected vertices is a \emph{geodesic} between them, and the \emph{distance} $d_G(u,v)$ is the length of one.
The \emph{diameter} of $G$ is the supremum of $d_G(u,v)$ over the pairs $u,v$ that are connected in $G$.
That supremum is $0$ when $V(G)=\emptyset$.
The diameter of a disconnected graph is therefore the supremum of the diameters of its components, and is not $+\infty$ merely because two vertices lie in different components; every graph whose diameter is named below is finite.

For $S\subseteq V(G)$, the subgraph \emph{induced} by $S$ is $G[S]$ with $V(G[S])=S$ and $E(G[S])=\{\{u,v\}\in E(G)\mid u\in S \text{ and } v\in S\}$.

The \emph{Boolean cube}, or \emph{hypercube}, $Q_n$ is the graph with $V(Q_n)=\zo^n$ in which $a\sim_{Q_n} b$ exactly when $\Delta(a,b)=1$, for the Hamming distance $\Delta$ of \Cref{def:2.4}.
It is connected, and $d_{Q_n}(a,b)=\Delta(a,b)$ for all $a,b\in\zo^n$.

Flipping the coordinates on which $a$ and $b$ differ, one at a time and in any order, gives a walk of length $\Delta(a,b)$.
No shorter walk exists: along a walk $a=z_0,\ldots,z_k=b$ each step satisfies $\Delta(z_{j-1},z_j)=1$, whence $\Delta(a,b)\leq\sum_{j=1}^{k}\Delta(z_{j-1},z_j)=k$ by the triangle inequality of \Cref{def:2.4}.
The diameter of $Q_n$ is therefore $n$.
\end{definition}

A graph isomorphism is a bijection between vertex sets that preserves and reflects adjacency.

\subsection{Native Machines and Simulation by Adjunction}\label{sec:2.2}

Fixed machine descriptions may include finite data.
We distinguish their description length from the native time taken to execute them.

\begin{definition}[The native machine model]\label{def:2.5}
Machines are deterministic multitape Turing machines.
One transition is one \emph{native step}, a step of the machine itself and not of any machine it simulates: a finite control reads the scanned symbols, changes state, writes on the writable tapes, and moves each head by at most one cell.
Program and auxiliary strings occupy separate read-only tapes.
Each tape head starts at its first cell, and a requested move left from that cell leaves the head there.
A finite binary string occupies cells one through its length; every later cell contains the blank symbol $\sqcup\notin\zo$.

The output tape starts blank and is never read by the machine.
Its head may move in either direction, and a transition writes at most one output symbol.
A \emph{normal halt} delivers the final binary output string, including the empty string.
Earlier transient output is not a halt: head movement, overwriting and erasure before the halt are part of the native computation.
A computation delivering a string of length $m$ takes at least $m$ steps.
A distinguished halting state delivers $\bot$, which is not a string and differs from $\eps$.

The work tapes have fixed finite hardwired binary data, distributed in a layout declared with the machine; other work cells start blank.
This initial data is present before the first transition and is not charged as runtime input or as time spent writing it.
The encoding below records both the data and its layout.
A program-only machine has no auxiliary input.
A \emph{conditional machine} also reads an auxiliary string; its unconditional reading is its run at the empty auxiliary string, and we write $V(p)=V(p,\eps)$ for it.
\end{definition}

To read a machine description before an arbitrary appended program, the encoding of machines must identify where the description ends.

\begin{definition}[Prefix-free sets and self-delimiting encodings]
\label{def:2.1}
A set $\Pi \subseteq \zo^{*}$ is \emph{prefix-free} if no element of $\Pi$ is a proper prefix of another.
An encoding is \emph{self-delimiting} if its range is prefix-free: a complete codeword cannot be the proper prefix of another.
The explicit encoding used below also admits a computable left-to-right parse which identifies a codeword's end.
\end{definition}

We fix an injective self-delimiting binary encoding of finite machine descriptions, the \emph{ambient encoding}, and write $\enc M$ for the code of $M$.
It records the finite control, its initial work data and, for a clocked machine (\Cref{def:2.11}), its clock.
The explicit representation is specified in \Cref{def:2.2}, and its length and computational properties are proved in \Cref{prop:2.3}.
The simulation and interpretation properties defined below measure uniformity in the length of this fixed code.
The counting argument of \Cref{sec:linear-overhead-fragmentation} uses only the code that \Cref{def:2.7} assigns to a source machine, not the bit pattern of the ambient encoding.

Surjectivity, adjunction and interpretation are different requirements.
The following definitions state exactly which of them each result assumes.

\begin{definition}[Efficient machine, efficient universal machine]\label{def:2.6}
A Turing machine $V$, taking a program and, in the conditional setting, an auxiliary string as well, is \emph{efficient} if, for every program $p$ and auxiliary string $y$, the configuration of $V(p,y)$ reached after $t$ steps is computable from $(p,y,1^{t})$ in time polynomial in $t+|p|+|y|$, the polynomial depending only on $V$.
A halting configuration is its own successor, so this is defined whether or not the computation has halted by step $t$.
Under the model just fixed the efficiency clause is automatic by \Cref{fact:2.9}.
We retain the clause because it fixes the quantity against which the configuration computations of \Cref{lem:3.3} are charged.
It is also a genuine hypothesis on any model in which the configuration after $t$ steps need not be computable in polynomial time.

An \emph{efficient universal machine} is a machine that is efficient and in addition \emph{universal}: for every $x \in \zo^{*}$ there is a program $p$ with $V(p,\eps) = x$.
For a machine taking a program alone the clause reads: for every $x \in \zo^{*}$ there is a $p$ with $V(p) = x$.
Here the word universal means only onto at the empty condition; it does not assert either of the simulation and interpretation properties defined next.
We also call this an efficient onto machine.
The words \emph{efficient} and \emph{universal} name separate requirements below, since efficiency is asked of some machines whose universality is not in question.

We write $V(p) = x$, or $V(p,y) = x$ in the conditional setting, \emph{within $t$ steps} if the computation halts with output $x$ after at most $t$ steps.
Machines are conditional except where stated, and we abbreviate $V(p) = V(p,\eps)$ throughout, so that the unconditional reading is the conditional one at the empty auxiliary string.
Program-only and conditional machines are distinguished where each construction is given.
\end{definition}

Universality asks that every string be the output of some program.
Simulation by adjunction additionally requires that every program of a machine $M$ be run unchanged after one fixed code that depends on $M$.

\begin{definition}[Simulation property]\label{def:2.7}
An efficient universal machine $V$ has the \emph{simulation property} if there are a polynomial $q$ and a map $M \mapsto \enc{M}_{V}$, computable in time polynomial in $|\enc{M}|$ and of length polynomial in $|\enc{M}|$, such that for every Turing machine $M$ and every string $z$: if $M(z)$ halts in $T$ steps then $V(\enc{M}_{V}z,\eps)$ halts with output $M(z)$ within $q(T,|z|,|\enc{M}_{V}|)$ steps, and if $M(z)$ diverges then so does $V(\enc{M}_{V}z,\eps)$.
We call $\enc{M}_{V}$ the \emph{code} of $M$ for $V$, and the map $M \mapsto \enc{M}_{V}$ the \emph{code map} of $V$; the code is independent of the raw suffix $z$ and of its length.
In this setting $M$ is the \emph{source machine} and $z$ the \emph{raw suffix}.
For the reference machine $U$ of \Cref{def:2.10}, the code of $M$ is the ambient encoding $\enc M$ itself; other machines use their own codes.
\end{definition}

Running $M$ on an unchanged program placed after a fixed code of $M$ is \emph{simulation by adjunction}, or \emph{adjunction} for short.
Simulation constrains the empty-condition branch; interpretation instead supplies the machine through the auxiliary string.

\begin{definition}[Interpretation property]\label{def:2.8}
A conditional efficient universal machine $V$ has the \emph{interpretation property} if there are a polynomial $q$ and a map $M \mapsto \enc{M}'_{V}$ subject to the same computability and length bounds as in \Cref{def:2.7}, such that for every Turing machine $M$ and every program $p$: if $M(p)$ halts in $T$ steps then $V(p,\enc{M}'_{V})$ halts with output $M(p)$ within $q(T,|p|,|\enc{M}'_{V}|)$ steps, and if $M(p)$ diverges then so does $V(p,\enc{M}'_{V})$.

The prime keeps this encoding apart from the simulation encoding of \Cref{def:2.7}.
The simulation and interpretation encodings need not coincide.
The polynomial $q$ may be taken nondecreasing in each argument: replace it by the sum of the absolute values of its coefficients times the same monomials.
We do so throughout.

The overhead is required to be uniform in $\enc{M}$, and not merely polynomial for each fixed $M$, because every use below applies the property to a machine built from the input of a reduction.
\Cref{def:2.7} is stated with the same uniformity, and its polynomial is normalized in the same way.
\Cref{fact:2.15} needs only the per-machine form, and uses that normalization.
\end{definition}

Surjectivity is asked at the empty auxiliary string only.
It could not be asked at every auxiliary string of a machine with the interpretation property of \Cref{def:2.8}.
Let $M_{c}$ ignore its input and print the string $c$.
If $V$ has the interpretation property of \Cref{def:2.8}, then $V(p,\enc{M_{c}}'_{V}) = M_{c}(p) = c$ for every program $p$, so the range of $V(\cdot,\enc{M_{c}}'_{V})$ is the single string $c$ and is not all of $\zo^{*}$.

We call an efficient universal machine with the simulation property of \Cref{def:2.7} an \emph{adjunction-universal machine}.
The name follows the term universal by adjunction, which Bauwens, Kozachinskiy and Shen record, quoting Downey and Hirschfeldt, for machines universal in their sense \cite[Sections 2 and 4]{BKS25}.
In their definition a machine is universal if, for every machine, some string placed before the unchanged programs of that machine makes it run them; that string need only exist, and no time bound is imposed.
Here the code $\enc{M}_{V}$ is computable from $M$ in polynomial time and the simulation has polynomial overhead, so an adjunction-universal machine satisfies an effective, time-bounded form of their universality.
Being onto, having the simulation property and having the interpretation property of \Cref{def:2.8} are together the \emph{reference conditions}.

\subsection{Full Native Fibers}\label{sec:full-native-fibers}

Fixing the output, program length and native deadline selects a set of feasible descriptions.
Its geometry is the subgraph induced by those descriptions in the program cube.

\begin{definition}[The full native fiber]\label{def:full-native-fiber}
For a machine $V$ of the model of \Cref{def:2.5}, a target $x\in\zo^*$ and integers $N,t\ge0$, define
\[
\mathcal F_V(x,N,t)
=Q_N[\{p\in\zo^N:V(p,\eps)\text{ normally halts with output }x\text{ within }t\text{ native steps}\}].
\]
For a program-only machine the auxiliary string is omitted.
We call $\mathcal F_V(x,N,t)$ the \emph{fiber} of $x$.
A distinguished halt contributes no vertex, including when $x=\eps$.
Every length-$N$ program is considered, however it begins.
A path keeps $V,x,N,t$ fixed and changes one program bit at each step.
The \emph{untimed fiber} of $x$ at length $N$ is the subgraph of $Q_N$ induced by the length-$N$ programs on which $V$ eventually halts normally with output $x$; it is the union of the fibers $\mathcal F_V(x,N,t)$ over all $t$.
\end{definition}

A prefix cylinder $h\zo^r$ consists of every string obtained by appending an arbitrary $r$-bit suffix to the fixed string $h$.
We call it the \emph{cylinder} of $h$.
The counting arguments for adjunction-universal machines in \Cref{sec:fragmentation} count the vertices of connecting paths or subgraphs, including programs outside the cylinder of a code, so they never assume that this cylinder is separated from the rest of the fiber.

\section{Encodings and Connectivity Complexity}\label{sec:3}\label{sec:7}

The complexity results need, beyond the definitions of \Cref{sec:2}, a concrete encoding, uniform clocks and a precise endpoint query.
This section provides that machinery with complete proofs, together with the numerical comparison of description lengths.
It ends with two results used in \Cref{sec:eventual-target-connectivity,sec:connected-target-sets}: every fixed-target connectivity language of the rigid reference $U$ is $\PSPACE$-complete (\Cref{cor:7.18}), and interpretation can be added to a machine without changing any fiber (\Cref{lem:interpretation-completion}).

\subsection{Explicit Encodings}\label{sec:ambient-encoding}

Our hardness results reduce from a problem about Boolean formulas, stated in \Cref{fact:2.27}.
We fix their syntax and truth-value conventions next.

\begin{definition}[Conjunctive normal form]
\label{def:2.20}
A \emph{literal} is a variable $v_i$ or its negation $\neg v_i$, a \emph{clause} is a disjunction of literals, and a formula is in \emph{conjunctive normal form} if it is a conjunction of clauses.
The letter $v$ is used for Boolean variables, and the letter $x$ is reserved for the target string.
A formula is in \emph{$3$-conjunctive normal form} if in addition every clause has at most three literals.
For a formula on the variables $v_1,\dots,v_n$, an assignment $a\in\zo^n$ sends $v_i$ to $a_i$ and $\neg v_i$ to $1-a_i$, a clause to the maximum of the values of its literals, and the formula to the minimum of the values of its clauses.
That last value is $\varphi(a)$.
The empty clause takes the value $0$ and the empty conjunction the value $1$.
An assignment $a\in\zo^n$ \emph{satisfies} $\varphi$ if $\varphi(a)=1$.
\end{definition}

We fix an integer $D\ge1$.
A polynomial clock has the form $c(k)=\sum_{i=0}^D c_i k^i$, with nonnegative integer coefficients and $c_0\ge1$; it grants at most $c(k)$ source transitions on an input of length $k$.
Its coefficients are represented in unary.
The complete clocked-machine convention is recorded below.
We first make the binary representation of controls, machines, clocked machines, tuples and formulas explicit.

The ambient code uses a parity-protected length header so that a one-bit change cannot open a different valid parse.

\begin{definition}[The ambient encoding]
\label{def:2.2}
Call the three strings $A_{0} = 000$, $A_{1} = 011$ and $A_{\mathrm{end}} = 101$ the \emph{legal blocks}; in the notation of \Cref{def:2.4} they satisfy $\wt(A_{0}) \equiv \wt(A_{1}) \equiv \wt(A_{\mathrm{end}}) \equiv 0 \pmod{2}$ and $\Delta(A_{i},A_{j}) = 2$ for $i \neq j$.
Write $\blk(e)$ for the string obtained from $e \in \zo^{*}$ by replacing each $0$ by $A_{0}$ and each $1$ by $A_{1}$.

Let $\rho$ be the concatenative raw encoding of finite objects, defined as follows.
Every raw encoding begins with a tag $\sigma$ of a fixed length $\lambda_{0} = O(1)$ naming the type of the object encoded, which is one of control, machine, clocked machine, tuple, and formula.
The tag makes $\rho$ injective across types as well as within each, and is displayed in the identities below.

A finite control $F$ is written as $\rho(F) = \sigma_{\mathrm{ctrl}} T_{F}$ for its transition table $T_{F}$.
The table uses a self-delimiting format of its own.
In it, a transition mentions only the tapes it touches and the states it names, each by a self-delimiting numeral, and the states of a table are numbered $1, \dots, S$ consecutively.
The format is prefix-free, so the raw control encodings form a prefix-free set, and that makes the machine clause $\rho(F,d) = \sigma_{\mathrm{mach}} \rho(F) d$ below injective and parsable in one left-to-right pass.
In particular $\enc{F}$ is defined.

The table carries an initial-state field, one self-delimiting numeral, ahead of the transitions, so that a wrapper redirects the entry point by rewriting a single field rather than by renumbering the inherited states.
Tapes, states, and tape symbols are each numbered consecutively from $1$.
A table with $k$ tapes, $S$ states, and an alphabet of $A$ symbols therefore has $k + S + A = O(|T_{F}|)$, which is the bound on the alphabet width that \Cref{fact:2.9} uses.

The two symbols $0$ and $1$ are numbered $1$ and $2$ and are written in a common fixed-width field, so that exchanging one for the other in a transition changes no field's width.
Declaring one further work tape, adding a bounded number of fresh states, rewriting the initial-state field, and adjoining a fixed gadget therefore enlarge the table by $O(\log(|T_{F}| + 2))$ bits and no more.
Rewriting an inherited bit-write in place changes no symbol field's width.

A machine with control $F$ and data string $d$ is written as $\rho(F,d) = \sigma_{\mathrm{mach}} \rho(F) d$.
A clocked machine with control $F$, data $d$, and clock coefficients $c_{0}, \dots, c_{D}$ as in \Cref{def:2.11} is written as
\begin{equation}
\label{eq:2.1}
\rho(F, d, c_{0}, \dots, c_{D}) = \sigma_{\mathrm{clk}} \rho(F) 1^{c_{0}} 0 1^{c_{1}} 0 \cdots 1^{c_{D}} 0 d,
\end{equation}
where the coefficient block is self-delimiting because it contains exactly $D+1$ zeros and $D$ is an absolute constant, so that the transition table, then the $D+1$ coefficients, then the data string are read off in one left-to-right pass.
The data string comes last in both cases, since it is written raw and carries no delimiter of its own; its extent is fixed by the length that the length header of \Cref{eq:2.2} announces.
For a clocked machine $M$ as in \Cref{eq:2.1}, $\rho(M)$ carries $1^{c_{i}}$ as a contiguous block for every $i \le D$.
\Cref{lem:2.12} reads these blocks.
The $D+1$ blocks are disjoint substrings of $\rho(M)$, so that $\sum_{i \le D} c_{i} \le |\rho(M)| \le |\enc{M}|$, and $|\rho(M)| = \lambda_{0} + |\rho(F)| + |d| + \sum_{i \le D} c_{i} + (D+1)$.
Emptying the data string of a clocked machine $M$ with control $F$ gives the clocked machine $F^{\flat} = (F, \eps, c_{0}, \dots, c_{D})$, so $\enc{F^{\flat}}$ is defined as well.
\Cref{prop:2.3} compares $|\enc{M}|$ with $|\enc{F^{\flat}}|$.
For an unclocked machine, $F^{\flat} = F$.

A tuple of $k \ge 2$ strings $u_{1}, \dots, u_{k}$ is written as
\[
\rho(u_{1}, \dots, u_{k}) = \sigma_{\mathrm{tup}} \sd(k) \sd(|u_{1}|) \cdots \sd(|u_{k-1}|) u_{1} \cdots u_{k},
\]
where $\sd(j) = \blk(\bin(j)) A_{\mathrm{end}}$ is the self-delimiting numeral of length $3(|\bin(j)| + 1)$ built from the legal blocks.
The header records the arity, because the lengths alone would not tell a left-to-right parse where the header stops.
With the arity in hand, the parse reads $k$, then the $k-1$ lengths, then the components.

For a pair this gives $|\rho(a,b)| = |a| + |b| + O(\log(|a|+|b|+2))$.
Arities are not assumed bounded, and none is needed: every numeral $\sd(j)$ has length $3(|\bin(j)| + 1) \ge 6$, so the header of a $k$-tuple occupies at least $6k$ bits and $k = O(|\rho(u_{1}, \dots, u_{k})|)$.
Appending one further component $u_{k+1}$ therefore lengthens the header by $|\sd(k+1)| - |\sd(k)| + |\sd(|u_{k}|)| = O(\log(|\rho(u_{1}, \dots, u_{k})| + 2))$ and the body by $|u_{k+1}|$, which is the estimate the wrappers below use.

The tuple rule is not applied to the pair $(F,d)$.
The transition table is already self-delimiting, so no length header is needed to separate the control from the data, and the data string is charged $|d|$ and no more, as the additive bounds below require.

A formula $\varphi$ in conjunctive normal form on $n$ variables with clauses $C_{1}, \dots, C_{m}$ is written as $\rho(\varphi) = \sigma_{\mathrm{cnf}} \sd(n) \sd(m) \rho(C_{1}) \cdots \rho(C_{m})$.
Each clause is written as its width in the numeral $\sd$ followed by its literals, and each literal as a sign bit followed by $\sd$ of its index.
The clause codes $\rho(C_{j})$ use a tagless local format.
They have no type tag, since the tag rule of the opening paragraph applies to the five listed types only, and they are parsed positionally inside $\rho(\varphi)$, where the width numeral of each clause announces where the next begins.
The parse is again left to right.

Finally, $|\rho(o)|$ is linear in the size of $o$ in every case, taking the size of a formula to be the length of that encoding.
The map $\rho$ is an injection.
Both $\rho$ and $\rho^{-1}$ are computable in polynomial time, and the range of $\rho$ is decidable in polynomial time by one left-to-right parse of the tag and of the self-delimiting fields that it announces.
The \emph{ambient encoding} of a finite object $o$ is
\begin{equation}
\label{eq:2.2}
\enc{o} = \blk\left(\bin\left(|\rho(o)|\right)\right) A_{\mathrm{end}} \rho(o) \parity(o), \qquad \parity(o) = \bigoplus_{i=1}^{|\rho(o)|} \rho(o)_{i},
\end{equation}
a block-encoded length header, the raw encoding, and one parity bit, where the empty exclusive-or is $0$, so that $\parity(o) = 0$ when $\rho(o) = \eps$.
A machine reading a codeword from the left reads the length header block by block until it meets $A_{\mathrm{end}}$, then reads the announced number of further bits together with the parity bit.
It rejects the string outright in any of six cases: some block is not legal; the block-decoded header $e$ is empty; $e$ has $|e| \ge 2$ and $e_{1} = 0$, so that $e \neq \bin(r)$ for the number $r$ it names; the total length differs from $3(|e|+1) + r + 1$; the parity does not check; or $\rho^{-1}$ is undefined on the announced span.

The leading-zero test is necessary.
Without it the pass accepts $\blk(e) A_{\mathrm{end}} \rho(o) \parity(o)$ for every $e$ whose value is $r = |\rho(o)|$, and so decides a strictly larger set than the range.

Take $e = 0\bin(r)$.
Every header block of the resulting string is legal, its decoded header is nonempty and names $r$, its total length is exactly $3(|e| + 1) + r + 1$, and its parity checks.

That string nonetheless lies outside the range.
The block parse of any codeword stops at its first $A_{\mathrm{end}}$ block, so a codeword beginning $\blk(0\bin(r)) A_{\mathrm{end}}$ would announce the header string $0\bin(r)$.
That string has a leading zero and length at least two, so it is $\bin(r')$ for no $r'$.
Only the clause forbidding a leading zero rejects it in the parse itself.
\end{definition}

When the hardwired data string encodes a tuple, its components occupy separate declared work tapes.
That layout is recorded in the finite control, so the ambient machine code determines the entire initial configuration.
A wrapper may put an added component on a fresh tape after the inherited tapes, preserving every inherited tape index.

\begin{proposition}[The ambient encoding is prefix-free, cheap, and rigid]
\label{prop:2.3}
The encoding of \Cref{def:2.2} is injective, has a prefix-free range, decidable and decodable in polynomial time, is computable in polynomial time, has length bounded by a polynomial in the size of the object encoded, and satisfies
\[
|\enc{M}| = \left|\enc{F^{\flat}}\right| + |d| + O(\log(|d|+2))
\]
for a machine $M$ presented as a control $F$ and a data string $d$, with an absolute implied constant, where $F^{\flat}$ is the clocked machine $(F, \eps, c_{0}, \dots, c_{D})$ obtained by emptying the data string, and $F^{\flat} = F$ for an unclocked machine; more generally $|\enc{o}| \le |\rho(o)| + 7 \lceil \log_{2} (|\rho(o)| + 2) \rceil$.
It is moreover \emph{rigid}: for every codeword $w$, every $w' \in \zo^{|w|}$ at Hamming distance one from $w$, and every $z \in \zo^{*}$, no codeword of the encoding is a prefix of $w'z$.
\end{proposition}

\begin{proof}
\emph{Prefix-freeness and the parse.} The header is a sequence of legal blocks terminated by the single occurrence of $A_{\mathrm{end}}$, since no block of $\blk(e)$ is $A_{\mathrm{end}}$, so it is self-delimiting and is read in one pass.
It announces $|\rho(o)|$, which fixes the total length of the codeword.
If one codeword is a prefix of another, the two are read by the same block parse, so they carry the same header, hence the same announced length, hence are equal.
No codeword is therefore a proper prefix of another.

Membership is decided, and $o$ recovered, by that one pass together with the parity check and the inversion of $\rho$, all in polynomial time.
\Cref{eq:2.2} is itself computed in polynomial time from $o$.
The encoding is injective, because $\rho$ is injective and because the header fixes the span of $\enc{o}$ that $\rho(o)$ occupies.
Its length is linear in $|\rho(o)|$, hence polynomial in the size of $o$.

\emph{Length.} Writing $r = |\rho(o)|$, the codeword has $3\left(|\bin(r)| + 1\right) + r + 1$ bits, and $|\bin(r)| = \lceil \log_{2}(r+1) \rceil$ for $r \ge 1$, while $|\bin(0)| = 1$.
In either case $|\bin(r)| \le L$ for $L = \lceil \log_{2}(r+2) \rceil$, so the excess of the codeword length over $r$ is $3|\bin(r)| + 4 \le 3L + 4 \le 7L$, the last step because $L \ge 1$.
That is the general bound.
For a machine the transition table is nonempty, so $r = \lambda_{0} + |\rho(F)| + |d| \ge 1$ there, and $|\rho(F)| \ge \lambda_{0} \ge 1$ as well.
The exact formula $|\bin(r)|=\lceil\log_2(r+1)\rceil$ for $r\ge1$ therefore applies to $\enc{F}$ and to $\enc{M}$ alike, giving
\[
|\enc{M}| - |\enc{F}| - |d| = \lambda_{0} + 3\left(\lceil \log_{2} (\lambda_{0} + |\rho(F)| + |d| + 1) \rceil - \lceil \log_{2} (|\rho(F)| + 1) \rceil\right).
\]
This lies between $\lambda_{0}$ and $\lambda_{0} + 3 \log_{2}(\lambda_{0} + |d| + 1) + 3$, by the inequality
\[
\lambda_{0} + |\rho(F)| + |d| + 1 \le (|\rho(F)| + 1)(\lambda_{0} + |d| + 1),
\]
and is therefore $O(\log(|d| + 2))$ with an absolute implied constant, since $\lambda_{0}$ is $O(1)$.
For a clocked machine we absorb the clock into the control.
Write
\[
\rho(F^{\flat}) = \sigma_{\mathrm{clk}} \rho(F) 1^{c_{0}} 0 \cdots 1^{c_{D}} 0, \qquad s = \left|\rho(F^{\flat})\right|,
\]
for the tag and the control together with the coefficient block of \Cref{eq:2.1}, so that $\rho(M) = \rho(F^{\flat}) d$ exactly, with no further tag.
By \Cref{eq:2.1} the string just displayed is $\rho$ of the clocked machine $(F, \eps, c_{0}, \dots, c_{D})$, so $\enc{F^{\flat}}$ is \Cref{eq:2.2} applied to an object of a declared type.
The derivation above then runs with $\lambda_{0}$ replaced by $0$ and gives
\[
|\enc{M}| - \left|\enc{F^{\flat}}\right| - |d| = 3\left(\lceil \log_{2}(s + |d| + 1) \rceil - \lceil \log_{2}(s+1) \rceil\right) \in \left[0, 3 \lceil \log_{2}(|d|+1) \rceil\right],
\]
the upper bound from $s + |d| + 1 \le (s+1)(|d|+1)$, so $|\enc{M}| = \left|\enc{F^{\flat}}\right| + |d| + O(\log(|d|+2))$ with an absolute implied constant.
This is the form the constructions below use, since the clock coefficients are part of what the caller supplies and not of the data.

\emph{Rigidity.} Let $w$ be a codeword, let $w'$ differ from $w$ in the single position $i$, and let $z$ be arbitrary.
Write $r = |\rho(o)|$ for the object $o$ that $w$ encodes and $k = |\bin(r)| + 1$ for the number of header blocks.
The codeword has $3k + r + 1$ bits, of which the first $3k$ form the header, the next $r$ are $\rho(o)$, and the last is the parity bit.
The position $i$ therefore lies in exactly one of these three regions, and the two cases below are exhaustive.

Suppose some codeword $u$ prefixes $w'z$.
Every codeword begins with a header of legal blocks, and both parses start at the first bit.
The blocks of $u$ therefore occupy the same positions as those of $w'$, and the two parses read the same positions until one of them stops.
Moreover, $w'$ agrees with $w$ off position $i$.

If $i$ lies in the header of $w$, let $B_{j}$ be the block containing it, so $j \le k$.
Every legal block has even Hamming weight, so the corrupted block has odd weight and is not legal.
The blocks before it are untouched, and none of them is $A_{\mathrm{end}}$: that block occurs in the header of $w$ only as the last one, which is $B_{j}$ or comes after it.

Concretely, the blocks $1, \dots, j-1$ of $w$ lie inside $\blk(\bin(r))$ and are therefore each $A_{0}$ or $A_{1}$.
The flip does not touch them, so they are also the blocks $1, \dots, j-1$ of $w'$, hence of $u$.
Since the header of a codeword terminates at its first $A_{\mathrm{end}}$ block, the parse of $u$ cannot stop before block $j$, and at block $j$ it meets an illegal block.
No such $u$ exists.

If $i$ lies in $\rho(o)$ or in the parity bit, then $w'$ has the same header as $w$: blocks $1, \dots, k-1$ are $A_{0}$ or $A_{1}$, and block $k$ is $A_{\mathrm{end}}$.
The header of $u$ therefore stops exactly at block $k$, and it is the header of $w$.
Hence $u$ announces the same length $r$ as $w$ and has the same total length $3k + r + 1$.
So $u$ is the prefix of $w'z$ of that length, which is $w'$ itself.
If $i$ lies in $\rho(o)$, the announced bits of $w'$ have the opposite parity to those of $w$ while the parity bit is unchanged.
If $i$ is the parity bit, the announced bits are unchanged and the parity bit is flipped.
Either way $w'$ fails the parity check and is not a codeword.
Therefore no codeword prefixes $w'z$.
\end{proof}

\subsection{Uniform Simulation and Native Clocks}\label{sec:clock-support}

Bounded native execution has a uniform interpreter.
We give the configuration construction because the later clocks depend on uniformity in the supplied machine code.

\begin{fact}[{Uniform simulation of multitape machines; folklore strengthening of \cite[Theorem 1.9 of the published edition]{AB09}, see also \cite{LV19}}]\label{fact:2.9}
There are a multitape Turing machine $\mathsf{Sim}$ and a polynomial $q_{\mathsf{Sim}}$ such that, for every multitape Turing machine $M$, every input $z$, and every $T$, the machine $\mathsf{Sim}$ on input $(\enc{M},z,1^{T})$ outputs the configuration of $M(z)$ after $T$ steps, in time at most $q_{\mathsf{Sim}}(T,|z|,|\enc{M}|)$, the polynomial $q_{\mathsf{Sim}}$ and its degree not depending on $M$.
The same holds for a machine taking its two inputs $p$ and $y$ on separate read-only tapes.
\end{fact}

\begin{proof}
The sources state the halting simulation with an overhead whose constant depends on $M$.
The constructions below need two things those statements do not give: the configuration after a prescribed number of steps, whether or not the computation has halted, and an overhead uniform in $|\enc{M}|$.
One construction supplies both, and we record it rather than cite it.

The simulator first lays out the initial configuration of $M(z)$, writing the hardwired data string $d$ across the simulated work tapes in the layout that $\rho(F)$ declares, as \Cref{def:2.5} requires of that configuration.
It then carries the $k$ tapes of $M$ interleaved on a single work tape, one track apiece, with a further track per tape marking that tape's head, and it keeps the step count in unary on a second work tape.
It reads the transition table off its own input rather than holding it in its control.
That is what turns a dependence on $M$ into a dependence on $|\enc{M}|$.

The number of tapes of $M$ and the width of its alphabet are both bounded by $|\enc{M}|$, so one column of the interleaving, carrying one cell of each tape together with the head markers, occupies $O(|\enc{M}|^{2})$ cells of the simulator.
The layout occupies $|d| \le |\enc{M}|$ columns before the first step, and the region in use after $T$ steps therefore spans $O(|\enc{M}|^{2}(T+|z|+|\enc{M}|))$ cells.

One simulated step sweeps that region and makes one lookup in the table.
The lookup costs $O(|\enc{M}|)$, so, including initialization and the final configuration output, the total is bounded by $O((T+1)(|\enc{M}|+1)^3(T+|z|+|\enc{M}|+1))$.
That is a polynomial in $T$, $|z|$ and $|\enc{M}|$ whose degree does not depend on $M$, which is what the fact asserts.
The simulator halts after simulating exactly $T$ steps, with the resulting configuration written out whether or not $M$ has halted.
\end{proof}

We now fix one machine with both the simulation property of \Cref{def:2.7} and the interpretation property of \Cref{def:2.8}.

\begin{definition}[The reference machine]\label{def:2.10}
The \emph{reference machine $U$} is defined as follows: if $y = \eps$ and $p = \enc{M}z$ for the unique machine encoding $\enc{M}$ that prefixes $p$, then $U(p,y)$ runs $M(z)$; if $y = \enc{M}$, it runs $M(p)$; and in every other case it diverges.
We call the first the \emph{simulation branch} and the second the \emph{interpretation branch}.
In both branches $\enc{M}$ ranges over the encodings of machines and of clocked machines alike, in the sense of \Cref{def:2.11}.
A clocked $M$ is run under its clock, so $U(p,\enc{M}) = \bot$ when that clock expires.
A prefixing codeword of any other type is not a machine encoding, and $U$ diverges.
Because its encoding is rigid (\Cref{prop:2.3}), we also call $U$ the \emph{rigid reference}.
\end{definition}

Such a machine exists and meets both requirements.
The parse is unique because $\enc{\cdot}$ is prefix-free, and the two branches do not collide because $\eps$ is not a codeword of \Cref{def:2.2}: every codeword there has length at least seven.
Universality holds directly, and not by way of either property, since the statements of both presuppose it.
For $x \in \zo^{*}$, let $M_{x}$ ignore its input and print $x$; then $U(\enc{M_{x}},\eps) = M_{x}(\eps) = x$.
Efficiency is automatic under the model fixed above.
Both properties hold with a single polynomial, by \Cref{fact:2.9}, whose overhead is uniform in $\enc{M}$.
That is exactly the uniformity the two definitions ask for.

The clock presentation below makes every construction that wraps a supplied clocked machine uniform in that machine.
It also specifies how a timeout differs from a normal empty output.

\begin{definition}[Clocked machine]\label{def:2.11}
Use the fixed absolute degree bound $D\ge1$ introduced above.
A clocked machine supplied as a decider (\Cref{def:2.19}) is written calligraphically as $\mathscr{D}$ to distinguish it from this degree bound.

A \emph{clocked machine} is a pair consisting of the code of a Turing machine and a clock $c(n) = \sum_{i \le D} c_{i} n^{i}$ of degree at most $D$, evaluated at the input length, so that a computation of the machine on an input $a$ is granted $c(|a|)$ steps.

The coefficients $c_{i} \in \mathbb{N}$ are written in unary inside the code, so that the encoding length of a clocked machine does not grow with the input length.
They are natural numbers, which is what makes $c$ nondecreasing on $\mathbb{N}$.
We adopt the convention $0^{0} = 1$, so that $c(0) = c_{0}$, and require $c_{0} \ge 1$, so that a clocked machine is granted at least one step on every input, the empty one included.

It is run under its clock.
If the clock expires it halts with a distinguished output $\bot \notin \zo^{*}$.
A timed-out computation is therefore a rejection when the machine is used as a decider, and it yields no string, hence no description, when the machine is used as a verifier.
Here a \emph{verifier} is a machine, clocked or not, whose normal outputs serve as targets, as opposed to a decider, whose outputs are the verdicts $0$ and $1$.
Since $\bot$ is not a string, $U(p,y) = \bot$ differs from every target, the empty target included.

Whenever a polynomial-time machine is supplied as input to one of the constructions below, it is presented in this form, and the relevant input length is given in unary as well.
The degree bound and the unary coefficients together give what the constructions need, namely one polynomial serving for every clocked machine at once.

A construction that wraps a supplied clocked machine must present its clock by that machine's own coefficients, and not by the value of its clock at a hardwired length.
A unary constant coefficient costs as many bits as the running time it names, so the wrapper's encoding would grow with that running time.
The fixed degree bound and the unary representation of the coefficients are both needed for a uniform polynomial bound in input length and code size.

Multiplying the supplied clock by a fixed constant and adding a linear polynomial preserves degree at most $D\ge1$.
Scaling and adding the unary coefficients takes polynomial time.
This is the closure used by the native wrappers below.
\end{definition}

The fixed degree and explicit unary coefficients bound every supplied clock by one polynomial in input length and code size.

\begin{lemma}[Uniform clock bound]\label{lem:2.12}
Let $M$ be a clocked machine with clock $c$ in the sense of \Cref{def:2.11}.
Then
\begin{equation}\label{eq:2.3}
c(n) \le (n+|\enc{M}|+D)^{D+2}
\end{equation}
for every $n \ge 0$, and $1^{c(n)}$ is computable from $\enc{M}$ and $1^{n}$ in time polynomial in $n+|\enc{M}|$.
\end{lemma}

\begin{proof}
The $D+1$ unary blocks $1^{c_{0}},\dots,1^{c_{D}}$ are disjoint substrings of $\rho(M)$ by \Cref{eq:2.1}, so $\sum_{i \le D} c_{i} \le |\rho(M)| \le |\enc{M}|$.
Also $|\enc{M}| \ge 1$, since a code is a nonempty string.
For every $n \ge 0$ and every $i \le D$ we have $n^{i} \le (n+1)^{i} \le (n+1)^{D}$, so
\[
c(n) = \sum_{i \le D} c_{i} n^{i} \le \left( \sum_{i \le D} c_{i} \right) (n+1)^{D} \le |\enc{M}|(n+1)^{D}.
\]
Write $\beta_{0} = n+|\enc{M}|+D$.
Then $|\enc{M}| \le \beta_{0}$ and $n+1 \le \beta_{0}$, the latter because $|\enc{M}|+D \ge 1$.
Multiplying gives
\[
c(n) \le |\enc{M}|(n+1)^{D} \le \beta_{0} \cdot \beta_{0}^{D} \le \beta_{0}^{D+2} = (n+|\enc{M}|+D)^{D+2},
\]
which is \Cref{eq:2.3}.
Since $D$ is an absolute constant, the right-hand side is a fixed polynomial in $n+|\enc{M}|$.
Parse the unary coefficients and the supplied string $1^n$.
For each $i=0,\ldots,D$, begin with $1^{c_i}$ and, $i$ times, replace every symbol $1$ by a copy of $1^n$ on a blank work tape.
The intermediate strings have length at most $|\enc{M}|(n+1)^D$; nested tape scans therefore implement each replacement in time polynomial in $n+|\enc{M}|$.
Concatenating the resulting $D+1$ strings produces $1^{c(n)}$ within the same kind of polynomial bound.
This includes $n=0$ and zero nonconstant coefficients; parsing and setup are charged independently of the output length.
The additive $D$ is a convenience rather than a necessity: the display above already gives $c(n) \le (n+|\enc{M}|)^{D+2}$, because $n+|\enc{M}| \ge n+1$ and $n+|\enc{M}| \ge |\enc{M}| \ge 1$.
\end{proof}

\subsection{Numerical Description-Length Comparison}\label{sec:numerical-invariance}

The numerical measure will be used only to compare machines; the connectivity questions retain every feasible description.

\begin{definition}[Time-bounded conditional Kolmogorov complexity]\label{def:2.13}
For $x, y \in \zo^{*}$ and a deadline $t \in \mathbb{N}$, the \emph{time-bounded conditional Kolmogorov complexity} of $x$ given $y$ is
\begin{equation}\label{eq:2.4}
\Kt(x | y) = \min \{ |p| \mid p \in \zo^{*} \text{ and } U(p,y) = x \text{ within } t \text{ steps} \},
\end{equation}
with the convention that the minimum of the empty set is $+\infty$.
\end{definition}

A subscript $V$ replaces the reference machine by $V$, and $\K^t_V(x)=\K^t_V(x|\eps)$ is the unconditional form.
The value is infinite when no such program exists.
In particular it is infinite whenever $|x|>t$, since a computation delivering $x$ takes at least $|x|$ steps (\Cref{def:2.5}).
The following truncation bound also bounds the length of a shortest program.

\begin{lemma}[Witness truncation]\label{lem:2.14}
For every machine $M$, every $x$, and every $t$, if $\Kt_{M}(x)$ is finite then $\Kt_{M}(x) \le t+1$.
\end{lemma}

The bound is folklore for models of the kind fixed in \cite[Section 7.1]{LV19}, which states no such theorem.
We verify it for the model of \Cref{sec:2.2}, since it depends on that model's head discipline and read-only tape layout.

\begin{proof}
Suppose $\Kt_{M}(x)$ is finite and let $p$ be a witness, so $M(p) = x$ within $t$ steps.
By the model of \Cref{sec:2.2} the program tape is read-only, its head begins on cell $1$, and it moves by at most one cell per step.
An induction on $j$ then gives that after $j$ steps the head is on a cell of index at most $j+1$.
The symbol read \emph{during} step $s$ is the one under the head after $s-1$ steps, hence lies at a cell of index at most $s$.
As the computation halts after some $j \le t$ steps, every program cell that it reads has index at most $t$, and a fortiori at most $t+1$.

Let $p'$ be the first $\min(|p|,t+1)$ bits of $p$, so $p' = p$ when $|p| \le t+1$ and $|p'| = t+1$ otherwise.
The two program tapes carrying $p$ and $p'$ agree on every cell of index at most $t+1$.
Beyond that index the first carries bits of $p$, and the second carries either bits of $p$ or the blank symbol.

We show by simultaneous induction on $s \le j$ that after $s$ steps the two computations are in the same state, have all their heads on the same cells, carry the same contents on every tape other than the program tape, and have their program-tape heads on a cell of index at most $s+1 \le t+1$.
The base case $s = 0$ is the common initial configuration.

For the inductive step, the transition taken depends only on the current state and on the symbols under the heads.
The symbols on the tapes other than the program tape agree by the inductive hypothesis.
The program symbol is read at an index at most $s+1 \le j \le t$, where the two program tapes agree, so the same transition is taken and all four clauses persist.
Hence $M(p') = M(p) = x$ within the same $j \le t$ steps, so $p'$ is a witness and $\Kt_{M}(x) \le |p'| \le t+1$.
\end{proof}

The usual resource-bounded invariance principle has the following form for adjunction.
Its per-machine polynomial is proved explicitly here; compare the particular efficient reference in \cite[Theorem 7.1.1]{LV19}.

\begin{fact}[{Invariance from adjunction}]\label{fact:2.15}
Let $V$ be an efficient universal machine with the simulation property of \Cref{def:2.7}.
For every machine $M$ there are a constant $c_{M}$ and a polynomial $q$ such that, for every string $x$ and every deadline $t$,
\[
\K^{q(t)}_{V}(x) \le \Kt_{M}(x) + c_{M},
\]
where $\Kt_{M}$ is the time-$t$ measure taken against $M$, in the notation just fixed.
\end{fact}

\begin{proof}
That the polynomial $q$ depends on $t$ alone is not obvious, since \Cref{def:2.7} grants an overhead polynomial in the running time of $M$ \emph{and} in the length of the simulated program.

Explicitly, let $q_{0}$ and $M \mapsto \enc{M}_{V}$ witness the simulation property of $V$, and put $c_{M} = |\enc{M}_{V}|$ and $q(t) = q_{0}(t,t+1,c_{M})$, a polynomial in $t$ for each fixed $M$.

If $\Kt_{M}(x) = +\infty$ the inequality is vacuous.
Otherwise let $p$ be a shortest witness, so that $M(p) = x$ within some $T \le t$ steps and $|p| = \Kt_{M}(x)$, which is at most $t+1$ by \Cref{lem:2.14}.
Then $V(\enc{M}_{V}p,\eps) = M(p) = x$ within $q_{0}(T,|p|,c_{M}) \le q(t)$ steps, since $q_{0}$ is nondecreasing in each argument.
Hence $\K^{q(t)}_{V}(x) \le |\enc{M}_{V}p| = \Kt_{M}(x) + c_{M}$.

Both $c_{M}$ and $q$ depend on $M$ and on $V$ and on nothing else.
That $q$ depends on $t$ alone, and not on the complexity, is supplied by \Cref{lem:2.14}.

Apply this in both directions to a pair of machines with the simulation property.
The measure $\K^t_V(x)$ is then determined by $x$ and $t$ up to an additive constant in the length budget and a polynomial change in the deadline, whichever of the two machines is used.
\end{proof}

\subsection{Implicit Graphs and Space Bounds}\label{sec:2.3}

For a general finite connected graph, the number of vertices gives a simple diameter bound.

\begin{lemma}[Diameter of a finite connected graph]
\label{lem:2.17}
A connected graph with $k\geq 1$ vertices has diameter at most $k-1$.
\end{lemma}

\begin{proof}
A walk of least length between two vertices repeats no vertex, since deleting the portion between two occurrences of one leaves a shorter walk, so it visits at most $k$ vertices and traverses at most $k-1$ edges.
\end{proof}

The computational input usually describes an exponentially larger graph through vertex and adjacency tests.

\begin{definition}[Implicit presentation]
\label{def:2.18}
Implicitness is a property not of a single finite graph, for which the conditions below hold by table lookup, but of a family presented by a string.
A family $\{G_w\}_{w\in\zo^*}$ of graphs is \emph{implicitly presented} if there are a polynomial $\mu$, a set $\mathscr{W}\subseteq\zo^*$ of \emph{well-formed} presenting strings, and algorithms $A_{\mathscr{W}}$, $A_V$ and $A_E$, each running in time polynomial in the total length of its arguments, such that
\begin{itemize}
\item $A_{\mathscr{W}}(w)$ decides whether $w\in\mathscr{W}$, and $G_w$ is the empty graph for $w\notin\mathscr{W}$;
\item every vertex of $G_w$ is a string of length at most $\mu(|w|)$;
\item $A_V(w,r)$ decides whether $r\in V(G_w)$; and
\item $A_E(w,r,r')$ decides whether $r\sim_{G_w} r'$.
\end{itemize}
The set $\mathscr{W}$ is what makes the family total on $\zo^*$, and it is where each presentation below records the shape its tests assume.
We say that $G_w$ is \emph{implicit} when the presenting family is clear from the context, as it is at every use below.
The \emph{st-connectivity} question for such a family asks, given $w$ and two strings $r$ and $r'$, whether both are vertices of $G_w$ and are connected in $G_w$.
\end{definition}

We use polynomial time and polynomial space with respect to the full binary input length, and polynomial-time many-one reductions.

\providecommand{\DSPACE}{\mathsf{DSPACE}}
\providecommand{\NSPACE}{\mathsf{NSPACE}}
\begin{definition}[Complexity classes and many-one reductions]
\label{def:2.24}
We write $\Ptime$ for the class of decision problems solvable by a deterministic Turing machine in time polynomial in the input length, and $\NP$ for those whose yes-instances admit certificates of polynomial length checkable in polynomial time.
Similarly, $\PSPACE$ denotes the class of decision problems solvable by a deterministic Turing machine in space polynomial in the input length, and $\NPSPACE$ the class of those solvable by a nondeterministic Turing machine in polynomial space.

For functions $s$ we write $\DSPACE(s)$ and $\NSPACE(s)$ for the deterministic and nondeterministic classes at space bound $s$, so that $\PSPACE=\bigcup_{k\geq 1}\DSPACE(n^k)$ and $\NPSPACE=\bigcup_{k\geq 1}\NSPACE(n^k)$.

A \emph{promise problem} is a pair $(\Pi_{\mathrm{yes}},\Pi_{\mathrm{no}})$ of disjoint sets of strings, and an algorithm \emph{solves} it if it accepts every element of $\Pi_{\mathrm{yes}}$ and rejects every element of $\Pi_{\mathrm{no}}$, its behavior on every other input being unconstrained.
A language $L$ is identified with the promise problem $(L,\zo^*\setminus L)$, so that the two readings agree on decision problems and every notion below applies to languages as a special case.
A promise problem \emph{lies in} $\PSPACE$ if some deterministic algorithm running in space polynomial in the input length solves it, and \emph{lies in} $\Ptime$ if some deterministic algorithm running in time polynomial in the input length solves it.

A \emph{polynomial-time many-one reduction} from $(\Pi_{\mathrm{yes}},\Pi_{\mathrm{no}})$ to $(\Sigma_{\mathrm{yes}},\Sigma_{\mathrm{no}})$ is a polynomial-time computable $f$ with $f(\Pi_{\mathrm{yes}})\subseteq\Sigma_{\mathrm{yes}}$ and $f(\Pi_{\mathrm{no}})\subseteq\Sigma_{\mathrm{no}}$, nothing being asked of $f$ outside $\Pi_{\mathrm{yes}}\cup\Pi_{\mathrm{no}}$.

A promise problem is \emph{$\PSPACE$-hard} if every language in $\PSPACE$, read as above, reduces to it in this sense, and $\PSPACE$-\emph{complete} if it is $\PSPACE$-hard and lies in $\PSPACE$.

If $(\Pi_{\mathrm{yes}},\Pi_{\mathrm{no}})$ reduces to $(\Sigma_{\mathrm{yes}},\Sigma_{\mathrm{no}})$ by $f$, and an algorithm $A$ of one of the two resource bounds solves the latter, then $A\circ f$ solves the former within the same bound, the output of $f$ being polynomially bounded in length.
Both classes are therefore closed downward under these reductions, promise problems included.

Reductions compose.
If $f$ reduces the first pair to the second and $g$ reduces the second to $(\Xi_{\mathrm{yes}},\Xi_{\mathrm{no}})$, then $g\circ f$ is computable in polynomial time and carries $\Pi_{\mathrm{yes}}$ into $\Xi_{\mathrm{yes}}$ and $\Pi_{\mathrm{no}}$ into $\Xi_{\mathrm{no}}$, so it reduces the first to the third.
In particular, reducing a $\PSPACE$-hard promise problem to a promise problem $\Xi$ makes $\Xi$ itself $\PSPACE$-hard, which is the form every hardness argument below takes.
\end{definition}

The next proof gives the finite-graph recursive argument in full.
The midpoint recursion follows Savitch's quadratic-space simulation method \cite{Sav70}; \cite[Theorem 4.14, Section 4.2.1]{AB09} gives a textbook exposition.
We need the stated endpoint tests and space bound for the graph presentation used here.

\begin{lemma}[Reachability in an implicit graph]
\label{lem:2.26}
Let $\{G_w\}_{w\in\zo^*}$ be implicitly presented in the sense of \Cref{def:2.18}, with vertex length bound $\mu$ and tests $A_V$ and $A_E$.
Deciding, given $w$ and two strings $r,r'$, whether both are vertices of $G_w$ and are connected in $G_w$ is possible in space polynomial in $|w|+\log(|r|+|r'|+2)$, hence in space polynomial in $|w|$ whenever $|r|$ and $|r'|$ are bounded by $2^{\mathrm{poly}(|w|)}$, as they are at every use below.
\end{lemma}

\begin{proof}
Increase the polynomial bound $\mu$ to a nonnegative integer-valued bound if necessary and put $m=\mu(|w|)$.
First check that $w$ is well formed, and reject otherwise.
Count each endpoint's length and reject it as soon as that count exceeds $m$.
The fields of the input triple $(w,r,r')$ and these counters can be read using space polynomial in $|w|$ plus $O(\log(|r|+|r'|+2))$; the fixed arity of the triple and its binary length fields require only logarithmic counters.
Once both endpoints pass, store them and run $A_V$ on each.
These tests have arguments of length polynomial in $|w|$.
Reject unless both are vertices.

For vertices $u,v$, define a recursive Boolean procedure $R_i(u,v)$.
At $i=0$, accept exactly when $u=v$ or $A_E(w,u,v)=1$.
For $i>0$, enumerate all strings $z$ of length at most $m$.
Discard $z$ unless $A_V(w,z)=1$.
For each remaining $z$, evaluate $R_{i-1}(u,z)$ and, if it accepts, evaluate $R_{i-1}(z,v)$.
Accept if both evaluations accept for some $z$, and reject when enumeration finishes otherwise.
The two recursive calls are performed sequentially, reusing their workspace.

Induction on $i$ shows that $R_i(u,v)$ accepts exactly when a walk of at most $2^i$ edges joins its endpoints.
The base case includes the zero-edge walk.
Two accepting recursive calls concatenate to a walk of at most $2^i$ edges.
Conversely, split a walk of length at most $2^i$ after its first $\min(2^{i-1},L)$ edges, where $L$ is its length.
The two resulting walks each have length at most $2^{i-1}$ and meet at a vertex considered by the enumeration.
This proves the induction.

There are at most $2^{m+1}-1$ possible vertices.
Every connecting walk can be shortened to a simple path, so $R_{m+1}(r,r')$ decides connectivity.
Each recursion frame stores a constant number of length-at-most-$m$ strings, an enumeration counter and its depth, using $O(m+1)$ cells.
There are at most $m+2$ simultaneous frames.
Let $P(|w|+m)$ be a fixed polynomial bound for the shared workspace of the vertex and adjacency tests.
These tests reuse that workspace.
Thus total space is
\[
O((m+1)^2)+P(|w|+m)
+O(\log(|r|+|r'|+2)),
\]
which has the asserted form.
Equal valid endpoints, $m=0$, and empty graphs are included in the same procedure.
If endpoint lengths are at most $2^{\operatorname{poly}(|w|)}$, the logarithmic term is polynomial in $|w|$ as well.
\end{proof}

\subsection{Predicates and the Reconfiguration Problem for Formulas}\label{sec:2.4}

Predicates may be supplied as clocked deciders.
Our hardness reductions use formulas with explicit satisfying endpoints.

\begin{definition}[Predicates and clocked deciders]
\label{def:2.19}
A \emph{predicate at length $n$} is a subset $R\subseteq\zo^n$, and a \emph{predicate} is a subset $R\subseteq\zo^*$.
A clocked machine $M$ in the sense of \Cref{def:2.11} \emph{decides $R$ at length $n$} if $M(a)=1$ for every $a\in R\cap\zo^n$ and $M(a)=0$ for every $a\in\zo^n\setminus R$, the outputs being the one-bit strings $1$ and $0$.

A timed-out computation outputs $\bot$ and therefore decides nothing, so a decider must finish within its clock at every input of that length.

It \emph{decides $R$}, without qualification, if $M(a)=1$ for every $a\in R$ and $M(a)=0$ for every $a\in\zo^*\setminus R$, that is, if it decides $R$ at every length at once.

A clocked machine runs in time polynomial in its input length by \Cref{lem:2.12}, so we use \emph{clocked machine} and \emph{clocked polynomial-time machine} interchangeably.
We call $R$ a \emph{clocked polynomial-time predicate at length $n$}, and $M$ a \emph{clocked polynomial-time decider} for it, when such an $M$ exists and is the object supplied.

The phrase \emph{presented by a clocked polynomial-time decider} means that the input to the construction at hand is the pair $\enc{M}$ and $1^n$, so that a deadline sufficient for running $M$ on any $a\in\zo^n$ may be written down in unary by \Cref{lem:2.12}.
Nothing is asked of $M$ at lengths other than $n$, which is why the length is named in the phrase \emph{clocked polynomial-time predicate at length $n$}.
\end{definition}

Write $|\varphi|=|\rho(\varphi)|$ for the size of a formula.
Whether a given assignment satisfies a given formula is decidable in time $O((|\varphi|+n+1)^3)$ on the multitape model of \Cref{sec:2.2}.
Here is an explicit polynomial bound, including $n=0$.
Put $S=|\varphi|+n+1$.
Scan the encoded formula from left to right, keeping binary counters for clause widths and literal indices.
All fields and counters needed for a valid formula have at most $O(S)$ positions.
Decoding each encoded bit, updating a counter and resetting its head can therefore be done with $O(S)$ transitions, for $O(S^2)$ total parsing time.

For each literal, first move the assignment head left $n$ times, which leaves it at cell one under the left-end rule, and then move right to the indexed variable.
There are at most $n$ further head moves, each implementable with an $O(S)$-time counter operation, followed by a read of that variable.
Thus the at most $O(|\varphi|)$ literal tests cost $O(|\varphi|nS)$ transitions in total.
Clause truth flags and the remaining scan counters cost $O(S^2)$.
Since $|\varphi|nS\le S^3$, these operations give the asserted bound.

If $n=0$, a valid formula has no literals; the same scan assigns value zero to an empty clause and value one to an empty conjunction, as required by \Cref{def:2.20}.
The positive term in $S$ also covers the final verdict write.

The difficulty in everything that follows lies in moving between two assignments, not in testing one.
The object that holds the reconfiguration structure is accordingly the set of satisfying assignments together with its adjacency, and not that set alone.

\begin{definition}[Solution set and solution graph]
\label{def:2.21}
For a Boolean formula $\varphi$ on $n$ variables, in conjunctive normal form as in \Cref{def:2.20} wherever a form is specified, the \emph{solution set} is
\begin{equation}
\label{eq:2.5}
R_\varphi=\{a\in\zo^n\mid \varphi(a)=1\},
\end{equation}
and the \emph{solution graph} of $\varphi$ is the graph with vertex set $R_\varphi$ in which two assignments are adjacent when they differ in exactly one variable.
Equivalently, it is the subgraph of the $n$-dimensional hypercube induced by $R_\varphi$.
More generally, for any $R\subseteq\zo^n$ the \emph{solution graph of $R$} is the subgraph of the hypercube induced by $R$, so that the notion applies to an arbitrary predicate and not only to one presented by a formula.
\end{definition}

The problem we reduce from asks about two designated vertices, rather than connectivity of the entire solution graph.

\begin{fact}[{Reconfiguration of satisfying assignments; \cite[Theorem 2.9 and Section 3.3 of the full version]{GKMP09}}]
\label{fact:2.27}
The following problem is $\PSPACE$-complete.
Given a Boolean formula $\varphi$ in conjunctive normal form on $n\geq 1$ variables and two satisfying assignments $s_{\mathrm{in}},s_{\mathrm{out}}\in\zo^n$ of $\varphi$, decide whether there is a sequence of satisfying assignments of $\varphi$, beginning at $s_{\mathrm{in}}$ and ending at $s_{\mathrm{out}}$, in which every two consecutive assignments differ in exactly one variable.
In the terminology of \Cref{def:2.16,def:2.21}, the question is whether $s_{\mathrm{in}}$ and $s_{\mathrm{out}}$ are connected in the solution graph of $\varphi$.
Hardness holds already when $\varphi$ is in $3$-conjunctive normal form, and the general form stated here follows, every $3$-conjunctive normal form being a conjunctive normal form.
\end{fact}

\subsection{The Full Fiber and Its Query Language}\label{sec:3.1}

The next definitions allow programs of every length up to a budget and programs of one exact length; the results below concern exact lengths only.

\begin{definition}[Adjacency models]\label{def:3.1}
Two strings $p, q \in \zo^{*}$ are \emph{flip-adjacent} if $|p| = |q|$ and they differ in exactly one position.
They are \emph{edit-adjacent} if one is obtained from the other by flipping, inserting, or deleting a single bit.
Flip-adjacency is the restriction of edit-adjacency to strings of equal length.
Indeed, an insertion or a deletion changes the length by exactly one, so if $|p| = |q|$ and $p$ and $q$ are edit-adjacent then the single edit is a flip and they differ in exactly one position.
Conversely, a flip is one of the three permitted edits.

Both relations are symmetric, since insertion and deletion are mutually inverse and a flip is its own inverse.
Both are irreflexive as well: a flip changes a position, and an insertion or a deletion changes the length.
The \emph{edit distance} between two strings is the least number of single-bit flips, insertions, and deletions carrying one to the other.
Edit-adjacency is therefore edit distance one.
\end{definition}

We apply these moves only to programs that meet the target and deadline constraints.

\begin{definition}[Description landscape]\label{def:3.2}
For $x, y \in \zo^{*}$, a deadline $t$, a length budget $\ell$, and a length $n$, the \emph{description landscape} is the set
\begin{equation}\label{eq:3.1}
\Land^{t}_{\ell}(x|y) = \{p \in \zo^{\le \ell} \mid U(p,y) = x \text{ within } t \text{ steps}\}.
\end{equation}
Its \emph{flip graph} $\Gland^{t}_{\ell}(x|y)$ has vertex set $\Land^{t}_{\ell}(x|y)$ and joins two descriptions when they are flip-adjacent; its \emph{edit graph} $\Eland^{t}_{\ell}(x|y)$ has the same vertex set and joins two descriptions when they are edit-adjacent.
The \emph{fixed-length landscape} is
\[
\Land^{t}_{=n}(x|y) = \{p \in \zo^{=n} \mid U(p,y) = x \text{ within } t \text{ steps}\},
\]
whose flip graph is written $\Gland^{t}_{=n}(x|y)$.
By the clause of \Cref{def:3.1} making flip-adjacency the restriction of edit-adjacency to equal lengths, its edit graph coincides with its flip graph, and so is not written separately.
When the machine is an efficient machine $V$ other than the reference machine we write $\Land^{t}_{\ell,V}(x|y)$ and $\Land^{t}_{=n,V}(x|y)$, and likewise for the graphs.
The subscript is dropped only where the text names the machine, as in ``with respect to $V$''; without a subscript or such a phrase the machine is $U$.
Omitting the condition, as in $\Land^{t}_{=n,V}(x)$, means the auxiliary string $\eps$, or no auxiliary input for a program-only machine, by the convention of \Cref{def:2.6}.
\end{definition}

The full native fiber $\mathcal F_V(x,N,t)$ of \Cref{def:full-native-fiber} is exactly the flip graph $\Gland^t_{=N,V}(x)$, and we call either one the fiber of $x$.

\begin{lemma}[Landscapes are implicit]\label{lem:3.3}
Let $V$ be an efficient machine, let $x, y \in \zo^{*}$, and let the deadline $t$, the length budget $\ell$, and the length $n$ be presented in unary.
Then the flip graph and the edit graph of $\Land^{t}_{\ell,V}(x|y)$ are implicit, as is the flip graph of the fixed-length landscape $\Land^{t}_{=n,V}(x|y)$, and likewise for the unconditional landscapes obtained by omitting the condition.
The presenting strings are $(x, y, 1^{t}, 1^{\ell})$ and $(x, y, 1^{t}, 1^{n})$ in the conditional case and $(x, 1^{t}, 1^{\ell})$ and $(x, 1^{t}, 1^{n})$ in the unconditional one, with vertex length bound $\mu(k) = k$ and with $\mathscr{W}$ the set of tuples of the shape displayed, in which the last two fields are unary.
Consequently the $st$-connectivity problem for any of these graphs lies in $\PSPACE$.
\end{lemma}

\begin{proof}
We verify the four conditions of \Cref{def:2.18}.
\Cref{prop:2.3} decides well-formedness: it recovers the fields of a tuple in polynomial time, and so tests both the arity and the unarity of the last two fields.
A string failing that test presents the empty graph.

A vertex is a string of length at most $\ell$, and $\ell$ is bounded by the input size because it is presented in unary, so vertices have polynomial length.

Membership of a candidate string $r$ in the vertex set is decided by testing $|r| \le \ell$ and then computing, from $(r, y, 1^{t})$, the configuration of $V(r,y)$ after $t$ steps.
That computation costs time polynomial in $t + |r| + |y|$, hence polynomial in the input length, because $V$ is efficient and $t$ is presented in unary.

The test accepts if and only if that configuration is halting, its state is not the distinguished one, and the contents of its output tape are exactly $x$.
This is correct for two reasons.
A halting configuration is its own successor by \Cref{def:2.6}, so the configuration after $t$ steps records the output of any computation that halted at some step $j \le t$.
And a halt in the distinguished halting state yields $\bot$, which differs from every string, the empty one included.

Adjacency of two candidate strings $r$ and $r'$ is decided as follows.
First run the membership test above on $r$ and on $r'$, rejecting unless both are vertices.
This is needed because \Cref{def:2.18} asks $A_{E}$ to decide adjacency in $G_{w}$, and a pair in which either member fails to be a vertex is not adjacent there.

For the flip graph, accept if and only if $|r| = |r'|$ and $|\{i : r_{i} \ne r'_{i}\}| = 1$, which is decided by one pass comparing the two strings position by position and counting the mismatches.

For the edit graph, accept if the flip test accepts, and otherwise test the two length cases.
If $|r'| = |r| + 1$, accept if and only if there are an index $j$ with $1 \le j \le |r'|$ and a bit $b$ with $r' = r_{1} \cdots r_{j-1} b r_{j} \cdots r_{|r|}$.
That is decided by trying each of the $|r'|$ values of $j$ and comparing the resulting string with $r'$.
If $|r| = |r'| + 1$, run the same test with the roles of $r$ and $r'$ exchanged.

In every remaining case, that is when $|r| = |r'|$ or $||r| - |r'|| \ge 2$, reject.
No single edit changes a length by more than one, and by \Cref{def:3.1} two strings of a common length are edit-adjacent only when they are flip-adjacent.

Each of these tests performs $O(|r| + |r'|)$ comparisons of $O(|r| + |r'|)$ symbols, so it runs in time polynomial in the total length of its arguments, as \Cref{def:2.18} requires.
The graphs are therefore implicit.

For the fixed-length landscape the only change is that the length test $|r| \le \ell$ is replaced by $|r| = n$, and $n$ is bounded by the input size because it too is presented in unary.
For the unconditional landscapes the auxiliary string is omitted from the simulation.
Neither change affects the remaining three conditions.
The final claim now follows from \Cref{lem:2.26}.
\end{proof}

To make connectivity a language, defined on every input string, we also say how inputs that are not well-formed queries are answered.

\begin{problem}[Unconditional connectivity]
\label{prob:7.1}
Fix a machine $V$, read unconditionally, so that $V(p)$ abbreviates $V(p,\eps)$ for a machine that takes a condition, by the convention of \Cref{def:2.6}.
The input is a tuple $(x, 1^{t}, p_{\mathrm{in}}, p_{\mathrm{out}})$ in which $x \in \zo^{*}$, the deadline $t$ is presented in unary, and $p_{\mathrm{in}}, p_{\mathrm{out}}$ are two programs of a common length $N$ lying in $\Land^{t}_{=N,V}(x)$.
The length $N$ is read off the two programs and is therefore bounded by the input length.
The question is whether $p_{\mathrm{in}}$ and $p_{\mathrm{out}}$ are connected in the flip graph $\Gland^{t}_{=N,V}(x)$.
We call this the \emph{unconditional reconfiguration connectivity problem with respect to $V$}.
\end{problem}

Write $\mathrm{Conn}(V)$ for this language, rejecting malformed tuples, nonunary deadlines, unequal endpoint lengths, and nonvertices.
We call $\mathrm{Conn}(V)$ the \emph{full connectivity language} of $V$, as opposed to its fixed-target restrictions $\mathrm{Conn}_x(V)$.
For fixed $x$, the language $\mathrm{Conn}_x(V)$ additionally rejects a tuple whose target field is not $x$.
Query length always means the entire tuple length, including $1^t$.

\begin{proposition}
\label{prop:7.9}
For every efficient machine $V$ in the sense of \Cref{def:2.6}, the unconditional reconfiguration connectivity problem with respect to $V$ is in $\PSPACE$.
\end{proposition}

\begin{proof}
A query of \Cref{prob:7.1} names the length $N$ only through its two programs, whereas the presenting string of the fixed-length landscape in \Cref{lem:3.3} carries $1^{N}$, so the algorithm first forms that string.
On an input $\iota$ of length $\ell$ it proceeds as follows.

First, it recovers the fields of $\iota$ as in the proof of \Cref{lem:3.3}, by \Cref{prop:2.3}, and rejects unless $\iota$ is a tuple $(x,\tau,p_{\mathrm{in}},p_{\mathrm{out}})$ of four strings.
It then rejects unless $\tau=1^{t}$ for some $t$, and unless $|p_{\mathrm{in}}|=|p_{\mathrm{out}}|$.
These are exactly the malformed tuples, nonunary deadlines and unequal endpoint lengths that $\mathrm{Conn}(V)$ rejects, and the tests take time polynomial in $\ell$.

Second, it puts $N=|p_{\mathrm{in}}|$ and writes the presenting string $w=(x,1^{t},1^{N})$ of the unconditional fixed-length landscape of \Cref{lem:3.3}.
Since $|x|$, $t$ and $N$ are each at most $\ell$, the string $w$ has length polynomial in $\ell$ and is written in time polynomial in $\ell$, and it is well formed in the sense of that lemma.
The graph it presents is $\Gland^{t}_{=N,V}(x)$.

Third, it runs the procedure of \Cref{lem:2.26} for the family of \Cref{lem:3.3} on $(w,p_{\mathrm{in}},p_{\mathrm{out}})$ and returns its answer.
That procedure accepts exactly when $p_{\mathrm{in}}$ and $p_{\mathrm{out}}$ are both vertices of $\Gland^{t}_{=N,V}(x)$, that is, both lie in $\Land^{t}_{=N,V}(x)$, and are connected in that graph.
It runs in space polynomial in $|w|+\log(|p_{\mathrm{in}}|+|p_{\mathrm{out}}|+2)$, hence in space polynomial in $\ell$.

An input passing the first step is therefore accepted if and only if both endpoints are vertices of the fiber at length $N$ and deadline $t$ and are connected in it, which is the definition of $\mathrm{Conn}(V)$.
The first two steps run in polynomial time and hence in polynomial space, and storing $w$ uses space polynomial in $\ell$, so the whole algorithm runs in polynomial space.
The argument uses $V$ only through \Cref{lem:3.3}, which needs only that $V$ is efficient; whether $V$ is onto plays no role.
\end{proof}

\subsection{Local Isolation and a Rigid Reference}\label{sec:local-reference-hardness}

A hard solution graph can sit inside the fiber of one machine, as the programs beginning with one header, whatever the machine's other programs do.
It suffices that no one-bit change inside the header keeps a program in the fiber.

\begin{proposition}[Local isolation suffices for hardness]
\label{prop:7.17}
Let $V$ be an efficient universal machine.
Suppose there is a polynomial-time computable map sending the presentation, as in \Cref{def:2.19}, of every clocked polynomial-time predicate $R \subseteq \zo^{n}$, at every $n \ge 1$, to a witness length $b' \ge n$, a length $N$, a target $x$, a deadline $1^{t}$, and a string $h$ with $|h| + b' = N$, such that, writing $R' = \{ w u \mid w \in R, u \in \zo^{b' - n} \}$ for the padding of $R$ to length $b'$,
\begin{itemize}
\item for every $v \in \zo^{b'}$, the program $h v$ outputs $x$ within $t$ steps if and only if $v \in R'$; and
\item for every $v \in \zo^{b'}$, every program at Hamming distance one from $h v$ that differs from it within the first $|h|$ bits fails to output $x$ within $t$ steps.
\end{itemize}
Then, at each such $R$, the map $v \mapsto h v$ is an isomorphism from the solution graph of $R'$ onto a union of connected components of the flip graph of $\Land^{t}_{=N}(x)$ with respect to $V$; and consequently the unconditional reconfiguration connectivity problem with respect to $V$ is $\PSPACE$-complete.
If moreover the map outputs one and the same target $x_{0}$ for every such $R$, then the fixed-target language $\mathrm{Conn}_{x_{0}}(V)$ is $\PSPACE$-complete as well.
\end{proposition}

\begin{proof}
Membership is \Cref{prop:7.9}.
For hardness, take a valid instance $(\varphi,s,t_0)$ of \Cref{fact:2.27}, with $n\ge1$ variables.
First present its solution predicate $R=R_\varphi$ as a clocked decider at length $n$.
Hardwire $\varphi$ and unary $n$ into a clocked decider which checks exact input length by reading at most $n+1$ cells and then uses the formula-evaluation procedure given after \Cref{def:2.19}.
Wrong-length inputs are rejected at that length check, even if their actual length is much larger.
Hence all its runs are bounded by one fixed polynomial in $|\varphi|+n+1$.

Choose an integer constant coefficient exceeding that bound and set all other clock coefficients to zero.
This is a permitted degree-zero clock, and its unary coefficient and complete decider code have polynomial length in the formula instance.
This construction uses a new formula decider; it does not encode the runtime of an arbitrarily supplied clocked machine at a hardwired length.
Apply the hypothesis to this presentation of $R$ to obtain $b',N,x,1^t,h$ in polynomial time.

The supplied integer $b'$ need not itself be polynomially bounded.
Put
\[
b''=\min\{b',\max(n,t+1)\},\qquad N''=|h|+b''.
\]
Both integers are computable in polynomial time from the supplied data, and $b''\le\max(n,t+1)$ is polynomially bounded because $1^t$ is supplied explicitly.

If $b''=b'$, both hypotheses remain unchanged.
Otherwise $b''\ge t+1$.
For every $v\in\zo^{b''}$, extend $v$ to $\widetilde v=v0^{b'-b''}$.
A program head starting at cell one and moving by at most one cell per transition cannot inspect any cell beyond $t+1$ during its first $t$ transitions.

The programs $hv$ and $h\widetilde v$ agree on all these cells, since $|hv|\ge b''\ge t+1$.
Their other initial tapes and states are identical.
Induction on the native transitions therefore gives identical configurations through time $t$, except for unread program cells.
In particular, their normal halting outputs by $t$ agree.
The same argument applies to $h'v$ and $h'\widetilde v$ for every header $h'$ differing from $h$ in one bit.
Since $b''\ge n$, the first $n$ bits of $v$ and $\widetilde v$ also agree.
Thus both hypotheses transfer to $b''$ and $N''$; no exponentially long extension is constructed by the reduction.

Write $R''=\{wu:w\in R, u\in\zo^{b''-n}\}$ and put $\widehat s=s0^{b''-n}$ and $\widehat t_0=t_0 0^{b''-n}$.
The first transferred hypothesis says that the target-$x$ vertices with prefix $h$ are exactly $hR''$.
A feasible flip within the last $b''$ positions is precisely a solution-graph edge of $R''$.
A flip within $h$ is infeasible by the second transferred hypothesis.
Hence $hR''$ is a union of components of the entire length-$N''$, deadline-$t$ fiber, and prefixing by $h$ is a graph isomorphism onto that union.

Padding preserves and reflects connectivity of the two zero-padded endpoints: a path in $R$ lifts with padding fixed, while projecting a path in $R''$ onto its first $n$ coordinates gives a walk in $R$ after stationary steps are removed.
Consequently $s$ and $t_0$ are connected in the solution graph of $\varphi$ if and only if $h\widehat s$ and $h\widehat t_0$ are connected in the entire target-$x$ fiber.
The identical edge argument applied directly to the original suffix length $b'$ proves the statement's isomorphism onto a union of components at length $N$.
The truncated map to $(x,1^t,h\widehat s,h\widehat t_0)$ is polynomial-time, including writing its endpoints.
Malformed formulas or nonsatisfying endpoints of the formula instance are recognized in polynomial time and sent to a fixed unequal-length no-query.
This proves that $\mathrm{Conn}(V)$ is $\PSPACE$-hard.

For the fixed-target assertion, suppose that the map outputs the target $x_{0}$ for every $R$, and let $f$ be the reduction just described.
Every query that $f$ outputs on a valid instance is $(x_{0},1^{t},h\widehat s,h\widehat t_{0})$, whose target field is $x_{0}$.
By the definition of $\mathrm{Conn}_{x_{0}}(V)$ after \Cref{prob:7.1}, a query with target field $x_{0}$ lies in $\mathrm{Conn}_{x_{0}}(V)$ if and only if it lies in $\mathrm{Conn}(V)$, and every query outside $\mathrm{Conn}(V)$ lies outside $\mathrm{Conn}_{x_{0}}(V)$.

Hence a valid yes-instance is sent into $\mathrm{Conn}_{x_{0}}(V)$, by the equivalence proved above.
A valid no-instance is sent to a query with target field $x_{0}$ outside $\mathrm{Conn}(V)$, hence outside $\mathrm{Conn}_{x_{0}}(V)$.
An invalid instance is sent to the fixed unequal-length no-query, which lies outside $\mathrm{Conn}(V)$ and so outside $\mathrm{Conn}_{x_{0}}(V)$.
Thus the same $f$ reduces the problem of \Cref{fact:2.27} to $\mathrm{Conn}_{x_{0}}(V)$, which is therefore $\PSPACE$-hard.
For membership, a query is accepted exactly when its target field equals the fixed string $x_{0}$, which is checked in polynomial time, and the algorithm of \Cref{prop:7.9} accepts it; this uses polynomial space.
\end{proof}

For the reference machine $U$, rigidity of the ambient encoding (\Cref{prop:2.3}) supplies the second hypothesis of \Cref{prop:7.17}: no machine encoding is a prefix of a program obtained from $\enc{M}z$ by changing one bit of its code $\enc{M}$.

\begin{corollary}[Every fixed target is hard for the rigid reference; previewed in \Cref{thm:preview-parameter}]
\label{cor:7.18}
For the machine $U$ of \Cref{def:2.10}, the language $\mathrm{Conn}_y(U)$ is $\PSPACE$-complete for every fixed target $y\in\zo^*$, including $\eps$.
Consequently $\mathrm{Conn}(U)$ is $\PSPACE$-complete, in particular already at the fixed target $00$.
\end{corollary}

\begin{proof}
Fix $y\in\zo^*$ and choose a fixed different string $r_y$: use $r_y=0$ if $y=\eps$, and $r_y=\eps$ otherwise.
Membership is \Cref{prop:7.9}, with the target field checked against this fixed $y$.
We verify \Cref{prop:7.17} with output target $y$.
Let a clocked decider $\mathscr D$ decide $R\subseteq\zo^n$, with supplied clock $q_{\mathscr D}$.

Construct $M$ so that it normally outputs $y$ on members of $R$ and $r_y$ on nonmembers at length $n$.
So that $\mathscr D$ may move its output head freely, $M$ does not let it write on the real output tape.
Instead, $M$ redirects the output writes and head moves of $\mathscr D$ to a fresh two-track work tape.
The first track carries a permanent marker on the first cell, recording the original output origin, and the second track starts blank.
The inherited program and work tapes, data and initial heads are unchanged.
The new transition table preserves the marker when writing and enforces the original left-end rule at that cell.
Since $\mathscr D$ never reads its output tape, these changes reproduce every inherited state and all native head movements, with at most a fixed constant factor in time.
In particular, overwrites and moves left before the first write are preserved.

On a normal halt, return the buffered-output head to its first-cell marker and inspect the symbol in its first cell.
At length $n$ it is the one-bit verdict guaranteed by \Cref{def:2.19}.
Only then write $y$ or $r_y$, according as that verdict is $1$ or $0$, on the untouched real output tape.
Distinguished halts remain distinguished.

A head visits at most $T+1$ cells in $T$ transitions, so returning to the marker costs at most $T$ further transitions.
Thus a fixed constant $c_y$, allowed to depend on the two fixed strings $y$ and $r_y$, gives a clock $c_y q_{\mathscr D}(k)+c_yk+c_y$ on inputs of length $k$, with degree at most $D$ and unary coefficients polynomially constructible from those of $\mathscr D$.
Off the promised length, any other normal output can be sent to a fixed default string after the same bounded return; no correctness is required there.
The table and data construction are polynomial in the supplied decider code.

Set $x=y$, $b'=n$, $h=\enc M$, and $N=|h|+n$.
The uniform clock bound and the fixed uniform native overhead of $U$ supply a common deadline $t$, writable in unary in polynomial time, by which $U(hv)=M(v)$ has halted for every $v\in\zo^n$.
The first hypothesis of \Cref{prop:7.17} now holds with $R'=R$.
If $h'$ differs from $h$ in one bit, \Cref{prop:2.3} says that no codeword of any length prefixes $h'v$.
At the empty condition $U$ therefore diverges on this program, proving the second hypothesis.
The map $\mathscr D\mapsto(b',N,x,1^{t},h)$ just described is computable in polynomial time, is defined at every $n\ge1$, and outputs the same target $x=y$ for every $R$.
The fixed-target clause of \Cref{prop:7.17} therefore applies with $x_{0}=y$ and shows that $\mathrm{Conn}_{y}(U)$ is $\PSPACE$-complete.

When $y=\eps$, the untouched blank output is admitted only on a normal halt; rejection normally outputs $r_y=0$, while distinguished halts and divergence remain infeasible.
Thus the empty target is included.
Since the hypotheses of \Cref{prop:7.17} hold for $U$, its main conclusion gives that $\mathrm{Conn}(U)$ is $\PSPACE$-complete; the case $y=00$ of the argument above gives the completeness of $\mathrm{Conn}_{00}(U)$.
\end{proof}

\subsection{Adding Interpretation Without Changing Native Fibers}\label{sec:interpretation-completion}

The constructions may first be program-only.
Interpretation can then be added without moving even the original deadlines.

\begin{lemma}[Interpretation completion]
\label{lem:interpretation-completion}
For every efficient onto program-only machine $W$, there is an efficient conditional machine $\widehat W$ with the interpretation property such that, for every $p,x,t$, the computation $\widehat W(p,\eps)$ normally halts with output $x$ within $t$ steps if and only if $W(p)$ does.
Thus every full native fiber, the full language $\mathrm{Conn}(\widehat W)=\mathrm{Conn}(W)$, and all unconditional measures $\K^t_{\widehat W}=\K^t_W$ are preserved at the original deadlines.
If $W$ has the simulation property, $\widehat W$ has it with the same codes and the same simulation polynomial.
In particular, any additive-length, polynomial-deadline measure agreement with $U$ is unchanged.
\end{lemma}

\begin{proof}
Let $q_0$ be the initial control state of $W$.
It cannot be a normal halting state, since the initially blank output tape would give the same output $\eps$ for every program, contradicting onto-ness.
The distinguished halting state would likewise preclude onto-ness.
Retain a native copy of all of $W$'s states, tapes, hardwired data, tape layout and transition rules.
Add the auxiliary tape and a disjoint set of work tapes for a native copy of the fixed reference machine $U$, with its declared initial data already in place.
The copies share the program and output tapes, and $U$ uses the new auxiliary tape as its condition.
Finite tape and state renamings do not change any transition count.

Give $\widehat W$ one fresh nonhalting entry state $q_*$.
For every tuple of symbols under the inherited heads, define the transition from $q_*$ as follows.
If the auxiliary first cell is blank, perform exactly the transition that $W$ performs from $q_0$ on those symbols, including its output action, head movements and next state in the inherited copy.
All added heads remain stationary and added work cells unchanged.
If that auxiliary cell is a bit, perform no inherited write or head movement and enter the native initial state of $U$.
This conditional lookup is one ordinary multitape transition: it reads the additional tape at the same time as the inherited tapes and applies the corresponding finite-table rule.
It does not combine two transitions of $W$.

On the empty condition, the first transition is therefore exactly $W$'s first transition, and each later inherited transition is unchanged.
If $W$ returns to its old initial state, it enters that inherited state rather than the fresh $q_*$, so it does not branch again.
This covers a first transition that writes output, normally halts, enters the distinguished halting state, or returns to $q_0$.
At time zero neither machine is normally halting.
Thus their unconditional string halts agree at every number of native steps, including zero.

On a nonempty condition, the first transition leaves all heads and shared tapes unchanged, so the computation is the native $U(p,y)$ computation after that one transition.
Use the nonempty interpretation encoding $\enc{M}$ of \Cref{def:2.10}.
The uniform polynomial interpretation overhead of $U$ gains just one transition; divergence and distinguished outcomes are retained.
Onto-ness follows from the unchanged empty branch, and efficiency follows from the fixed native model.
If $W$ has the simulation property, that entire empty branch and all its native deadlines are unchanged, proving the optional assertion.
The pointwise equality of unconditional halting predicates gives identical fibers and identical full connectivity languages, including malformed-query rejection.
Taking the same minimum over the same feasible programs gives $\K^t_{\widehat W}(x)=\K^t_W(x)$ for every $x,t$, and hence preserves the stated measure comparisons without changing any deadline or length constant.
\end{proof}

\section{Fragmentation Forced by Adjunction}\label{sec:7.7}\label{sec:fragmentation}

An adjunction-universal machine must run every program of every source machine after that source's code, so a fixed source prescribes the outputs of all programs in the cylinder of its code; we call these programs \emph{mandatory descriptions}.
This section first proves, by counting, that every adjunction-universal machine disconnects the fibers of almost all targets at linear overhead.
It then shows that two machines satisfying the reference conditions can disagree about connectivity at one early linear deadline and program length $2n+3$, where $n$ is the target length.
Next it proves, for every adjunction-universal machine, that most fibers at that same length have exponentially many components at every deadline beyond one polynomial deadline.
It closes with a second family of separated mandatory descriptions that trades program length against the number of components.

\subsection{Persistent Fragmentation at Linear Overhead}
\label{sec:linear-overhead-fragmentation}

A single code supplies two distant descriptions of every target.
A path joining them may pass through programs outside the cylinder of that code.
Counting the outside vertices needed to join these pairs gives a bound that is independent of all other programs of the machine.
We call this counting the packing argument.

\begin{theorem}[Persistent fragmentation at linear overhead; previewed in \Cref{thm:preview-fragmentation}]
\label{thm:linear-overhead-fragmentation}
Let $V$ have the simulation property.
There are an integer $d\ge0$ and a nondecreasing polynomial $Q_V$ such that, for every integer $m\ge1$, putting $N=m+d+1$ gives the following conclusions.
Every target in $\zo^m$ has a nonempty fiber by deadline $Q_V(N)$.
There is one set $B_m\subseteq\zo^m$ with
\[
|B_m|\ge\left(1-\frac{2(2^d-1)}{m+2}\right)2^m
\]
such that $\mathcal F_V(x,N,t)$ is disconnected for every $x\in B_m$ and every $t\ge Q_V(N)$.
\end{theorem}

\begin{proof}
Fix a native linear-time machine $F$ which, on input $bz$ with $b\in\zo$, prints $z$ if $b=0$ and $\overline z$ if $b=1$, and enters the distinguished halting state on the empty input.
A scan, return and copy implement $F$ within $a(r+1)$ transitions on every input of length $r$, for a fixed integer $a$.
Let $h=\enc{F}_V$ be its code for $V$, put $d=|h|$, and let $q_V$ be the nondecreasing polynomial in its simulation property.

At suffix length $m+1$, exactly two programs in the cylinder $h\zo^{m+1}$ eventually print a given $x\in\zo^m$: they are $h0x$ and $h1\overline x$.
Adjunction applies to every raw suffix, so these are all the target-$x$ programs in that cylinder, and both halt within
\[
q_V(a(m+2),m+1,d).
\]
One nondecreasing polynomial $Q_V(N)$ majorizes this expression for every $m$, since $N=m+d+1$ and $d,a$ are fixed.
Hence every target fiber is nonempty by that common deadline.

Let $C_m$ be the set of targets $x$ whose two endpoints $h0x$ and $h1\overline x$ are connected in the untimed fiber of $x$ at length $N$.
This does not assume that the whole untimed fiber is connected.
For each $x\in C_m$, choose a simple connecting path.
Its only vertices in $h\zo^{m+1}$ are its endpoints, because those two are the only target-$x$ programs in that cylinder.
Since $m+1\ge2$, the endpoints are not adjacent.
If $d=0$, there is no vertex outside the cylinder and no connecting path, so $C_m$ is empty.
Assume $d\ge1$.
The endpoints $h0x$ and $h1\overline x$ differ in all $m+1$ suffix coordinates, so the path changes each of them at least once, and it must leave and return to the common code $h$, which needs at least two flips inside $h$.
It has at least $m+3$ edges and therefore at least $m+2$ internal vertices, all outside the cylinder.

Different targets' paths share no vertex, because a deterministic program has only one normal output.
There are exactly $(2^d-1)2^{m+1}$ vertices outside the cylinder.
Thus
\[
(m+2)|C_m|\le(2^d-1)2^{m+1}.
\]
The same inequality holds when $d=0$.
Take $B_m=\zo^m\setminus C_m$ and divide by $2^m$ to obtain the stated cardinality bound.
Both endpoints for every $x\in B_m$ are present by $Q_V(N)$.
A connecting path at any later finite deadline would be a path in the untimed fiber, contrary to the definition of $B_m$.
Thus this one set works simultaneously for all later deadlines.
For small $m$ the right-hand side of the bound may be negative, and the statement is then vacuous.
The case $m=0$ is excluded because $h0$ and $h1$ would be adjacent.
\end{proof}

The argument does not need the cylinder to be separated from the rest of the fiber; it counts the outside vertices instead.
Sharper families of separated mandatory descriptions are developed in the rest of this section, and the contrasting eventual-connectivity construction in \Cref{sec:eventual-target-connectivity}.

\subsection{A Common Early Deadline}\label{sec:early-reference-separation}

A machine that prints the first $m$ bits of each program of odd length $2m+1$ has large connected fibers, and adding a parity test makes every such fiber edgeless.
Adding a delayed simulation branch to each gives two machines satisfying all three reference conditions, and none of that branch's programs halts by the common early deadline.

\begin{definition}[The prefix machines]
\label{def:7.38}
For a string $p$ of odd length $2m+1$ write $\mathrm{pre}(p)$ for the prefix of $p$ of length $m$.
Define
\[
V_2(p) =
\begin{cases}
\mathrm{pre}(p) & \text{if $|p|$ is odd},\\
\bot & \text{otherwise,}
\end{cases}
\qquad
V_3(p) =
\begin{cases}
\mathrm{pre}(p) & \text{if $|p|$ is odd and $\wt(p)$ is even},\\
\bot & \text{otherwise,}
\end{cases}
\]
with $\bot$ the output of the distinguished halting state of \Cref{sec:2.2}.
\end{definition}

These two native routines have linear running time.
Their universality below means surjectivity in the convention of \Cref{def:2.6}; adjunction and interpretation are established only for the machines $W_2$ and $W_3$ of \Cref{thm:uniform-reference}, which are built from these two.
At a fixed odd length, the fiber of $V_2$ is a subcube and that of $V_3$ is the even-weight half of the same subcube.

\begin{theorem}[The prefix-machine separation]
\label{thm:7.39}
Put $N(n) = 2n+1$ and let $t(n)$ be a deadline large enough that $V_2$ and $V_3$ of \Cref{def:7.38} halt on every program of length $N(n)$, which the proof shows is achieved by some $t(n) = O(n)$.
The machines $V_2$ and $V_3$ of \Cref{def:7.38} are efficient universal machines in the sense of \Cref{def:2.6}, and for every $x$ with $|x| \ge 1$ the fiber $\Gland^{t(|x|)}_{=N(|x|),V_i}(x)$ has at least two vertices for $i=2,3$, is connected for $V_2$, and is edgeless, hence disconnected, for $V_3$.
\end{theorem}

\begin{proof}
\emph{Both machines are efficient and universal.} Each scans its program, keeping its length parity and, for $V_3$, its weight parity in the finite control.
During that scan it writes a unary full-length counter and one marker on a second counter for every completed pair of input bits.
For a program of odd length $2m+1$, the second counter consequently has exactly $m$ markers.
The full-length counter permits a return scan to the program's first cell; after resetting the second counter's head, the machine copies one input bit per marker and halts.
At a rejected length or parity it returns $\bot$ instead.
All scans, resets and copying take at most $c(|p|+1)$ steps for one constant $c$ fixed by the two finite controls, including the empty program.

The efficiency clause of \Cref{def:2.6} holds by \Cref{fact:2.9}.
At length $N(n)=2n+1$ the bound is $c(2n+2)=O(n)$ for $n\ge1$, so this common bound and every larger one meet the theorem's hypothesis.

For universality, let $x \in \zo^*$ and put $n = |x|$.
Every $p = xz$ with $|z| = n+1$ has odd length $2n+1$ and $\mathrm{pre}(p) = x$, so $V_2(p) = x$.
Among those $z$ there is one with $\wt(xz)$ even, since flipping the last bit of $z$ changes that parity, and for such a $z$ also $V_3(p) = x$.
At $n = 0$ the program $p = 0$ has length $1$, prefix $\eps$ and even weight, so both machines print $\eps$.

\emph{The fiber of $V_2$.} Fix $x$ with $n = |x| \ge 1$ and put $N = 2n+1$.
A string $p$ of length $N$ satisfies $V_2(p) = x$ exactly when $\mathrm{pre}(p) = x$, that is exactly when $p = xz$ for some $z \in \zo^{n+1}$.
The computation takes linear time, so $\Land^{t(n)}_{=N}(x) = x\zo^{n+1}$ with respect to $V_2$.
That set has $2^{n+1} \ge 2$ elements.

Two of them, $xz$ and $xz'$, are flip-adjacent exactly when $\Delta(xz,xz') = 1$, which since they share the prefix $x$ holds exactly when $\Delta(z,z') = 1$.
The fiber is therefore isomorphic to the Boolean cube $Q_{n+1}$ of \Cref{def:2.16} under $z \mapsto xz$, and is connected.

\emph{The fiber of $V_3$.} At the same $x$, $N$ and deadline, $V_3(p) = x$ holds exactly when $p = xz$ with $\wt(xz)$ even, that is exactly when $\wt(z) \equiv \wt(x) \pmod{2}$.
Exactly half of the strings $z \in \zo^{n+1}$ have each weight parity, since flipping the last bit is an involution exchanging the two halves, so the fiber has $2^n \ge 2$ elements.

Every element has even weight, while a single flip changes the weight by exactly one and so makes it odd.
Hence no two elements are flip-adjacent, the fiber is edgeless on at least two vertices, and it is disconnected.
\end{proof}

The additional simulation and interpretation properties can be imposed without losing a separation at one common linear deadline.
A delay in the simulation branch leaves the programs that run the prefix machines as the only descriptions present at that deadline.

\begin{theorem}[Uniform separation under the reference conditions; previewed in \Cref{thm:H}]
\label{thm:uniform-reference}
There are conditional machines $W_2,W_3$ satisfying \Cref{def:2.6,def:2.7,def:2.8}, an integer $n_0\ge1$, and a common deadline $t(n)=O(n)$ such that, for every $x\in\zo^n$ with $n\ge n_0$, at length $N(n)=2n+3$ the fiber $\Gland^{t(n)}_{=N(n),W_2}(x)$ is connected on $2^{n+1}$ vertices and the fiber $\Gland^{t(n)}_{=N(n),W_3}(x)$ is edgeless on $2^n$ vertices.
The deadline depends only on $n$, and the measures $\K^t_{W_2}$ and $\K^t_{W_3}$ agree up to additive constants and polynomial changes of the deadline.
\end{theorem}

\begin{proof}
Each $W_i$ dispatches on its inputs.
A nonempty auxiliary string is handed, with the program, to the reference machine $U$, which supplies interpretation.
At the empty auxiliary string, a program $00r$ runs $V_i$ on $r$, a program $11r$ runs $U$ on $r$ after a delay quadratic in the program length, and every other program yields $\bot$.
At the chosen linear deadline only $00$ programs can have halted with a string output, so the fibers are those of $V_i$ with $00$ prepended.

\emph{Native copies of the fixed machines.}
Fix the linear-time implementations of $V_2,V_3$ from \Cref{thm:7.39} and the native reference $U$ from \Cref{def:2.10}.
Each $W_i$ contains finite-control copies of these fixed machines, on disjoint work tapes carrying their declared initial data and layout.
Its preparatory routines use additional tapes and leave those inherited work tapes and their heads unchanged.
A native copy reproduces each inherited transition, including its output action and all head movements, in one transition after a finite tape and state renaming.
Its states are disjoint from the preparatory states, so returning to an inherited initial state does not repeat the preparation.

First inspect the auxiliary first cell without moving any head or writing output.
If it is a bit, enter a native conditional copy of $U$ on the unchanged inputs $p,y$.
That copy starts with its original initial configuration and therefore executes $U(p,y)$ after the single dispatch transition.
Every code $\enc M$, which serves below as the interpretation encoding, is nonempty, whereas the empty condition begins with a blank.
Thus this branch implements interpretation and never applies at the empty condition.

At the empty condition, inspect the first two program bits.
Programs shorter than two bits, and prefixes $01$ or $10$, enter the distinguished halting state.
On $00r$, copy $r$ onto an initially blank private tape, without adding a delimiter, and return its head to cell one.
The copied tape contains exactly $r$, followed by blanks.
Run the native copy of $V_i$, using this tape as its program input and the real output tape as its output.
The copied program tape is physically writable, but the inherited input rules never write it, so it has precisely the read-only behavior required by $V_i$.
Its physical left end also gives the original left-end rule.
Preparation and the native $V_i$ routine take at most $c(|p|+1)$ transitions, for one common fixed integer $c\ge1$.

On $11r$, put $L=|p|$ and first execute an enforced delay of at least $L^2$ transitions, without output or halting.
Scan the program to record $L$ in unary on two fresh tapes, then make $L$ traversals of a block of $L$ marked cells, resetting its head between traversals.
This takes $O((L+1)^2)$ transitions and includes the required $L^2$ transitions.
After the delay, copy $r$ to an initially blank private tape and reset its head to cell one.
Run a second native copy of $U$ with this tape as its program input and with the unchanged empty auxiliary tape.
Its work data, work heads and real output tape are still in their initial configurations.
As above, its input rules leave the copied string unchanged, and its left boundary is the physical first cell.

The resulting computation is exactly the native $U(r,\eps)$ computation after preparation.
Normal outputs, distinguished halts, divergence, overwrites and erasures are all preserved.
In particular, no $11$ program normally halts before its enforced delay finishes, even for target $\eps$.

\emph{The reference conditions.}
Efficiency follows from the fixed native model and \Cref{fact:2.9}.
For every string $x$, choose $r$ with $U(r,\eps)=x$.
Then $W_i(11r,\eps)=x$ after finite preparation, proving onto-ness, including at $x=\eps$.
For adjunction use
\[
\enc M_{W_i}=11\enc M.
\]
This code is polynomial-time computable and has length $|\enc M|+2$.
On every raw suffix $z$, the copied input to $U$ is exactly $\enc Mz$.
The delay and copying take polynomial time in $|\enc M|+|z|+1$.
If $M(z)$ halts in $T$ steps, the uniform simulation polynomial of $U$ bounds the remaining native execution in $T,|z|,|\enc M|$.
Adding the preparation gives one uniform polynomial in $T,|z|,|\enc M_{W_i}|$; divergence is preserved by the native copy.
For interpretation take $\enc M'_{W_i}=\enc M$.
The nonempty-condition branch runs $U(p,\enc M)=M(p)$ with one extra transition and hence has the required uniform polynomial overhead and divergence behavior.
\Cref{fact:2.15}, applied in both directions, now gives the numerical measure comparison.

\emph{The entire graphs at one deadline.}
Put $N(n)=2n+3$ and $t(n)=c(N(n)+1)$.
Choose $n_0=c+1$.
For $n\ge n_0$ we have $N=N(n)\ge c+1$, and
\[
N^2-c(N+1)=N(N-c)-c\ge N-c>0.
\]
Thus every $00$ computation at that length has halted by $t(n)$, whereas no $11$ computation has halted by then.
All other programs produce no string.
For any $x\in\zo^n$, the vertex sets of the two fibers are therefore
\[
\Land^{t(n)}_{=N(n),W_2}(x)=\{00xz:z\in\zo^{n+1}\},
\]
\[
\Land^{t(n)}_{=N(n),W_3}(x)=\{00xz:z\in\zo^{n+1}\text{ and }\wt(xz)\equiv0\pmod2\}.
\]
In the first, $z\mapsto00xz$ preserves and reflects one-bit adjacency, giving $Q_{n+1}$ on $2^{n+1}$ vertices.
In the second, flipping the last bit of $z$ exchanges the two parities, so exactly $2^n$ suffixes are feasible.
Every feasible program has even weight, while a one-bit flip changes parity, so this graph is edgeless.
Since $n\ge n_0\ge1$, it has at least two vertices and is disconnected.
The same length and deadline work for every target of length $n$.
\end{proof}

The deadline in \Cref{thm:uniform-reference} is part of the statement.
At larger deadlines, programs in the delayed simulation branch may enter the fibers.
The theorem consequently does not assert persistent disagreement at every larger deadline, and it gives no complexity separation between machines satisfying the reference conditions.

For every adjunction-universal machine, the two-endpoint bound of \Cref{thm:linear-overhead-fragmentation} already disconnects the fibers of almost all targets at its own length $n+d+1$, where $n$ is the target length, at every deadline beyond one polynomial deadline.
We now prove such fragmentation at the length $2n+3$ of the early separation, again for every adjunction-universal machine, and then force many components.

\subsection{Fragmentation at the Exact Common Length}\label{sec:balanced-fragmentation}

The next construction forces fragmentation at the same length $2n+3$ as the early separation, where $n$ is the target length.
It uses many separated mandatory descriptions in each target fiber.

Write $\mathbb F_2$ for the field with elements zero and one, with addition and multiplication modulo two.
A binary matrix acts on binary column vectors using these operations.
Its rank is the largest number of linearly independent rows, its kernel is the set of vectors it maps to zero, and the preimage set of $x$ is the set of vectors it maps to $x$.
The construction uses one standard finite-field existence fact.
A polynomial is monic when its leading coefficient is one, and is irreducible when it has no factorization into polynomials of smaller positive degree.

\begin{fact}[Binary extension fields; {\cite[Section 19.2, Theorem 19.12]{Sho09}}]
\label{fact:binary-extension-fields}
For every integer $s\ge1$ there is a monic irreducible polynomial $f\in\mathbb F_2[X]$ of degree $s$.
The polynomial residues modulo $f$ form a field with $2^s$ elements, represented uniquely by binary polynomials of degree less than $s$.
\end{fact}

Addition in this field is coefficientwise addition modulo two.
Multiplication is polynomial multiplication followed by division by $f$ and retention of the remainder.
Both operations take polynomial time in $s$ by the elementary algorithms.
The parity-check idea below is classical; see the constructions of Bose and Ray-Chaudhuri and of Hocquenghem \cite{BRC60,Hoc59}.
We prove the precise matrix and distance parameters used here.

\begin{lemma}[Balanced maps with separated preimage sets]
\label{lem:balanced-syndrome-map}
For integers $r\ge n\ge1$, put
\[
s=\left\lceil\log_2(r+1)\right\rceil,
\qquad e=\left\lfloor\frac{n}{s}\right\rfloor.
\]
One can construct in deterministic time polynomial in $r+n$ a binary matrix $H_{n,r}$ of rank $n$, with $n$ rows and $r$ columns, such that every preimage set of $w\mapsto H_{n,r}w$ has exactly $2^{r-n}$ elements and minimum Hamming distance at least $e+1$.
The distance assertion is vacuous for a singleton preimage set.
\end{lemma}

\begin{proof}
\emph{Constructing the field.}
Enumerate the $2^s$ monic binary polynomials of degree $s$.
For each candidate test divisibility by every monic polynomial of degrees $1,\ldots,\lfloor s/2\rfloor$.
A reducible degree-$s$ polynomial has a factor in this range: in any factorization into two positive degrees, one degree is at most $s/2$.
Thus the first candidate passing all tests is irreducible, and \Cref{fact:binary-extension-fields} guarantees that one is found.
For $s=1$ the test list is empty and every monic linear polynomial is irreducible.
There are fewer than $2^{s/2+1}$ trial divisors, and each division takes polynomial time in $s$.
The full search therefore takes $O(2^{3s/2+1}\operatorname{poly}(s))$ time.
Since $2^s<2(r+1)$, this is polynomial in $r$.
Construct the field $\mathbb F=\mathbb F_2[X]/(f)$ using the polynomial representations in the fact.

\emph{The parity checks.}
Choose $r$ distinct elements $\alpha_1,\ldots,\alpha_r$ of $\mathbb F$ by enumerating their binary representations.
There are enough because $2^s\ge r+1$.
Form the $e$-by-$r$ field matrix with entry $\alpha_i^j$ in column $i$ and row $j$, for $j=0,\ldots,e-1$; when $e\ge1$, its zeroth row consists of ones, including at $\alpha_i=0$.
Expand each field entry into its $s$ binary coordinates.
This gives at most $es\le n$ binary rows and a binary matrix $B$ with $r$ columns.
For a binary vector $w$, the condition $Bw=0$ means exactly
\[
\sum_{i=1}^r w_i\alpha_i^j=0
\quad\text{in }\mathbb F,
\qquad 0\le j<e.
\]

Suppose a nonzero such vector has weight $q\le e$.
Let $i_1,\ldots,i_q$ be its support.
The first $q$ field rows on these $q$ columns form an invertible matrix.
To see this, a dependence among its rows would give a nonzero polynomial of degree at most $q-1$ vanishing at the $q$ distinct points $\alpha_{i_1},\ldots,\alpha_{i_q}$.
A nonzero polynomial of degree at most $q-1$ has at most $q-1$ roots: division by $X-a$ removes a root $a$ and lowers the degree by one, and induction proves the bound.
Thus its rows are independent, and a square matrix with independent rows is invertible by elimination over the field.
The first $q$ displayed equations say that this square matrix maps the all-ones vector $(w_{i_1},\ldots,w_{i_q})$ to zero, contradicting its invertibility.
Every nonzero vector in the binary kernel of $B$ therefore has weight at least $e+1$.
When $e=0$, this conclusion is just the trivial positive-weight bound.

\emph{Full rank and balanced preimages.}
Use binary Gaussian elimination to take a basis of the rows of $B$.
Append standard unit rows not yet in their span until there are exactly $n$ independent rows, giving $H_{n,r}$.
This succeeds because $n\le r$: if a proper row span contained every unit row it would already contain every binary row vector.
Each appended row imposes another zero equation, so the new kernel is contained in the old one and retains its minimum-distance bound.

Rank $n$ gives $n$ pivot columns; setting the other $r-n$ coordinates freely and solving for the pivots yields exactly $2^{r-n}$ preimages of each right-hand side $x\in\mathbb F_2^n$.
Thus the map is onto and balanced.
The difference of two vectors in one preimage set is a nonzero kernel vector, whose weight is their Hamming distance, proving the asserted separation.
Field arithmetic, matrix formation, elimination and the unit-row extensions all take polynomial time in $r+n$.
The search cost above shows that the same holds for the complete deterministic construction.
\end{proof}

The next elementary bound counts every vertex of a connecting subgraph, including vertices outside the cylinder of a code.
It therefore needs no assumption, such as the rigid encoding of the reference machine $U$, about programs outside that cylinder.
Its proof uses a spanning tree of a nonempty finite connected graph, which is a subgraph on all its vertices with a unique simple path between each pair of vertices.
In this section $D$ denotes a lower bound on Hamming distances, not the clock degree bound of \Cref{def:2.11}.

\begin{lemma}[Connecting separated terminals]
\label{lem:separated-terminals}
If a finite connected subgraph $G$ of a Boolean cube contains $k\ge2$ specified vertices, called terminals, whose pairwise Hamming distances are at least $D$, then
\[
2(|V(G)|-1)\ge kD.
\]
In particular, $|V(G)|\ge kD/2$.
\end{lemma}

\begin{proof}
Root a tree at one specified vertex, initially taking only that vertex and no edges.
If some vertex of $G$ has not been reached, a path from the root to it has a first edge leaving the reached set.
Adjoin that edge and its previously unreached endpoint.
The new vertex has only its one attaching edge in the constructed subgraph, so this operation preserves the unique-path property.
After $|V(G)|-1$ additions the resulting tree spans $G$ and has $|V(G)|-1$ edges.

A one-vertex tree has a closed walk of length zero; adjoining a leaf inserts a two-edge detour at its neighbor.
Induction over these additions therefore gives a closed walk from the root using every tree edge twice, of length $2(|V(G)|-1)$.
List the $k$ specified vertices in the order of their first appearances on that walk, the root first.
Cut the walk at these $k$ first appearances.
The last of the resulting $k$ portions runs from the first appearance of the last-listed specified vertex back to the root.
Each of the resulting $k$ portions joins two distinct specified vertices, so it has at least $D$ edges: every cube edge changes just one coordinate.
The portions together are the whole walk, proving the inequality.
\end{proof}

The connected-subgraph bound also limits how many of a family of vertex-disjoint graphs can have few components.
Here a component counts when it meets the graph's specified vertices, and few means fewer than half as many as there are specified vertices.
We state this count separately because both constructions below, the balanced preimage sets and the Reed--Muller family, use it.

\begin{lemma}[Counting components from separated vertices]
\label{lem:separated-component-count}
Let $X$ be a finite set, and let $G_x$, for $x\in X$, be subgraphs of the same Boolean cube $Q_N$ whose vertex sets are pairwise disjoint.
Let $W\ge2$ and $D\ge1$ be integers.
Suppose each $G_x$ contains a specified set $P_x$ of $W$ vertices whose pairwise Hamming distances are at least $D$.
Write $c_x$ for the number of connected components of $G_x$ meeting $P_x$.
Then
\[
\left|\{x\in X:c_x<W/2\}\right|<\frac{4\cdot2^N}{DW}.
\]
For every other $x\in X$, the graph $G_x$ has at least $W/2$ connected components.
\end{lemma}

\begin{proof}
Put $Y=\{x\in X:c_x<W/2\}$.
If $Y$ is empty, the strict bound holds because $4\cdot2^N/(DW)>0$.
This also covers an empty index set $X$.
Suppose therefore that $Y$ is nonempty, and fix $x\in Y$.

At most $c_x$ specified vertices lie in components containing just one specified vertex, since each such component accounts for one of the $c_x$ components.
Consequently the number $a_x$ of specified vertices in components containing at least two satisfies
\[
a_x\ge W-c_x>W/2.
\]

For each such component $C$, let $k_C\ge2$ be its number of specified vertices.
It is a finite connected subgraph of $Q_N$, and those vertices still have pairwise distances at least $D$.
Thus \Cref{lem:separated-terminals} gives $|V(C)|\ge k_CD/2$.
The components of $G_x$ have disjoint vertex sets, so summing these bounds yields
\[
|V(G_x)|\ge\frac D2\sum_{C:k_C\ge2}k_C
=\frac D2a_x>\frac{DW}{4}.
\]

The graphs indexed by different $x$ also have disjoint vertex sets, all inside the $2^N$-vertex cube.
Summing the strict inequality over the nonempty set $Y$ gives
\[
|Y|\frac{DW}{4}<\sum_{x\in Y}|V(G_x)|\le2^N.
\]
Division by the positive number $DW/4$ proves the bound.
For $x\notin Y$, at least $c_x\ge W/2$ components meet $P_x$; these are components of the entire graph, proving the last assertion.
\end{proof}

We can now keep the program length of the earlier separation while allowing every later deadline.
The balanced construction forces exponentially many components for most targets at every sufficiently large target length.

\begin{theorem}[Exponential multiplicity at every large target length; previewed in \Cref{thm:preview-fragmentation,thm:H}]
\label{thm:eventual-fragmentation}
Let $V$ be an adjunction-universal machine.
There are a constant $a_V>0$, a nonnegative integer $C_V$, an integer $n_0\ge C_V+1$ and a nondecreasing polynomial $P_V$ such that, for every $n\ge n_0$, there is a set $\mathcal B_n\subseteq\zo^n$ with
\[
|\mathcal B_n|\ge
\left(1-a_V\frac{\log_2(n+1)}{n}\right)2^n
\]
for which the untimed fiber of $x$ at length $2n+3$, induced by
\[
\{p\in\zo^{2n+3}:V(p,\eps)=x\},
\]
has at least $2^{n-C_V}$ connected components for every $x\in\mathcal B_n$.
The fiber $\Gland^t_{=2n+3,V}(x)$ has at least that many components for every such $x$ and every $t\ge P_V(n)$.
In particular, all these graphs are nonempty and disconnected.
The same set $\mathcal B_n$ works for all these later deadlines; no algorithm for recognizing it is asserted.
\end{theorem}

\begin{proof}
\emph{One fixed polynomial-time machine.}
Define $F$ independently of $V$.
On input $z$, it first determines the full length $L$ and parses $z=1^a0w$; it rejects, meaning that it enters the distinguished halting state, if there is no zero.
Write $r=|w|$ and compute
\[
n=\frac{L+a-3}{2}.
\]
On the programs built below, this value of $n$ is the target length.
If this is not an integer satisfying $r\ge n\ge1$, return $\bot$.
Otherwise construct the matrix of \Cref{lem:balanced-syndrome-map} and output $H_{n,r}w$.
Every part of this algorithm, including its scans and the field search, takes polynomial time in $L+1$, since $r\le L$ and accepted parameters have $n\le r$.
Implement it as a native multitape machine and fix a nondecreasing polynomial $R(L)$ bounding its complete actual running time at every length, including rejected inputs.
Its finite machine code is fixed.

\emph{Alignment without self-reference.}
Let $u=\enc{F}_V$ be the code of $F$ for $V$ given by the simulation property, and put $d=|u|$.
For sufficiently large $n$, set
\[
N=2n+3,
\qquad r=2n+2-2d,
\qquad p(w)=u1^d0w\quad(w\in\zo^r).
\]
These programs all have length $d+(d+1)+r=N$.
The input seen by $F$ has length $L=N-d$ and unary prefix parameter $a=d$, so it computes exactly the intended $n=(L+d-3)/2$ and $r$.
The unary parameter supplies the code length as ordinary input, so the fixed machine $F$ never computes its own code.
Take $n$ large enough that $r\ge n+1$.
Then every target $x\in\zo^n$ has exactly
\[
k=2^{r-n}\ge2
\]
mandatory descriptions in the cylinder of $u1^d0$, namely the $p(w)$ with $H_{n,r}w=x$.
Put
\[
s=\left\lceil\log_2(r+1)\right\rceil,
\qquad D=\left\lfloor\frac{n}{s}\right\rfloor+1.
\]
For each fixed target $x$, the suffixes $w$ of its mandatory programs form one preimage set of $H_{n,r}$, so by \Cref{lem:balanced-syndrome-map} any two distinct ones differ in at least $D$ coordinates.
The programs share the prefix $u1^d0$, so any two distinct mandatory programs outputting $x$ are at Hamming distance at least $D$.

Let $q_V$ be a nondecreasing polynomial majorant for the simulation overhead in \Cref{def:2.7}.
Because $L\le N$, every mandatory program has finished by
\[
P_V(n)=q_V\left(R(2n+3),2n+3,d\right),
\]
enlarging this expression to an integer-valued nondecreasing polynomial if necessary.
This is a bound on the native transitions of $V$, and it does not assume that the source and the simulator take the same number of steps.

\emph{Counting in the untimed fiber.}
For $x\in\zo^n$, let
\[
S_x=\{p\in\zo^N:V(p,\eps)=x\text{ eventually}\},
\]
and let $G_x$ be the subgraph of $Q_N$ induced by $S_x$, which is the untimed fiber of $x$ at length $N$.
These untimed fibers have pairwise disjoint vertex sets, because a deterministic program has only one normal output.
Apply \Cref{lem:separated-component-count} to them, their $k$ mandatory programs per target, and the distance $D$ above.
Let $\mathcal C_n$ be the targets for which fewer than $k/2$ components meet those programs.
The lemma gives
\[
|\mathcal C_n|<\frac{4\cdot2^N}{Dk}.
\]

Since $k=2^{r-n}$ and $N-r=2d+1$, division by $2^n$ yields
\[
\frac{|\mathcal C_n|}{2^n}
<\frac{4\cdot2^{N-r}}{D}
=\frac{2^{2d+3}}{D}
<2^{2d+3}\frac{s}{n}.
\]
Here $D=\lfloor n/s\rfloor+1>n/s$ and $n,s$ are positive.
As $r\le2n+2$, we have
\[
s\le\log_2(2n+3)+1
\le\log_2(n+1)+\log_2 3+1
\le4\log_2(n+1),
\]
where the last inequality uses $\log_2(n+1)\ge1$.
Thus $a_V=2^{2d+5}$ gives the stated size bound for $\mathcal B_n=\zo^n\setminus\mathcal C_n$.
Put $C_V=2d$ and choose $n_0\ge\max\{1,2d+1\}$, so $r\ge n+1$ throughout.
For every $x\in\mathcal B_n$, at least $k/2$ untimed components meet its mandatory programs, and
\[
\frac{k}{2}=2^{n+1-2d}\ge2^{n-C_V}\ge2.
\]
This proves the untimed assertion and its nonemptiness and disconnection consequences.

\emph{Every later deadline.}
Fix $x\in\mathcal B_n$ and $t\ge P_V(n)$.
Every untimed component just counted contains a mandatory program, which is already a vertex of the fiber at deadline $P_V(n)$.
A path in the time-$t$ fiber would also be a path in $G_x$, so it cannot join vertices from different such untimed components.
The entire time-$t$ fiber therefore has at least $k/2\ge2^{n-C_V}$ components.
The set $\mathcal B_n$ was defined using the untimed fibers alone, so this holds simultaneously for every later deadline.
The count includes paths that change bits of the code $u$, so it never assumes that the cylinder of $u1^d0$ is separated from the other programs of the fiber.
\end{proof}

The theorem does not force disagreement between two machines at later deadlines.
Instead, it shows that the connectedness of the $W_2$ fibers in \Cref{thm:uniform-reference}, asserted at its specified early deadline, does not persist, for most targets, to every later deadline.
For every adjunction-universal machine, and so for $W_2$, most targets at the same program length have disconnected whole fibers once all the mandatory descriptions above have appeared.
Adding feasible vertices can create new components, even though an existing path never disappears when the deadline increases.
This distinguishes entire-fiber connectedness from connectivity of two fixed endpoints.

\paragraph{\emph{Counting the Forced Components.}}

The next family of separated mandatory descriptions, built from Reed--Muller codes, gives an explicit tradeoff between program length, component multiplicity and the exceptional target fraction.
We use classical Reed--Muller evaluation codes; see, for example, \cite[Section~2.3 of the full version]{ASW15}.
The needed distance bound and the complete machine construction are proved below.

A multilinear polynomial has exponent at most one for each variable in every monomial.
For integers $0\le j<k$, the coefficient vector of a multilinear polynomial in $k$ variables of degree at most $j$ lists, in a fixed order, its coefficients on all monomials of degree at most $j$.
Its evaluation vector lists its values at all binary inputs, in one fixed order.
The Reed--Muller code of degree $j$ consists of the evaluation vectors, of length $2^k$, of all multilinear polynomials in $k$ variables of degree at most $j$ over $\mathbb F_2$.
There are $\sum_{i=0}^j\binom{k}{i}$ monomials of degree at most $j$, which is the number $b$ of coefficients below.

\begin{theorem}[Many components under adjunction; previewed in \Cref{thm:preview-fragmentation}]
\label{thm:component-multiplicity}
Let $V$ be an adjunction-universal machine.
There are a constant $d\ge0$ and a nondecreasing polynomial $P_V$ with the following property.
For any integers $k\ge1$ and $0\le j<k$, put
\[
m=2^k,\qquad b=\sum_{i=0}^j\binom{k}{i},\qquad
D=2^{k-j},\qquad N=m+b+d.
\]
Except for at most the fraction $\min\{1,4\cdot2^d/D\}$ of the $m$-bit targets $x$, the untimed fiber of $x$ at length $N$, induced by
\[
S_{x,N}=\{p\in\zo^N:V(p,\eps)=x\},
\]
has at least $2^{b-1}$ connected components.
The same targets have at least that many components in $\Gland^t_{=N,V}(x)$ for every $t\ge P_V(m)$.
One constant $d$ and one polynomial work for all the displayed pairs $(k,j)$.
\end{theorem}

\begin{proof}
For a coefficient vector $w\in\zo^b$, let $E_{k,j}(w)$ be the length-$m$ evaluation vector of its multilinear polynomial of degree at most $j$ over $\mathbb F_2$, in a fixed ordering of the monomials and of the $k$-bit evaluation points.
Distinct coefficient vectors give evaluation vectors at distance at least $D$.
Evaluation at a point is linear over $\mathbb F_2$ in the coefficients, so $E_{k,j}(w)\oplus E_{k,j}(w')=E_{k,j}(w\oplus w')$, and the distance between the evaluation vectors of $w\ne w'$ is the weight of $E_{k,j}(g)$ for the nonzero coefficient vector $g=w\oplus w'$.
Write the polynomial of $g$ as $\sum_{B}g_BX_B$, the sum over the sets $B$ of at most $j$ variables, where $X_B$ is the product of the variables in $B$, with $X_\emptyset=1$.
Choose a set $A$ with $g_A=1$ and $|A|=a$ maximal among the sets with nonzero coefficient, so $a\le j$.

Fix values $c$ of the $k-a$ variables outside $A$, and sum the polynomial over the $2^a$ evaluation points that agree with $c$ outside $A$.
Write $y_i$ for the value such a point gives to a variable $i\in A$.
For one monomial $X_B$, its value at such a point is the product of the fixed values $c_i$ for $i\in B\setminus A$ times the product of the $y_i$ for $i\in B\cap A$.
The second product is one at exactly $2^{a-|B\cap A|}$ of these points, namely those with $y_i=1$ for every $i\in B\cap A$, so the sum of $X_B$ over the points is that first product times $2^{a-|B\cap A|}$, reduced modulo two.
If $B$ does not contain $A$, then $|B\cap A|<a$, the count is even, and the sum is zero.
If $B$ contains $A$ and has nonzero coefficient, then $|B|\le a$ by the maximality of $a$, so $B=A$, and the sum is $2^{0}=1$ times the empty product, which is one.

Summing over $B$ gives that the sum of the polynomial over the $2^a$ points equals $g_A=1$.
Hence at least one of these points has evaluation one.
The $2^{k-a}$ choices of $c$ give disjoint sets of points, so the evaluation vector has weight at least $2^{k-a}\ge2^{k-j}=D$.
This also covers a nonzero constant polynomial, for which $A$ is empty and the one point is $c$ itself.

Define one fixed polynomial-time machine $F$.
It first reads its entire input and obtains its length $L$.
Reject if $L<3$.
If $L=m+b$ for a pair $(k,j)$ as displayed in the theorem, the inequalities $1\le b<m$ give
\[
2^k<L<2^{k+1},
\]
so $k=\lfloor\log_2L\rfloor$ is determined by $L$.
Compute the binomial prefix sums and find the unique $j<k$ for which $b=L-2^k$ is the displayed sum; reject if none exists or $k<1$.
The sums strictly increase with $j$.
Parse the input as $w a$, with $|w|=b$ and $|a|=m$, and output
\[
a\oplus E_{k,j}(w),
\]
where $\oplus$ is coordinatewise addition modulo two.
There are at most $m$ monomials and $m$ evaluation points.
Enumerating the monomials as subsets of the $k$ variables and evaluating each of them at every point takes $O(mbk)$ elementary operations, hence polynomial time on a fixed multitape implementation.
The binomial calculations use $O(k)$-bit integers.
Consequently, for fixed constants $A_0,c_0\ge1$, one bound $A_0(L+1)^{c_0}$ covers the native running time of $F$ at every input, including rejected lengths.
The pair $(k,j)$ is read off the input length, so the code of $F$ is the same for every pair.

Let $h=\enc{F}_V$ be the code of this one $F$ for $V$, and put $d=|h|$.
For a fixed pair $(k,j)$ as displayed and each $x\in\zo^m$, all the programs
\[
\mathcal P_x=
\{h w (x\oplus E_{k,j}(w)):w\in\zo^b\}
\]
print $x$.
Each has length $N$, and there are $W=2^b$ of them, one for each $w$.
Any two distinct programs in $\mathcal P_x$ are at Hamming distance at least $D$, already in their last $m$ positions.
By the polynomial simulation bound and $L<2m$, all these programs, for every degree and target, halt by one nondecreasing polynomial deadline $P_V(m)$.

Count components in the untimed fiber induced by $S_{x,N}$, including paths that change bits of the code $h$.
For this fixed pair $(k,j)$, the graphs for distinct $m$-bit targets have disjoint vertex sets, because $V$ is single-valued.
Their respective sets $\mathcal P_x$ each contain $W=2^b$ specified vertices, separated by at least $D$.
Apply \Cref{lem:separated-component-count}.
Fewer than $4\cdot2^N/(DW)$ targets have fewer than $W/2$ components meeting their specified programs.
Dividing by the $2^m$ possible targets gives
\[
\frac{4\cdot2^N}{DW2^m}
=\frac{4\cdot2^d}{D},
\]
since $N=m+b+d$ and $W=2^b$.
The exceptional fraction is therefore at most $\min\{1,4\cdot2^d/D\}$.
Every remaining untimed fiber has at least $W/2=2^{b-1}$ components meeting $\mathcal P_x$.

All the specified programs have halted by $P_V(m)$.
At every $t\ge P_V(m)$, each of those untimed components consequently has a vertex in the time-$t$ fiber.
A path in the time-$t$ fiber is a path in the untimed fiber, so different counted components cannot be joined at that deadline.
The same targets therefore have at least $2^{b-1}$ components at every later deadline.
The constant code and polynomial deadline still work for every displayed pair $(k,j)$, as established by the single source above.
\end{proof}

Taking the evaluation degree near half the number of variables turns the preceding component count into an exponential one.

\begin{corollary}[Exponential component multiplicity; previewed in \Cref{thm:preview-fragmentation}]
\label{cor:exponential-components}
For every adjunction-universal machine $V$, there are $d$ and $P_V$ such that for each $m=2^k$ with $k\ge2$, some length $N<2m+d$ has the following property.
Except for a fraction at most $\min\{1,4\cdot2^d/\sqrt m\}$ of the $m$-bit targets $x$, the untimed fiber induced by $S_{x,N}$ has at least $2^{m/2-1}$ components, and so does the fiber $\Gland^t_{=N,V}(x)$ for every $t\ge P_V(m)$.
\end{corollary}

\begin{proof}
Take $j=\lfloor k/2\rfloor$ in \Cref{thm:component-multiplicity}.
By binomial symmetry, $b=m/2$ for odd $k$, and $b=(m+\binom{k}{k/2})/2$ for even $k$.
Thus $b\ge m/2$, $b<m$ and $D=2^{\lceil k/2\rceil}\ge\sqrt m$, giving all the assertions.
\end{proof}

For comparison, take degree $j=2$ with $k\ge3$.
Then $b=1+k+\binom{k}{2}$ and $N=m+O_V(\log^2m)$, and all but $O_V(1/m)$ of the targets have at least $2^{b-1}=m^{(\log_2m+1)/2}$ components.
In the exponential case, any labeling that gives distinct components distinct labels needs at least $2^{m/2-1}$ labels, hence at least $m/2-1$ bits, which is linear in $N$; a labeling with only polynomially many possible values in $N$ is therefore impossible.
This counting bound does not exclude a polynomial-time component test, since such a test may use polynomially many bits.
\Cref{thm:component-multiplicity} supplies no computational hardness reduction.

\section{Eventual Connectivity and Uniform Hardness}\label{sec:eventual-target-connectivity}

The fragmentation bounds hold at program lengths chosen from the target length.
We construct a machine satisfying the three reference conditions for which every fixed target has empty or connected fibers at every deadline once programs are sufficiently long.
At shorter lengths, its fibers still make its full connectivity language $\PSPACE$-complete.

Recall the full native fiber $\mathcal F_V(x,N,t)$ of \Cref{def:full-native-fiber}.
This graph includes every program of length $N$, however it begins.
A program is \emph{admitted} at deadline $t$ when it is a vertex of this graph.
The construction rests on a coloring of a cube.
Here a coloring is any map from cube vertices to a set of colors; adjacent vertices may have the same color.
The class of a color is its preimage, and it is connected when the subgraph induced by that class is connected.
The construction needs a coloring of the cube in which every color class is connected and every vertex has a neighbor of every color.
\Cref{lem:connected-syndrome-coloring} gives one, by modifying the Hamming-syndrome coloring of \cite{CKTV15}; we prove it in full because both the connectedness of its classes and the neighbor of every color are used below.

\begin{lemma}[A connected coloring with a neighbor of every color]
\label{lem:connected-syndrome-coloring}
Let $n\ge3$, let $K=2^a$ be the largest power of two with $2K-1\le n$, and put $k=K-1$.
Let $H$ be the binary $a\times k$ matrix whose columns are the nonzero vectors of $\mathbb F_2^a$, in increasing numerical order.
Write an $n$-bit string as $p=(u,v,b,w)$, with $|u|=|v|=k$ and $|b|=1$, and define
\[
D_n(p)=(1-b)Hu+bHv.
\]
The colors are the vectors of $\mathbb F_2^a$.
Every color class is nonempty and connected.
For every $p$ and every color $j$, some one-bit neighbor $q$ satisfies $D_n(q)=j$.
Each vertex of color $j$ is joined to a fixed representative of that color by a path of at most $n+1$ flips inside its color class.
The color $D_n(p)$ is computable in polynomial time by a routine whose head movements and number of transitions depend only on $n$.
\end{lemma}

\begin{proof}
Call $Hu$ and $Hv$ the syndromes of the registers $u$ and $v$, and call $b$ the selector.
The color is the syndrome of the selected register ($u$ if $b=0$, $v$ if $b=1$), and the other register is ignored.
Vector addition is coordinatewise modulo two.
The columns of $H$ include all unit vectors.
Choose $r_j$ using those unit columns so that $Hr_j=j$.
Start with selector zero and syndrome $Hu=j$.
Change the ignored register $v$ to $r_j$, then flip the selector to one.
Change the now-ignored $u$ to $r_j$, flip the selector back, and clear $w$.
Every intermediate vertex has color $j$.
This reaches $(r_j,r_j,0,0^{|w|})$ in at most
\[
2k+2+|w|=n+1
\]
flips.
From selector one, first change $u$ to $r_j$ and flip the selector to zero.
Then change $v$ to $r_j$ and clear $w$.
At most $n$ flips suffice.
These paths prove connectedness and the representative bound.
They also exhibit a vertex of every color.

For the neighbor property, let $y$ be the currently selected syndrome.
If $y\ne j$, flip the coordinate of the selected register whose column is $y+j$.
That column exists and the new syndrome is $j$.
If $y=j$, flip a bit of the ignored register.
Such a bit exists because $k\ge1$.
Thus the neighbor is distinct from $p$ also in this case.
Every chosen neighbor has the opposite parity to $p$.

For the computation, scan both registers and compute both syndromes with fixed-width $a$-bit loops.
At position $i$, the column is the $a$-bit representation of $i$.
Use fixed-duration bit operations, with dummy operations when an input bit is zero.
The counters and scan lengths depend only on $n$ and the current position.
A fixed-width multiplexing loop reads both syndromes and the selector, following the same head schedule for either selector value.
All these are ordinary polynomial-time multitape routines; their stored bits vary with $p$, but their durations and head movements do not.
\end{proof}

For each $n\ge3$, take $a$ and $K(n)=2^a$ as in \Cref{lem:connected-syndrome-coloring} for this $n$.
In the machine below, $n$ is the length of a program's \emph{body}, the part after its two leading mode bits.
The machine prints each color as a binary string, its \emph{label}, defined next.
For a color vector $j$, let $i$ be its integer value and let $\lambda_a(j)$ be the binary representation of $i+1$ with its leading one removed.
Thus increasing color integers give the strings ordered first by length and then lexicographically (shortlex order), beginning $\eps,0,1,00,\ldots$.
The palette $\mathcal P_n$ consists of the first $K(n)$ such strings.
For $n<3$ put $\mathcal P_n=\emptyset$, since at such a body length the machine below prints no color.
The label map is injective at fixed $n$, and its outputs have length at most $a$.

\begin{theorem}[Whole fibers become connected at every deadline; previewed in \Cref{thm:preview-eventual}]
\label{thm:eventual-connected-fibers}
There is an efficient onto machine $V$ with the simulation property such that, for every target $x$, one can compute a threshold
\[
N_x\le2^{|x|+2}+1
\]
for which $\mathcal F_V(x,N,t)$ is empty or connected for every $N\ge N_x$ and every native deadline $t$.
Every nonempty such graph has diameter at most $2N+4$.
The interpretation property can be added to $V$ without changing any fiber $\mathcal F_V(x,N,t)$ or any halting time on the empty auxiliary input.
\end{theorem}

\begin{proof}
Fix an efficient onto program-only machine $A$ with the simulation property, which we call the base machine.
Any such machine will do; \Cref{thm:uniform-hard-fixed-easy-reference} takes the program-only realization of the reference machine $U$, which runs $U$ with its auxiliary tape fixed empty.
A program for $V$ of length $N\ge2$ consists of two mode bits followed by a body $p$ of length $n=N-2$.
Write $\pi(p)$ for the parity of its Hamming weight, and call $p$ even or odd accordingly.
For $n\ge3$, keep the following outputs:
\[
\begin{array}{c|cc}
\text{mode}&\pi(p)=0&\pi(p)=1\\
\hline
00&A(p)&A(p)\\
01&A(p)&\lambda_a(D_n(p))\\
10&\lambda_a(D_n(p))&A(p)\\
11&\lambda_a(D_n(p))&\lambda_a(D_n(p)).
\end{array}
\]
Programs in mode $00$ are base programs, and programs in mode $11$ are core programs.
Programs in modes $01$ and $10$ are helpers.
A helper in mode $01$ or $10$ is a source when it runs $A$, and a router when it prints a color label.
If $n<3$, only the base runs $A$ and every other mode enters the distinguished halting state.
Programs shorter than two bits also reject.
These rules cover every program, including arbitrary bodies that are not valid reference codes.

\emph{One common color clock.}
For $n\ge3$, the common front end reads the mode, scans the body, records $n$ and its parity, and computes both syndromes and the selected color with the length-dependent schedule of \Cref{lem:connected-syndrome-coloring}.
During these scans it also copies the body onto an initially blank private tape.
It writes exactly the body bits, no delimiter or marker; all later cells remain blank.
A length counter returns that tape's head to cell one in $n$ left moves.

All scans and finite branch tests have equal duration for the different bit values, padding a branch when necessary.
The resulting front-end duration depends only on $n$.
The work tapes and heads reserved for the native copy of $A$, with its fixed initial data and layout, have not been touched, and the real output tape is still blank.

To print a color of integer value $i$, first store $i$ in exactly $a+1$ binary positions, with a leading zero.
Compute $i+1$ by a full fixed-width increment scan, padding the steps after the carry disappears as well, and reset the head in a fixed $a+1$ moves.
Thus this addition also has duration depending only on $a$.

Scan those positions from the most significant end, skipping leading zeros and the first one, and print each subsequent bit.
A one-bit flag records whether the first one has been passed.
Use the same fixed number of transitions at every position: before that first one, the output head stays still and no output is written; afterwards the bit is written and the output head moves right.
The output tape is never read, so these different output actions do not alter the loop duration.
Pad the finite flag cases to that common duration.
There are exactly $a+1$ iterations, including when the resulting label is $\eps$.

Thus every core and router program, which we together call the \emph{color programs}, of every color at this body length, normally halts at one common time $C(n)$.
That time has a fixed polynomial upper bound in $n$.
Define these color routines first, and then choose a fixed nondecreasing integer polynomial $B(n)$ with $B(n)>C(n)$ for every $n\ge3$.
The color routines do not use $B$, so this choice is not circular.

\emph{One shared native simulation.}
Every program that runs $A$, whether a base program or a source helper, enters the same prelude, ignoring its old mode and parity flags.
Compute $J=B(n)$ in binary on a fresh tape and decrement the counter to zero.
Fixed-width scans give exactly $J$ iterations, each taking at least one and at most $O(\log_2(J+2))$ transitions, including its zero test and head reset.
The delay is therefore at least $B(n)$ and has polynomial upper cost in $n$.
For the finitely many lengths $n<3$, use the same body-copy preparation and any fixed additional delay on the base.

After preparation, enter a finite-control copy of the fixed native machine $A$.
Its program input is the private tape containing exactly $p$, with its head at cell one; its work data, layout and work heads are the original initial ones, and its output is the untouched real output tape.
Although the private input tape is writable in $V$, the copied input rules never write it.
Its physical first cell implements the original left-end rule, including for an empty body.

Finite tape and state renamings reproduce every native transition of $A$ in one transition, with the same writes, erasures and head movements.
The copy's states are disjoint from the front-end states, so a return to its old initial state does not repeat the preparation.
No final output copy or test of transient writes is performed: a normal halt delivers precisely the final output configuration of $A$.
Distinguished halts remain distinguished, and divergence remains divergence.

Write $S(n)$ for the complete prelude duration before that native copy starts.
It depends only on $n$ and has one fixed nondecreasing integer-polynomial upper bound $P(n)$, including body copying, syndrome computation, counter evaluation and the delay.
For $n\ge3$ it satisfies $S(n)\ge B(n)>C(n)$.
If $A(p)$ normally halts after $u$ transitions, the base $00p$ and its \emph{matching source helper}, which is $01p$ for even $p$ and $10p$ for odd $p$, both halt at exactly
\[
T_{\mathrm b}(p)=T_{\mathrm s}(p)=S(n)+u.
\]
They start the same copy from identical relevant configurations at the same time; their original program tapes differ only in the ignored mode bits.
This equality includes an initial normal halt, with $u=0$ and output $\eps$, if such a halt occurs.
No count of $u$ is needed in the construction.

\emph{Every deadline.}
By the threshold computation below, $N\ge N_x$ forces $x\in\mathcal P_{N-2}$, so palette targets are the only case.
Fix $n\ge3$ and $x\in\mathcal P_n$.
Before $C(n)$ its whole fiber is empty: all color programs halt at $C(n)$, and every normal simulation halt occurs after $C(n)$.
At and after $C(n)$, the fiber contains every core program $11q$ whose color has label $x$; these form a copy of one color class of \Cref{lem:connected-syndrome-coloring} and are therefore connected.
Every router $01q$ or $10q$ in the fiber differs in one mode bit from the core program $11q$, which prints the same label and is present from $C(n)$ on.

An admitted source $01p$ has even $p$ and $A(p)=x$.
The coloring lemma supplies a one-bit neighbor $q$ with label $x$; its parity is odd, so $01q$ is a router.
Hence
\[
01p\rightarrow01q\rightarrow11q
\]
is an admitted target-$x$ path: its last two vertices already halted at $C(n)$.
An admitted source $10p$ instead has odd $p$.
Its color-$x$ neighbor $q$ has even parity, so $10q$ is a router, and the admitted path is
\[
10p\rightarrow10q\rightarrow11q.
\]
An admitted base $00p$ has its matching source helper admitted at exactly the same time: it is $01p$ for even $p$ and $10p$ for odd $p$.
That first mode flip followed by the relevant two-edge path attaches the base to the core.
All cases preserve the original target, program length and deadline.
They cover every noncore vertex, so the entire fiber is connected at every deadline from $C(n)$ onward.

Each vertex attaches to the core in at most three flips.
Every core vertex reaches its fixed representative in at most $n+1$ flips, all within the complete core already present at that deadline.
Thus two admitted vertices have a connecting walk of length at most
\[
3+(n+1)+(n+1)+3=2n+8=2N+4.
\]
Removing repeated portions cannot increase that length, giving the diameter bound.

\emph{Threshold and reference properties.}
The shortlex rank of $x$, which is the integer value of the color labeled $x$, is $i(x)=2^{|x|}-1+\operatorname{val}(x)$.
Let $K_x$ be the least power of two at least $\max\{2,i(x)+1\}$.
For $|x|\ge1$,
\[
2^{|x|}\le i(x)+1=2^{|x|}+\operatorname{val}(x)\le2^{|x|+1}-1,
\]
so $K_x\le2^{|x|+1}$.
For $x=\eps$, $i(x)=0$ and $K_x=2$, giving the same bound.
Put $N_x=2K_x+1\le2^{|x|+2}+1$; in particular $N_\eps=5$.
If $N\ge N_x$, then $n=N-2\ge2K_x-1\ge3$, so $K(n)\ge K_x>i(x)$ and the actual palette contains $x$.
All these integer operations are computable from $x$.

For adjunction use $\enc M_V=00\enc M_A$.
On every raw suffix $z$, the copied body is exactly $\enc M_Az$, with $n=|\enc M_A|+|z|$.
If $M(z)$ halts in $T$ steps, let $q_A$ be the nondecreasing uniform simulation polynomial of $A$ and let $h'=|\enc M_V|$.
The complete native runtime of $V$ is at most
\[
P(h'+|z|)+q_A(T,|z|,h'),
\]
because $|\enc M_A|\le h'$ and the native copy adds no slowdown after preparation.
This is one polynomial in the required three arguments, independent of $M$.
Prepending $00$ also preserves the polynomial computability and length bounds of the code map.
The base gives onto-ness, and efficiency follows from the fixed native implementation.
Finally \Cref{lem:interpretation-completion} adds interpretation while preserving every unconditional output predicate at its original deadline and retaining this code map.
\end{proof}

\Cref{fig:admission-attachments} summarizes the attachment path and the native-time comparisons used in the proof.

\begin{figure}[htbp]
\centering
For $n\ge3$, let $p$ be even and $A(p)=x\in\mathcal P_n$.
Choose a one-bit neighbor $q$ with $\lambda_a(D_n(q))=x$.
Write $T_{\mathrm s}(p)$ and $T_{\mathrm b}(p)$ for the normal halt times of the matching source helper and base.
\[
\begin{array}{ccccccc}
00p&\rightarrow&01p&\rightarrow&01q&\rightarrow&11q\\
\text{base}&&\text{source helper}&&\text{router}&&\text{core}\\
T_{\mathrm b}(p)&&T_{\mathrm s}(p)&&C(n)&&C(n)
\end{array}
\]
\[
C(n)<B(n)\le T_{\mathrm s}(p)=T_{\mathrm b}(p).
\]
\caption{Attachment inside the palette.
Arrows mark the order of attachment along undirected one-bit edges; each next vertex has halted no later than the preceding one.
For odd $p$, exchange the helper modes $01$ and $10$.
All color programs of this body length halt at the common time $C(n)$.
The displayed base and its matching source helper appear together after that time.
For a target outside the palette, every admitted source helper is a leaf adjacent to its simultaneously admitted base.
This is the leaf structure used in \Cref{thm:uniform-hard-fixed-easy-reference}.}
\label{fig:admission-attachments}
\end{figure}

The threshold may depend on the target without preventing a polynomial algorithm for each fixed target.
The explicit threshold also gives one uniform algorithm with a large target-dependent factor.

\begin{corollary}[Every fixed target is polynomial-time decidable; previewed in \Cref{thm:preview-eventual}]
\label{cor:reference-fixed-target-easy}
Let $V$ be an efficient machine such that each target $x$ has a length $N_x$ for which $\mathcal F_V(x,N,t)$ is empty or connected for every $N\ge N_x$ and every deadline $t$.
Then $\mathrm{Conn}_x(V)\in\Ptime$ for every $x$.
For the machine of \Cref{thm:eventual-connected-fibers}, the full language has a uniform algorithm running in
\[
2^{2^{O(|x|)}}\operatorname{poly}(\text{query length}).
\]
\end{corollary}

\begin{proof}
Fix $x$ and hardwire its finite threshold $N_x$.
First check syntax, equal endpoint lengths and feasibility of both endpoints by bounded simulation at the supplied unary deadline.
Reject infeasible queries.
For a common length $N\ge N_x$, feasibility implies connectivity.
For $N<N_x$, enumerate the finite cube of at most $2^{N_x}$ programs, simulate each at the given deadline, and search its induced graph.
The number of vertices is a constant for this fixed target, while bounded simulation is polynomial in query length.
No uniformly computable threshold is required for this fixed-target assertion.

For the explicit construction, compute $N_x$ using integers with $O(|x|)$ bits.
A cube search below that threshold costs $2^{O(N_x)}$ times a polynomial in query length.
The threshold bound in \Cref{thm:eventual-connected-fibers} gives the stated uniform running time.
It does not give a polynomial bound independent of $x$.
\end{proof}

Uniform hardness survives when the base machine $A$ is the reference machine $U$ of \Cref{def:2.10}.
The reduction adapts the proof of \Cref{cor:7.18}, but the argument that the encoded solution graph is a union of fiber components must now account for the helper programs.
For a target outside the palette, each source helper in the fiber is a leaf attached to its base program (\Cref{fig:admission-attachments}), so the helpers join no two components of the subgraph induced by the base programs.
We call these components base components.

\begin{theorem}[Uniform hardness with every fixed target easy; previewed in \Cref{thm:preview-eventual}]
\label{thm:uniform-hard-fixed-easy-reference}
The machine in \Cref{thm:eventual-connected-fibers} can be chosen so that its full connectivity language is $\PSPACE$-complete while $\mathrm{Conn}_x(V)\in\Ptime$ for every fixed target $x$.
It satisfies all three reference conditions and retains the length thresholds and diameter bounds of that theorem.
The hardness reduction from an input of length $L$ in \Cref{fact:2.27} uses a target of length $O(\log(L+2))$, while its entire query, including the unary deadline, has polynomial length.
\end{theorem}

\begin{proof}
Take the base machine $A$ to be the program-only realization of $U$, which runs $U$ with its auxiliary tape fixed empty.
Fix a body length $m$ and a target $y\notin\mathcal P_m$; in this proof $m$ denotes a body length, and $n$ is reserved for the number of formula variables.
Let $Q_y$ be the length-$m$ bodies for which $U(p)$ eventually normally outputs $y$.
Write $\alpha(p)=01$ for even $p$ and $10$ for odd $p$.
For $m\ge3$, the untimed fiber of $y$ at program length $m+2$ has exactly the vertices
\[
\{00p:p\in Q_y\}\cup\{\alpha(p)p:p\in Q_y\}.
\]
Every router and core prints a palette label and hence cannot print $y$.
Two base vertices are adjacent exactly when their bodies are neighbors.
Within this fiber, each helper is adjacent only to its base.
Flipping a body bit of a helper reverses the parity, so the result is a router.
The two helper modes differ in two bits, so no helper is adjacent to another helper with the same body.
Flipping a mode bit of $\alpha(p)p$ gives either the base $00p$ or the core program $11p$.

The shared native simulation in \Cref{thm:eventual-connected-fibers} makes each base and its matching helper halt at exactly the same time.
Thus at every finite deadline each admitted base has its helper admitted as a leaf attached to it, no other helper is admitted, and the helpers join no two base components.
For $m<3$ there are no helpers and the same assertion about base components is immediate.

Reduce from the explicit-endpoint solution-connectivity problem of \Cref{fact:2.27}.
Let $(\varphi,s,t_0)$ be a valid instance on $n\ge1$ variables, of encoding length $L\ge n+|\varphi|$ and $L\ge1$.
For every integer $b\ge1$, put $y_b=1^b$.
Construct a total native verifier $M_{\varphi,b}$ which checks exact input length $n$, prints $y_b$ on satisfying assignments, and prints $\eps$ otherwise.
Its formula, unary length, target and initial layout are explicit data.
The length check reads at most $n+1$ cells, and the formula-evaluation procedure of \Cref{sec:2.4} gives fixed polynomial bounds in $L+b+1$ for construction and all length-$n$ runs.
Put $h_b=\enc{M_{\varphi,b}}$ and $m_b=|h_b|+n$.
By \Cref{prop:2.3}, there are fixed constants $C,e\ge1$ such that
\[
m_b\le C(L+b+1)^e
\]
for every $L,b$ in this construction.
Choose one fixed integer $\kappa>e$ with
\[
3^{\kappa-e}>(C+1)(2\kappa+1)^e.
\]
Such an integer exists because the exponential in $\kappa$ eventually exceeds that fixed polynomial.
Set $u=L+2\ge3$ and choose directly
\[
b=\kappa\lceil\log_2u\rceil.
\]
The ceiling is computable from the binary length of $u$ and a power-of-two test.
Since $\log_2u>1$, this gives
\[
\kappa\log_2u\le b\le \kappa(\log_2u+1)<2\kappa\log_2u,
\qquad L+b+1\le(2\kappa+1)u.
\]
Consequently
\[
m_b+1\le(C+1)(2\kappa+1)^e u^e<u^{\kappa}\le2^b.
\]
The strict inequality uses $u^{\kappa-e}\ge3^{\kappa-e}$ and the choice of $\kappa$.
This calculation accounts for the code's dependence on the target before any code is built.
Only one verifier and one code are constructed, with $b=O(\log(L+2))$ and all their lengths polynomial in $L$.
For $m_b\ge3$, every palette label has length at most $\log_2K(m_b)<b$, because $K(m_b)\le(m_b+1)/2<2^b$.
Hence $y_b$ lies outside the actual palette; for $m_b<3$ that palette is empty.

Abbreviate the chosen code and body length by $h,m$.
Let $T_M(L+b+1)$ be a fixed polynomial bound for the verifier runs on length-$n$ inputs.
The uniform simulation polynomial of $U$, evaluated at $T_M(L+b+1)$, $n$ and $|h|$, gives a polynomial upper bound $u_{\max}$ on the running time of $U(hv)$ for every $v\in\zo^n$.
Let $P(m)$ be the polynomial majorant for the complete shared prelude in \Cref{thm:eventual-connected-fibers}.
Take
\[
\tau=P(m)+u_{\max}.
\]
The native copy adds exactly the underlying $U$ transitions after this prelude.
Thus every satisfying base and its source helper have halted by $\tau$.
This deadline is polynomial in $L$, so $1^\tau$ is polynomial-time writable.
Base programs $00hv$ with unsatisfying $v$ output $\eps\ne y_b$ and are infeasible at this target at every deadline.

Flipping one of the last $n$ bits of $00hv$ keeps the program in the fiber exactly when the flipped assignment also satisfies $\varphi$.
A flip inside the code gives $00h'v$.
By the full rigidity clause of \Cref{prop:2.3}, no codeword of any length prefixes $h'v$, so $U$ diverges on that body.
A mode flip can reach only the source-helper leaf $\alpha(hv)hv$; the other neighbor is a router and cannot print $y_b$.
That leaf has no other target-$y_b$ neighbor by the parity argument above.
The base programs $00hv$ with $v$ satisfying, together with their helper leaves, therefore form a union of components of the entire deadline-$\tau$ fiber.
Deleting the leaves recovers exactly the solution graph of $\varphi$.

Thus $s$ and $t_0$ are connected in the solution graph of $\varphi$ if and only if $00hs$ and $00ht_0$ are connected in $\mathcal F_V(y_b,m+2,\tau)$.
The reduction outputs $(y_b,1^\tau,00hs,00ht_0)$.
Malformed source instances or infeasible source endpoints map to a fixed malformed no-query.

We argue directly instead of applying \Cref{prop:7.17}, whose second hypothesis fails here: a mode-bit flip of $00hv$ can give a feasible helper.
Membership is \Cref{prop:7.9}.
Apply \Cref{lem:interpretation-completion} to retain the full language, all unconditional deadline graphs and the code map while adding interpretation.
The fixed-target conclusion follows from \Cref{cor:reference-fixed-target-easy}.
\end{proof}

The hardness reduction varies its target and uses lengths below that target's eventual threshold.
The construction does not give a machine whose full connectivity language is not $\PSPACE$-hard, and therefore does not settle \Cref{op:reference-full-hardness}.
Nor does it assert that every target has empty or connected fibers at every program length.

The exponential dependence of the threshold on target length cannot be replaced by a uniform subexponential bound.
This follows by retuning the fixed source in the packing argument of \Cref{thm:eventual-fragmentation}.
The resulting obstruction occurs at an exponentially large program length for most targets of every sufficiently large length.
It concerns connectivity of the whole fiber, rather than the complexity of testing its components.

\begin{theorem}[Exponential lengths with permanent fragmentation; previewed in \Cref{thm:preview-eventual}]
\label{thm:exponential-fragmentation-length}
Let $V$ have the simulation property of \Cref{def:2.7}.
There are constants $c_V>0$, $h$ and $m_0$, and a polynomial $Q_V$, with the following property.
For every $m\ge m_0$ there are a length $N_m$ and a set $\mathcal B_m\subseteq\zo^m$ satisfying
\[
2^{c_Vm-2}\le N_m<2^{c_Vm}+h,
\qquad |\mathcal B_m|>\frac{3\cdot2^m}{4}.
\]
For every $x\in\mathcal B_m$ and every native deadline $t\ge Q_V(N_m)$, the entire graph $\mathcal F_V(x,N_m,t)$ is nonempty and disconnected.
The same set $\mathcal B_m$ works for every such deadline.
\end{theorem}

\begin{proof}
We use \Cref{lem:balanced-syndrome-map,lem:separated-terminals}.
For integers $b\ge m\ge1$, the first lemma constructs a rank-$m$ binary matrix $H_{m,b}$ in polynomial time.
Put $s=\lceil\log_2(b+1)\rceil$.
Every output of the linear map $w\mapsto H_{m,b}w$ from $\zo^b$ to $\zo^m$ has exactly $2^{b-m}$ preimages, and distinct preimages have Hamming distance at least
\[
D=\left\lfloor\frac ms\right\rfloor+1.
\]
The second lemma bounds the size of a connected cube subgraph containing separated specified vertices, even when their connecting paths leave the specified set.

\emph{One source before taking its code.}
Define a fixed native machine $F$ as follows.
It scans its entire raw input and parses it at the first zero as $1^a0w$.
Write $b=|w|$.
If there is no zero, or if $b=0$, it enters the distinguished halting state.
Otherwise it computes, in binary,
\[
s=\lceil\log_2(b+1)\rceil,\qquad K_a=2^{2a+4},
\qquad r=b\bmod K_a,\qquad m=K_as+r.
\]
It rejects unless $1\le m\le b-1$.
After this guard, it constructs $H_{m,b}$, multiplies it by $w$ over $\mathbb F_2$, and writes exactly the resulting $m$ bits, including leading zeros, before normally halting.

This is a polynomial-time machine on every raw input.
The binary representation of $K_a$ has $2a+5$ bits; no unary block of that size is allocated.
Computing the other integers and testing the guard takes polynomial time in the input length.
The matrix is constructed only when $m<b$, so its construction, multiplication and fixed-width output also take polynomial time.
Let $R(L)$ be one nondecreasing integer polynomial bounding the complete native runtime on every input of length $L$, including rejected inputs.
There is no hardwired variable data: the work tapes start blank, and the output tape remains blank until the final left-to-right write.
Invalid cases enter the distinguished halting state rather than output $\eps$.

Let $u=\enc{F}_V$ be the code for $V$ of this already defined source.
Let $d=|u|$, put $h=2d+1$, and fix $K=2^{2d+4}$.
For each raw $w$ consider the program
\[
p(w)=u1^d0w.
\]
The source receives precisely the raw suffix $1^d0w$, and thus uses $a=d$ and the fixed constant $K$.
Its full program length is $N=b+h$.
The order of these choices is essential: first define $F$, then obtain $u$, then supply $d$ as an ordinary raw-input parameter.
No parameter-dependent recomputation of the code or circular bound on a code is used.
If $d=0$, the raw prefix is simply $0$, and $h=1$.

Take a nondecreasing polynomial majorant $q_V$ for the simulation overhead in \Cref{def:2.7}.
The source input length is $b+d+1\le N$.
Hence every successful mandatory program $p(w)$ has normally halted by the common native deadline
\[
Q_V(N)=q_V(R(N),N,d),
\]
after integer-polynomial majorization if needed.
This includes all source scans, setup and emission.
It bounds physical transitions of $V$ and is polynomial in program length $N$; it need not be polynomial in the much shorter target length.

\emph{Every sufficiently large target length.}
Given a sufficiently large $m$, write $m=Ks+r$ with $s=\lfloor m/K\rfloor$ and $0\le r<K$.
Let $\delta$ be the least nonnegative residue of $r-2^{s-1}$ modulo $K$, and choose
\[
b=2^{s-1}+\delta.
\]
Take $m_0=K(K+1)$, so $m\ge m_0$ gives $s\ge K+1\ge17$, since $K\ge16$.
For every integer $s\ge7$ we have $2^{s-1}\ge s^2$.
At $s=7$ this is $64\ge49$.
If it holds at $s$, then $2^s\ge2s^2\ge(s+1)^2$, because $2s^2-(s+1)^2=s^2-2s-1\ge0$ for $s\ge7$.
Consequently
\[
2^{s-1}\ge s^2\ge(K+1)s>K(s+1),
\]
where the last inequality follows from $(K+1)s-K(s+1)=s-K>0$.
Then
\[
2^{s-1}\le b\le2^{s-1}+K-1\le2^s-1.
\]
Therefore $\lceil\log_2(b+1)\rceil=s$.
Indeed, $b+1>2^{s-1}$ gives $\lceil\log_2(b+1)\rceil\ge s$, and $b+1\le2^s$ gives $\lceil\log_2(b+1)\rceil\le s$.
Also $b\bmod K=r$, so the source computes exactly the prescribed output length $m$.
Its guard succeeds, since
\[
m\le K(s+1)-1\le b-1.
\]
Put $N_m=b+h$.
For every $m$-bit target there are exactly $k=2^{b-m}\ge2$ mandatory programs outputting it.
For each fixed $m$-bit target, all its mandatory programs are admitted by $Q_V(N_m)$ and have pairwise distances at least
\[
D=\left\lfloor\frac{Ks+r}{s}\right\rfloor+1
=K+\left\lfloor\frac rs\right\rfloor+1\ge K+1.
\]

\emph{The full untimed cube.}
For a target $x$, let $S_{x,N_m}$ be the set of all length-$N_m$ programs eventually normally outputting $x$.
Let $\mathcal C_m$ be the set of $m$-bit targets whose mandatory programs all lie in one component of the full induced graph $Q_{N_m}[S_{x,N_m}]$.
For each such target, that component has at least $kD/2$ vertices by \Cref{lem:separated-terminals}.
The components chosen for different targets are disjoint, since a deterministic program has at most one normal output.
They all lie in the same cube, so
\[
|\mathcal C_m|\frac{kD}{2}\le2^{N_m}.
\]
Using the definitions of $k,N_m,h,K$ gives
\[
\frac{|\mathcal C_m|}{2^m}
\le\frac{2^{h+1}}D
\le\frac{2^{2d+2}}{K+1}
=\frac{K}{4(K+1)}<\frac14.
\]
Thus $\mathcal B_m=\zo^m\setminus\mathcal C_m$ has the required strict three-quarter bound.
This counts the entire untimed fiber.
Its paths may change the code $u$, the unary parameter, the parse or any other coordinate; no isolation of a code cell is assumed.

Fix $x\in\mathcal B_m$ and $t\ge Q_V(N_m)$.
All its mandatory programs are feasible at this deadline, yet at least two lie in different components even in the untimed fiber.
No deadline subgraph can connect them.
The entire fiber is therefore nonempty and disconnected, with the same set $\mathcal B_m$ at every later deadline.
Finally, $m/K=s+r/K<s+1$, so $m/K-2<s-1$ and
\[
N_m\ge b\ge2^{s-1}>2^{m/K-2}.
\]
The upper bound follows from $b\le2^s-1<2^s\le2^{m/K}$:
\[
N_m=b+h<2^{m/K}+h.
\]
Taking $c_V=1/K$ proves the bounds.
\end{proof}

The set $\mathcal B_m$ need not be recognizable.
Its size nevertheless forces a lower bound on any assignment of eventual-connectivity thresholds.

\begin{corollary}[The exponential order of eventual thresholds is necessary; previewed in \Cref{thm:preview-eventual}]
\label{cor:exponential-eventual-threshold}
Suppose $V$ has the simulation property and integer-valued thresholds $N_x$ such that $\mathcal F_V(x,N,t)$ is empty or connected for every $N\ge N_x$ and every deadline $t$.
For every sufficiently large $m$, more than three quarters of the targets $x\in\zo^m$ satisfy
\[
N_x\ge2^{c_Vm-2}
\]
for the constant $c_V>0$ of \Cref{thm:exponential-fragmentation-length}.
In particular no uniform upper envelope $N_x\le2^{o(|x|)}$ is possible.
Together with \Cref{thm:eventual-connected-fibers}, exponential growth in target length is therefore necessary for every adjunction-universal machine with such thresholds and is achieved by one machine satisfying the reference conditions.
The exponent constants are not asserted sharp.
\end{corollary}

\begin{proof}
For each target $x\in\mathcal B_m$ in \Cref{thm:exponential-fragmentation-length}, the nonempty disconnected fiber at length $N_m$ forces $N_x>N_m$.
Otherwise that length would be covered by the assumed eventual property at every deadline.
Since the thresholds are integers, $N_x\ge N_m+1$, and the displayed lower bound follows.
A uniform subexponential envelope would have base-two logarithm divided by $m$ tending to zero, contradicting this bound for all sufficiently large $m$.
The exponential upper envelope is supplied by \Cref{thm:eventual-connected-fibers}.
\end{proof}

\subsection{Dependence of Target-Parameterized Complexity}
\label{sec:target-parameter}

For a query of full encoding length $\ell$, take $k=|x|$ as the parameter.
A language is fixed-parameter tractable in $k$ if one algorithm decides it in time $f(k)\ell^c$, for a computable function $f$ and a constant $c$ independent of the query and parameter.
The next corollary compares two machines whose full connectivity languages are both $\PSPACE$-complete; they differ in how the running time can depend on this parameter.

\begin{corollary}[Target-length complexity depends on the machine; previewed in \Cref{thm:preview-parameter}]
\label{cor:target-parameter-reference}
There are two machines satisfying all three reference conditions whose full connectivity languages are $\PSPACE$-complete.
For the machine of \Cref{thm:uniform-hard-fixed-easy-reference}, connectivity is fixed-parameter tractable in target length, with time $2^{2^{O(k)}}\ell^{O(1)}$.
The rigid reference machine $U$ has a $\PSPACE$-complete restriction at target $00$.
Consequently, if $\Ptime\ne\PSPACE$, its connectivity language is not fixed-parameter tractable in target length.
\end{corollary}

\begin{proof}
The first machine has the uniform algorithm in \Cref{cor:reference-fixed-target-easy}, whose polynomial exponent is independent of $x$, and $\PSPACE$-completeness of the full language by \Cref{thm:uniform-hard-fixed-easy-reference}.
The rigid reference satisfies all three properties by \Cref{def:2.10}.
\Cref{cor:7.18}, at the fixed two-bit target $y=00$, gives that $\mathrm{Conn}_{00}(U)$ is $\PSPACE$-complete, and \Cref{prop:7.9} gives membership of the full language.
If an algorithm for $U$ had time $f(k)\ell^c$, its restriction to $k=2$ would run in polynomial time, making a $\PSPACE$-complete language polynomial-time decidable.
This would imply $\Ptime=\PSPACE$.
\end{proof}

For the first machine, any fixed-parameter algorithm must depend more than single-exponentially on the target length, unless $\Ptime=\PSPACE$.
This holds because the reduction of \Cref{thm:uniform-hard-fixed-easy-reference} uses targets of logarithmic length.

\begin{corollary}[No single-exponential target dependence; previewed in \Cref{thm:preview-parameter}]
\label{cor:no-single-exponential-target}
If $\Ptime\ne\PSPACE$, the machine of \Cref{thm:uniform-hard-fixed-easy-reference} admits no algorithm with running time $2^{O(|x|)}\ell^{O(1)}$ on its full connectivity language $\mathrm{Conn}(V)$.
\end{corollary}

\begin{proof}
For a source input of length $L$, the reduction in \Cref{thm:uniform-hard-fixed-easy-reference} produces a query of length polynomial in $L+2$ and a target of length at most $c_0\log_2(L+2)$, for one fixed constant $c_0$.
An algorithm with running time at most $2^{a|x|+b}\ell^c$, for fixed constants $a,b,c$, would therefore decide the $\PSPACE$-complete source in time polynomial in $L+2$.
This implies $\Ptime=\PSPACE$, proving the conditional exclusion.
It is not a matching double-exponential running-time lower bound.
\end{proof}

\section{Targets with Connected Fibers}
\label{sec:connected-target-sets}

The preceding results distinguish connectivity of a whole fiber from the complexity of deciding connectivity between two of its vertices.
We now ask which targets have connected fibers at every length and every native deadline.
The quantification over deadlines is essential: a disconnected fiber at one deadline is already a counterexample, even if later programs join its components.

For a machine $V$ define
\[
G(V)=\{x\in\zo^*: \mathcal F_V(x,N,t)\text{ is empty or connected for every }N,t\in\mathbb N\}.
\]
Here $\mathcal F_V$ includes every length-$N$ program, and $\mathbb N$ includes zero.
We call the elements of $G(V)$ the \emph{globally connected targets} of $V$, shortened in this section to \emph{connected targets}.
We first prescribe finite sets and then prove effective restrictions on arbitrary $G(V)$.

\subsection{Prescribing Finitely Many Easy Targets}\label{sec:fixed-target-complexity}

We call a target $x$ \emph{easy} for $V$ when its fixed-target connectivity problem $\mathrm{Conn}_x(V)$ lies in $\Ptime$.
For a fixed finite set $S$ of targets, each target of $S$ gets a block of programs, called hubs, that print it whatever their remaining bits are, and these hubs halt before any program that simulates a base machine $A$.
For every other target, the fiber at long enough lengths and late enough deadlines is an exact copy of a fiber of $A$ at a shorter length and a shifted deadline.

\begin{lemma}[Completion on a finite set of targets]
\label{lem:finite-target-completion}
Let $A$ be an efficient onto program-only machine with the simulation property, and let $S=\{x_1,\ldots,x_K\}$ be a fixed finite set of distinct strings.
There is an efficient onto program-only machine $V_S$ with the simulation property and constants $r\ge1$, $B\ge0$, $L\ge1$, having the following properties.
For every $x\in S$, every fiber $\mathcal F_{V_S}(x,N,t)$ is empty or connected, and $\mathrm{Conn}_x(V_S)\in\Ptime$.
For $x\notin S$, the fiber is empty when $N<r$ or $t<B$.
At every other length and deadline, deleting the first $r$ bits is an isomorphism from its entire fiber onto the fiber of $A$ at length $N-r$ and deadline
\[
\left\lfloor\frac{t-B}{L}\right\rfloor.
\]
All constants and the machine may depend on $A,S$.
The finite description of $V_S$ and the constants $r,B,L$ are computable from the finite description of $A$ and the list of the strings of $S$.
\end{lemma}

\begin{proof}
Put $r=\max(1,K)$, and let $e_i$ be the length-$r$ word with a single one in position $i$, for $1\le i\le K$.
Write every program of length at least $r$ as $aw$ with $|a|=r$, and call $a$ its control header.
The machine first reads exactly the $r$ bits of the control header; if a blank is encountered earlier, it enters the distinguished halting state.
At control $0^r$, run $A$ on the unchanged suffix $w$.
The control $e_i$ instead prints $x_i$ independently of $w$.
Every other control enters the distinguished halting state.
For $K=0$ there are no $e_i$ controls, and only control $0$ runs the base machine.
We call the programs with control $0^r$ base programs, and the programs with control $e_i$ the hubs of $x_i$.

We specify the exact halting times, since a hub must be present as soon as a base vertex is admitted; the hub halting time $H$ chosen next is internal to this proof.
Parsing the control header takes a fixed number of transitions on every program with at least $r$ bits.
A hub routine writes its fixed string, then waits until one common normal-halting time $H$.
Choose $H$ large enough for all these finite routines, including one that prints the empty string.
If $K=0$, choose any fixed parsing-compatible $H$.
Choose the base setup and padding time $B$ strictly larger than $H$.

When $V_S$ runs $A$ on the suffix $w$, its program head treats physical cell $r+1$ as the first cell of $A$'s program.
That simulated first cell needs its own left boundary.
After the control parse, the physical program head is at cell $r+1$, which represents cell one of $A$'s program.

A new work tape has a sentinel at its first cell, and its head follows the simulated program head.
At the sentinel, a requested left move keeps both heads fixed, exactly as \Cref{def:2.5} requires at the virtual first cell.
Away from it, both heads move together.
The physical program head can therefore never enter the control prefix.
Disjoint additional tapes handle the parse; the inherited tapes and hardwired data of $A$ have their declared initial contents and head positions when simulation starts.
The sentinel and this fixed setup are included in $B$.
This also simulates the empty suffix at $N=r$, whose first virtual cell is blank.

Because $A$ is fixed, compile its transition table into fixed-length blocks of $L$ native transitions, padding every transition case to the same positive integer $L$.
Perform its output write within the corresponding block and keep the output head stationary on the other transitions of that block.
No final variable-length copy is used.
A normal halt of $A(w)$ after exactly $T$ transitions thus becomes a normal halt of $V_S(0^rw)$ after exactly $B+LT$ transitions.
An initially normal-halting machine is handled at $B$, covering $T=0$.
Distinguished halts remain distinguished, and divergence remains divergence.
This fixes the entire deadline behavior, not merely an upper bound.

If $\enc M_A$ is the code of $M$ for $A$, use $0^r\enc M_A$ for $V_S$.
It runs every raw suffix $z$ as $A(\enc M_Az)$, with no delimiter restriction.
Its length and construction time are polynomial in the ambient code of $M$, since $r$ is fixed with $S$.
If $q_A$ is the uniform simulation bound of $A$ and $\widehat q_A$ is a nondecreasing polynomial majorant of it, then $V_S$ halts within
\[
B+L\widehat q_A(T,|z|,|0^r\enc M_A|),
\]
since $|\enc M_A|\le |0^r\enc M_A|$.
Thus one polynomial works for every source machine and suffix.
Onto-ness follows from the base branch, and efficiency from the fixed native model.

Fix $x_i\in S$ and $N\ge r$.
At deadlines below $H$ the fiber is empty.
At $H\le t<B$ it is exactly the cube $\{e_iw:|w|=N-r\}$.
From deadline $B$ onward it contains that cube and the base vertices $0^rw$ whose computations have already produced $x_i$ and normally halted.
Every such base vertex has the adjacent hub vertex $e_iw$, because the controls differ in one bit.
The cube is connected by flipping its free suffix bits one at a time.
Consequently the whole fiber is connected whenever nonempty, including its zero-dimensional cube when $N=r$.
No hub of another target prints $x_i$, because the listed targets are distinct, and every control other than $0^r$ and the $e_j$ rejects.
For $N<r$ every program rejects.
An arbitrary query for this target is therefore decided by syntax, equal-length and endpoint-feasibility tests, which run in time polynomial in the query length, since the deadline is given in unary.

For a target outside $S$, no hub is feasible.
At $t\ge B$ and $N\ge r$, the exact clock gives the full vertex set
\[
\{0^rw: |w|=N-r, A(w)=x\text{ normally by }
\lfloor(t-B)/L\rfloor\}.
\]
Every feasible program has that control, so no path can change it.
Deleting it preserves and reflects every one-bit edge, giving the stated full-fiber isomorphism.
The earlier deadlines and shorter lengths admit no vertex.

\emph{Computing the construction.}
These finite choices are computable from the given finite description of $A$ and the explicit list of target strings.
Supply that list by its cardinality $K$ and the lengths and bits of its $K$ strings, also allowing $K=0$.
Parse and copy $A$'s finite control, declared work tapes, initial data and layout, and compute $r=\max(1,K)$.
Give the inherited simulation states disjoint names from the control-parse and padding states, so returning to an old initial state cannot restart preparation.
The control tests, suffix-origin sentinel and fixed setup use finitely many prescribed states and transitions.

Synthesize the fixed printers and count their finite execution lengths, including the control parse; padding their maximum gives a computable common $H$.
For $K=0$, choose a parsing-compatible $H$ directly, since there are no printers.
Enumerate the finitely many transition cases of $A$ and synthesize the corresponding finite native blocks, including their boundary tests and output actions.
Their maximum length gives a computable positive $L$, and padding makes all blocks have that length.
A normal or distinguished halt is kept as an internal control flag during that padding; the actual halting state is entered only at the end of its block.
The setup routine has a computable finite length as well.
Choose an integer $B$ exceeding both that length and $H$, and pad the setup to exactly $B$ transitions.

These syntactic operations produce the finite description of $V_S$ and the constants $r,H,B,L$, while retaining the exact clock $B+LT$.
They do not require running $A$ to discover whether it is onto or extracting its code map or global simulation polynomial.
Those properties are used only as the stated promise when verifying the constructed machine's properties.
The conclusion is computability of this constructor; no polynomial bound on its running time is asserted.
\end{proof}

The exceptional finite set may be prescribed while every other fixed target remains hard.
Hardness outside $S$ comes from the rigidity of the reference machine $U$ through \Cref{cor:7.18}, not from the simulation property alone.

\begin{theorem}[Prescribing finitely many easy targets; previewed in \Cref{thm:preview-connected-targets}]
\label{thm:finite-target-reference}
For every fixed finite set $S\subseteq\zo^*$, there is a machine $V$ satisfying the three reference conditions of \Cref{def:2.6,def:2.7,def:2.8} such that:
\begin{itemize}
\item for every $x\in S$, every fiber $\mathcal F_V(x,N,t)$ is empty or connected and $\mathrm{Conn}_x(V)\in\Ptime$;
\item for every fixed target $y$ outside $S$, the language $\mathrm{Conn}_y(V)$ is $\PSPACE$-complete.
\end{itemize}
In particular, $\mathrm{Conn}(V)$ is $\PSPACE$-complete.
If $\Ptime\ne\PSPACE$, the set of targets whose restrictions are in $\Ptime$ is exactly $S$.
The encodings witnessing the simulation and interpretation properties belong to the constructed machine and are not required to be the fixed ambient encoding.
\end{theorem}

\begin{proof}
By \Cref{cor:7.18}, $\mathrm{Conn}_y(U)$ is $\PSPACE$-complete for every fixed target $y$, including $\eps$.
Use the program-only realization of $U$, which runs $U$ with its auxiliary tape fixed empty.
Its rules always read blank on that tape, so they can be compiled without changing any transition count.
Apply \Cref{lem:finite-target-completion} with this base machine $A=U$ and the prescribed $S$.

Fix $y\notin S$.
For a source query in $\mathrm{Conn}_y(U)$, first send malformed encodings or a target field different from $y$ to a fixed unequal-length no-query at target $y$.
Otherwise prefix both endpoints by $0^r$ and replace the deadline $t$ by $B+Lt$.
The lemma gives an exact full-fiber isomorphism at that deadline, so the reduction preserves both answers.
Its output length is polynomial, including the unary deadline, because $B,L,r$ are fixed.
The corresponding restriction of $V_S$ is therefore $\PSPACE$-hard and belongs to $\PSPACE$ by \Cref{prop:7.9}.
The same lemma gives the connected fibers and polynomial-time restrictions for the targets in $S$.

Finally apply \Cref{lem:interpretation-completion} to $V_S$.
It is efficient and onto and has the code map just established, so every hypothesis of that lemma holds.
Its completion retains that code map and every unconditional halting predicate at its original physical deadline, and adds interpretation with the required polynomial encoding and overhead.
Hence all three reference conditions and all target conclusions hold for the same completed machine $V$.
Some target lies outside the finite set $S$, and its $\PSPACE$-complete restriction makes $\mathrm{Conn}(V)$ $\PSPACE$-hard; membership remains \Cref{prop:7.9}.
If $\Ptime\ne\PSPACE$, none of the complete outside restrictions belongs to $\Ptime$, while all restrictions inside $S$ do.
This proves the final exact-set assertion.
\end{proof}

The theorem has the quantifier order: for every fixed finite $S$ there is a machine $V$.
It does not give one machine with all target restrictions easy, nor a machine whose full connectivity language is not $\PSPACE$-hard.
Whether every machine satisfying the three reference conditions has a $\PSPACE$-complete full connectivity language therefore remains open (\Cref{op:reference-full-hardness}).

\subsection{Exact Finite Sets of Connected Targets}
\label{sec:exact-finite-connected-targets}

The construction of \Cref{lem:finite-target-completion} also controls $G(V)$ exactly: every finite set of targets is the set of globally connected targets of some machine satisfying the three reference conditions.
The finite-hub construction realizes each such set exactly, without assuming a separation between complexity classes.

\begin{theorem}[Exact finite connected-target sets; previewed in \Cref{thm:preview-connected-targets}]
\label{thm:exact-finite-connected-targets}
For every fixed finite $S\subseteq\zo^*$ there is a machine $V$ satisfying the three reference conditions of \Cref{def:2.6,def:2.7,def:2.8} with $G(V)=S$.
The assertion includes $S=\emptyset$ and sets containing $\eps$.
\end{theorem}

\begin{proof}
First fix a target $y$ and construct a native machine $E_y$ which normally outputs $y$ exactly on the inputs $00$ and $11$, and diverges on every other input.
An exact length test reads the blank after the second bit.
Write $h_y=\enc{E_y}$ and choose a finite native deadline $D_y$ admitting both successful $U$-computations.
The programs $h_y00,h_y11$ have equal length and Hamming distance two.

Each is isolated in the entire target-$y$ fiber of $U$ at length $|h_y|+2$ and deadline $D_y$.
A suffix flip gives $01$ or $10$, which diverges.
A flip inside $h_y$ gives a program with no codeword prefix of any length by \Cref{prop:2.3}, so the unconditional reference machine diverges on it.
These exhaust every neighbor.
Other programs elsewhere in the cube cannot connect either isolated vertex.
This also holds for $y=\eps$, since divergence produces no normal empty output.

Apply \Cref{lem:finite-target-completion} to the program-only realization of $U$ and to $S$, with constants $r,B,L$.
Every target in $S$ then belongs to $G(V_S)$.

For each $y\notin S$, at length $r+|h_y|+2$ and deadline $B+LD_y$, the lemma identifies the entire target-$y$ fiber with the disconnected reference fiber just constructed.
Hence $y\notin G(V_S)$, and $G(V_S)=S$.

Finally apply \Cref{lem:interpretation-completion}.
It preserves every unconditional fiber at its actual native deadline and retains adjunction, so the completed machine satisfies all three reference conditions and has the same set $G$.
\end{proof}

\subsection{Effective Restrictions on Connected Targets}
\label{sec:effective-connected-targets}

A set is \emph{computably enumerable}, abbreviated c.e., if an algorithm lists its elements, allowing repetitions; it is \emph{co-c.e.} if its complement is c.e.
Fix a standard effective indexing $(W_e)_{e\in\mathbb N}$ of c.e. sets of binary strings.
An infinite set $A$ is \emph{effectively immune} here if a total computable function $b$ satisfies $|W_e|\le b(e)$ whenever $W_e\subseteq A$.
The finite-subset bound below holds whether or not $G(V)$ is infinite.

\begin{theorem}[Packing a c.e. family of connected targets; previewed in \Cref{thm:preview-connected-targets}]
\label{thm:ce-connected-target-packing}
Let $V$ have the simulation property.
There is a total computable map $e\mapsto h_e$ to codes for $V$, with $k_e=|h_e|$, such that, if $W_e\subseteq G(V)$, then for every integer $n\ge2$,
\[
(n+1)\min\{|W_e|,2^{n-1}\}\le(2^{k_e}-1)2^n.
\]
The minimum is understood as $2^{n-1}$ when $W_e$ is infinite.
\end{theorem}

\begin{proof}
From $e$ construct one machine $M_e$.
On a nonempty input $ab$, with first bit $a$, let $i$ be the binary value of $b$ when $a=0$, and of $\overline b$ when $a=1$.
Suppress repeated outputs in the enumeration of $W_e$ and wait for its $(i+1)$-st distinct element.
Print that element and halt if it appears, and diverge otherwise.
On the empty input diverge.
This is one effectively constructible machine for all input lengths.
The computable code map gives $h_e=\enc{M_e}_V$, independent of $n$.

Fix $n\ge2$ and write $L=\min\{|W_e|,2^{n-1}\}$.
Call a rank $i$ available if $W_e$ has at least $i+1$ distinct elements.
For each available rank $i<2^{n-1}$, let $s_i$ be the $(i+1)$-st distinct element of $W_e$ and $b_i$ the $(n-1)$-bit representation of $i$.
Exactly two eventual target-$s_i$ programs lie in the cylinder $h_e\zo^n$: they are $h_e0b_i$ and $h_e1\overline{b_i}$.
A program of the cylinder whose suffix encodes an unavailable rank diverges.
If $L=0$, the asserted bound is immediate.
Otherwise there are finitely many successful endpoint computations, so their actual native halting times have a finite maximum $T$.
No algorithm for computing $T$, or deciding which ranks are available, is needed.

Assume $W_e\subseteq G(V)$.
For each available rank $i<2^{n-1}$, the fiber $\mathcal F_V(s_i,k_e+n,T)$ contains both endpoints and is connected, since $s_i\in G(V)$.
Choose a simple path between its two endpoints.
Since those endpoints are its only vertices in the cylinder, all internal vertices lie outside it.
Their suffixes are complements, so at least $n$ suffix flips occur.
Leaving and returning to a nonempty code requires two more flips, hence at least $n+1$ internal outside vertices.
If $k_e=0$, no outside route exists and $L>0$ is impossible.
Different outputs' paths are disjoint.
Counting the $(2^{k_e}-1)2^n$ outside programs therefore gives the inequality in every case.
\end{proof}

The bound becomes finite and effective after choosing one suffix length as a function of the code.
No efficiency bound is asserted for the resulting computation.

\begin{corollary}[A computable bound on c.e. subsets; previewed in \Cref{thm:preview-connected-targets}]
\label{cor:effective-connected-target-bound}
Put
\[
a(k)=\max\{2,2^{k+1}-2\},\qquad
B(k)=\left\lfloor\frac{(2^k-1)2^{a(k)}}{a(k)+1}\right\rfloor.
\]
For the codes in \Cref{thm:ce-connected-target-packing}, the function $e\mapsto B(k_e)$ is total computable, and $W_e\subseteq G(V)$ implies $|W_e|\le B(k_e)$.
In particular, $G(V)$ contains no infinite c.e. subset.
\end{corollary}

\begin{proof}
For every $k\ge0$ we have $a(k)+1>2(2^k-1)$.
Therefore
\[
B(k)<2^{a(k)-1}.
\]
Apply \Cref{thm:ce-connected-target-packing} at $n=a(k_e)$.
It yields
\[
\min\{|W_e|,2^{a(k_e)-1}\}\le B(k_e)<2^{a(k_e)-1}.
\]
Thus the minimum cannot be saturated; $W_e$ is finite and its cardinality is at most $B(k_e)$.
The conclusion includes $k_e=0$, for which $B(k_e)=0$.
The machine constructor, code map, code length and displayed integer operations are all computable, proving total computability of the bound.
\end{proof}

Globally connected targets have zero density even when sampled in an arbitrary computable order, rather than by output length.

\begin{theorem}[Zero density along computable lists; previewed in \Cref{thm:preview-connected-targets}]
\label{thm:connected-target-list-density}
Let $V$ have the simulation property and let $(s_i)_{i\in\mathbb N}$ be any total computable injective sequence of binary strings.
Put $C(q)=|\{i<q:s_i\in G(V)\}|$.
There is a constant $k$, namely the length of the code for $V$ of a machine built from the list, such that, whenever $n\ge2$ and $2^{n-2}<q\le2^{n-1}$,
\[
\frac{C(q)}q\le\frac{4(2^k-1)}{n+1}.
\]
Consequently $C(q)/q\rightarrow0$ as $q\rightarrow\infty$.
\end{theorem}

\begin{proof}
Use the machine of the proof of \Cref{thm:ce-connected-target-packing}, except that the program of rank $i$ prints $s_i$, computed by the list algorithm, instead of the $(i+1)$-st distinct element of $W_e$.
Its code has one fixed length $k$ across all $n$.
At suffix length $n$ the cylinder holds the targets $s_i$ with $i<2^{n-1}$, and all their computations halt, since the list is total.
Injectivity gives exactly two cylinder programs per such target.
Choose paths only for those targets $s_i$ that belong to $G(V)$, at a common finite deadline admitting all endpoints.
The same outside-cylinder count gives
\[
(n+1)C(2^{n-1})\le(2^k-1)2^n.
\]
For $2^{n-2}<q\le2^{n-1}$, monotonicity gives $C(q)\le C(2^{n-1})$ and $2^n<4q$.
Combining these inequalities proves the bound.
Every integer $q\ge2$ belongs to one such interval, and its corresponding $n$ tends to infinity with $q$.
The right-hand side therefore tends to zero.
\end{proof}

For a set $A\subseteq\mathbb N$, asymptotic density zero means $|A\cap\{0,\ldots,q-1\}|/q\to0$.
A \emph{computable permutation} of $\mathbb N$ is a total computable bijection $\pi:\mathbb N\to\mathbb N$.
The set $A$ has \emph{intrinsic density zero} if $\pi(A)$ has asymptotic density zero for every such permutation.

We recall the injective-sampling characterization of intrinsic density zero.

\begin{fact}[{Computable injective sampling; \cite[Corollary 4.3 of the accepted preprint]{Ast15}}]
\label{fact:intrinsic-zero-sampling}
A set $A\subseteq\mathbb N$ has intrinsic density zero if and only if, for every total computable injection $f:\mathbb N\to\mathbb N$,
\[
\frac{|\{i<q:f(i)\in A\}|}{q}\to0
\qquad(q\to\infty).
\]
\end{fact}

For a computable bijection $\beta:\mathbb N\to\zo^*$, we say that $G(V)$ has intrinsic density zero under $\beta$ when $\beta^{-1}(G(V))$ does.

\begin{corollary}[Intrinsic density zero; previewed in \Cref{thm:preview-connected-targets}]
\label{cor:connected-target-intrinsic-density}
Let $V$ have the simulation property.
For every computable bijection $\beta:\mathbb N\to\zo^*$, the set $\beta^{-1}(G(V))\subseteq\mathbb N$ has intrinsic density zero.
\end{corollary}

\begin{proof}
Fix $\beta$, put $A=\beta^{-1}(G(V))$, and let $\pi:\mathbb N\to\mathbb N$ be any computable permutation.
Its inverse is computable: on input $i$, search $j=0,1,\ldots$ until $\pi(j)=i$; totality makes every test terminate, and bijectivity guarantees exactly one successful value.
Thus $s_i=\beta(\pi^{-1}(i))$ defines a total computable sequence of binary strings, and it is injective because $\pi^{-1}$ and $\beta$ are injective.
For every $j\in\mathbb N$ we have $j\in A$ exactly when $\beta(j)\in G(V)$.
Hence, for every $q\ge1$,
\[
|\pi(A)\cap\{0,\ldots,q-1\}|
=|\{i<q:\pi^{-1}(i)\in A\}|
=|\{i<q:s_i\in G(V)\}|,
\]
where the first equality holds because $i\in\pi(A)$ exactly when $\pi^{-1}(i)\in A$.
By \Cref{thm:connected-target-list-density}, applied to the sequence $(s_i)$, the right-hand side divided by $q$ tends to zero as $q\to\infty$.
So $\pi(A)$ has asymptotic density zero.
Since $\pi$ was an arbitrary computable permutation, $A$ has intrinsic density zero.
\end{proof}

The same conclusion follows from \Cref{fact:intrinsic-zero-sampling}: for a total computable injection $f:\mathbb N\to\mathbb N$, the sequence $s_i=\beta(f(i))$ is total computable and injective, and $f(i)\in\beta^{-1}(G(V))$ exactly when $s_i\in G(V)$, so \Cref{thm:connected-target-list-density} gives the limit that the fact requires.

The rate in \Cref{thm:connected-target-list-density}, which depends on the code length $k$, is the quantitative conclusion of our packing argument.
Effective immunity, concerning c.e. subset cardinalities, is a separate restriction; see \cite[Definitions 1.1--1.2]{Case25} for that established terminology.

A nonempty disconnected fiber is a finite witness that its target lies outside $G(V)$, and enumerating such witnesses shows that $G(V)$ is co-c.e.
Neither this argument nor the preceding packing proves that $G(V)$ is infinite.

\begin{proposition}[Computability of the connected-target set; previewed in \Cref{thm:preview-connected-targets}]
\label{prop:connected-target-coce}
For every native machine $V$, the set $G(V)$ is co-c.e.
If $V$ has the simulation property and $G(V)$ is infinite, then $G(V)$ is effectively immune.
\end{proposition}

\begin{proof}
For any triple $(x,N,t)$, simulate all finitely many length-$N$ programs for $t$ native transitions.
Test which have normally halted with output $x$, excluding distinguished halts, and decide connectivity of their finite induced graph.
These calculations decide whether this triple witnesses a nonempty disconnected fiber.
Enumerating all triples and listing their witnessing targets enumerates exactly the complement of $G(V)$.
The effective-immunity assertion, conditional on infinitude, is \Cref{cor:effective-connected-target-bound} with the computable bound $e\mapsto B(k_e)$.
\end{proof}

\phantomsection
\addcontentsline{toc}{section}{References}
\begingroup
\makeatletter
\let\addcontentsline\@gobblethree
\makeatother
\newcommand{\etalchar}[1]{$^{#1}$}

\endgroup

\end{document}